\documentclass{article} 

\usepackage{amsthm} 

\usepackage[ruled]{algorithm2e}

\SetAlFnt{\small}
\SetAlCapFnt{\small}
\SetAlCapNameFnt{\small}
\SetAlCapHSkip{0pt}
\IncMargin{-\parindent}

\usepackage{epsfig}
\usepackage[hang,nooneline]{subfigure}
\usepackage{amsmath}
\usepackage{amsfonts}
\usepackage{amssymb}
\usepackage{graphicx}
\usepackage{epstopdf}
\usepackage{amsmath}
\usepackage{amssymb}
\usepackage{caption}
\usepackage{float}
\usepackage{epstopdf}
\usepackage{comment}
\usepackage{graphicx} 
\usepackage{booktabs} 
\usepackage{perpage} 
\usepackage{caption}
\usepackage{float}
\usepackage{nameref}
\usepackage{verbatim}
\usepackage{hyperref}

\usepackage{color}

\newcommand*{\mathcolor}{}
\def\mathcolor#1#{\mathcoloraux{#1}}
\newcommand*{\mathcoloraux}[3]{%
  \protect\leavevmode
  \begingroup
    \color#1{#2}#3%
  \endgroup
}

\usepackage[svgnames]{xcolor}

\usepackage{xcolor}
\usepackage{amsmath}

\newcommand{\icol}[1]{
  \left(\begin{smallmatrix}#1\end{smallmatrix}\right)%
}

\newcommand{\irow}[1]{
  \begin{smallmatrix}(#1)\end{smallmatrix}%
}

\newcommand{\iroww}[1]{
  \begin{smallmatrix}#1\end{smallmatrix}%
}

\newenvironment{myenumerate}{
\begin{enumerate}
 \setlength{\itemsep}{1pt}
 \setlength{\parskip}{0pt}
 \setlength{\parsep}{0pt}}{\end{enumerate}
}

\newenvironment{myitemize}{
\begin{itemize}
 \setlength{\itemsep}{1pt}
 \setlength{\parskip}{0pt}
 \setlength{\parsep}{0pt}}{\end{itemize}
}

\theoremstyle{plain}

\theoremstyle{definition}
\newtheorem{defn}{Definition}[section]

\newtheorem{thm}{Theorem}[section]

\newtheorem{cond}[thm]{Condition}
\newtheorem{obs}[thm]{Observation}

\theoremstyle{remark}

\begin{document}


\title{\vspace{-2.5cm}Cross-boundary Behavioural Reprogrammability Reveals Evidence of Pervasive Universality}
\author{J\"{U}RGEN RIEDEL\\
Universit\"{a}t Oldenburg, Germany;\\
Algorithmic Nature Group, LABORES, Paris, France \medskip \\
HECTOR ZENIL\footnote{Co-first author. JR and HZ conceived the project, wrote the code, analyzed the data, and wrote the paper. and data available online at \url{https://github.com/algorithmicnaturelab/reprogrammingcapabilitiesofCAs}}\\
Department of Computer Science, University of Oxford, UK;\\
Algorithmic Dynamics Lab,
Unit of Computational Medicine,\\ Karolinska Institute \& SciLifeLab, Stockholm, Sweden;\\
Algorithmic Nature Group, LABORES, Paris, France}


\date{}

\maketitle

\newtheorem{mytheorem}{Theorem}
\newtheorem{myobservation}{Observation}

\begin{abstract}
We exhaustively explore the reprogrammability capabilities and the intrinsic universality of the Cartesian product $P \times C$ of the space $P$ of all possible computer programs of increasing size and the space $C$ of all possible compilers of increasing length such that $p \in P$ emulates $p^\prime \in P$ with $T|p^\prime|=|p|$ under a coarse-graining transformation $T$. Our approach yields a novel perspective on the complexity, controllability, causality and (re)programmability discrete dynamical systems. We find evidence that the density of (qualitatively different) computer programs that can be reprogrammed grows asymptotically as a function of program and compiler size. To illustrate these findings we show a series of behavioural boundary crossing results, including emulations (for all initial conditions) of Wolfram class 2 Elementary Cellular Automata (ECA) by Class 1 ECA, emulations of Classes 1, 2 and 3 ECA by Class 2 and 3 ECA, and of Classes 1, 2 and 3 by Class 3 ECA, along with results of even greater emulability for general CA (neighbourhood $r=3/2$), including Class 1 CA emulating Classes 2 and 3, and Classes 3 and 4 emulating all other classes (1, 2, 3 and 4). The emulations occur with only a linear overhead and can be considered computationally efficient. We also found that there is no hacking strategy to compress the search space based on compiler profiling in terms of e.g. similarity or complexity, suggesting that no strategy other than exhaustive search is viable. We also introduce emulation networks, derive a topologically-based measure of complexity based upon out- and in-degree connectivity, and establish bridges to fundamental ideas of complexity, universality, causality and dynamical systems.
\end{abstract}

\vspace{5mm}





\section{Introduction}

The concept of \textit{Turing universality} is the most important concept and property both in practice and theory in computer science. However, proofs of universality are very difficult, if not impossible, and little is known about the incidence of computational universality among all `possible' computer programs yet one can find statistical evidence and explore certain aspects in an asymptotic fashion when dealing with uncomputable challenges~\cite{hamkins}

While time complexity results related to the type of emulation reported here may apply only to the particular chosen model of computation, given the parallel computing capabilities of cellular automata, the results on universality and the reprogramming capabilities of computer programs are independent of the choice of model (thanks to the Turing-equivalence of these models~\cite{neumann,smith,durand}). Indeed, for every cellular automaton of size $n$, there is a Turing machine proportional to $n$ that computes the same function~\cite{smith,nks}, so results relevant to cellular automata reported here may be relevant to any set of computing formalisms independently of the choice of (Turing-complete) model.

The chief advantage of using cellular automata (CA) as a model of computer programs is that they reveal their code in a very visual fashion and making cellular automata transparent to immediate inspection and easier grasping.


By finding the right initial condition (a compiler) between any pair of computer programs, we show that even the simplest computer programs are able to emulate the most complex ones (and conversely), mediated only by the specific choice of initial conditions (compiler) at the beginning of the computation effectively reprogramming it to behave like the other computer program for all initial conditions (under the coarse-graining).

We found that when exhaustively exploring larger spaces of compilers, each computer program is, in average, capable of emulating an increasing number of other qualitatively different computer programs in its own rule space, a feature called intrinsic universality strictly stronger than Turing universality.

We find that if $C$ is the (semi-computable) set of initial conditions and $c\in C$ a compiler of size $m$, and $A$ and $B$ the sets of emulating and emulable computer programs respectively (via cellular automata) for fixed program size $n$, then $A$ and $B$ have asymptotic density 1 for increasing compiler size (initial condition) $m$. Thus the number of computer programs that can be reprogrammed is of density 1, suggesting a phenomenon of pervasive intrinsic universality. 

No particular parameter choice plays any particular role in the final results other than the ones that cannot be avoided when dealing with uncomputable problems (e.g. whether 2 programs are actually computing the same function). For the choice of rule spaces, any $k$-state CA can be reduced to 2 states in a larger neighbourhood range rule space, and while asymmetric neighbourhood ranges can introduce some biases in the symmetry classes of equivalence that we used to reduce the number of cellular automata in every space (to reduce over-counting trivial repetitions), the reduction is only accessory and averaging over different (both symmetric and asymmetric) neighbourhood ranges   minimizes the bias. The introduction and concept of a \textit{Wolfram class} (later replaced by a formal definition based on quantitative complexity measures used for cases in which Wolfram classes have not been agreed/estimated before) is only used for the purposes of analyzing the results (the behaviour of programs). The choice of enumeration was minimized by (1) taking a natural choice (increasing length size) and (2) taking a complete rule space for every program size.

\section{Notation, Concepts and Methods}

\begin{defn} A {\it cellular automaton} (CA) is a tuple $\langle S, (\mathbb{L}, +), T, f \rangle$ 
with a set $S$ of states (or colours), a lattice $\mathbb{L}$ with a binary operation $+$, a neighbourhood template $T$, and a local rule $f$.
\end{defn}

The {\it set of states} $S$ is a finite set with elements $s$ taken from a finite alphabet $\sum$ with at least two elements. It is common to take an alphabet composed of all integers modulo $s$: $\sum = \mathbb{Z}_s = \{0,...,s-1\}$.  An element of the lattice $i\in \mathbb{L}$ is called a cell. The lattice $\mathbb{L}$ can have $D$ dimensions and can be either infinite or finite with cyclic boundary conditions.

\begin{defn}
The {\it neighbourhood template} $T=\langle\eta_1,...,\eta_m\rangle$ is a sequence of $\mathbb{L}$. In particular, the neighbourhood of cell $i$ is given by adding the cell $i$ to each element of the template $T$: $T=\langle i+\eta_1,...,i+\eta_m\rangle$. 
Each cell $i$ of the CA is in a particular state $c[i] \in S$. A {\it configuration} of the CA is a function $c: \mathbb{L} \rightarrow S$. The {\it set of all possible configurations} of the CA is defined as $S_\mathbb{L}$.~\cite{Powley2009}
\end{defn}

The {\it evolution of the CA} occurs in discrete time steps $t=0,1,2,...,n$. The transition from a configuration $c_t$ at time $t$ to the configuration $c_{(t+1)}$ at time $t+1$ is induced by applying the local rule $f$. The local rule is to be taken as a function $f: S^{|T|} \rightarrow S$ which maps the states of the neighbourhood cells of time step $t$ in the neighbourhood template $T$ to cell states of the configuration at time step $t+1$:

\begin{equation}
c_{t+1}[i]=f\left(c_t[i+\eta_1],...,c_t[i+\eta_m]\right )
\end{equation}
The general transition from configuration to configuration is called the {\it global map} and is defined as: $F: S^\mathbb{L} \rightarrow S^\mathbb{L}$.

In the following we will consider only 1-dimensional (1-D) CAs. The lattice can be either finite, i.e., $\mathbb{Z}_N$, having the length $N$, or infinite, $\mathbb{Z}$. In the 1-D case it is common to introduce the {\it radius} of the neighbourhood template which can be written as $\langle -r,-r+1,...,r-1,r \rangle$ and has length $2 r+1$ cells. With a given radius $r$ the local rule is a function $f: \mathbb{Z}_{|S|}^{{|S|}^{(2r+1)}} \rightarrow \mathbb{Z}_{|S|}$ with $\mathbb{Z}_{|S|}^{{|S|}^{(2r+1)}}$ rules. Three cases of 1-D CA will be studied further in this paper. We study CA which have states taken from the set $\mathbb{Z}_2$  and have different ranges. We have CA with range $r = 1/2$, which have the neighbourhood template $\langle 0,1 \rangle$, meaning that the neighbourhood comprises the centre cell and one cell to the right. We will call these \textit{Primitive Cellular Automata} (PCA). We have the so called \textit{Elementary Cellular Automata (ECA)} with radius $r=1$ (closest neighbours), having the neighbourhood template $\langle -1,0,1\rangle$, meaning that the neighbourhood comprises a central cell, one cell to the left and one to the right. We also have what we will call \textit{General Cellular Automata} (GCA) as called in~\cite{nks}, with radius $r=3/2$, i.e. they have the neighbourhood template $\langle -1,0,1,2 \rangle$, meaning that the neighbourhood comprises the central cell, one cell to the left and two to the right. The rule space for PCA contains $2^{2^{2}}=16$ rules, the rule space for ECA contains $2^{2^{3}}=256$ rules, and the GCA rule space contains $2^{2^{4}}=65\,536$ rules. Here we only consider non-equivalent rules subject to the operations complementation, reflection, conjugation and joint transformation (reflection and conjugation together) (see Sup. Mat.). For example, the number of reduced rules for ECA is 88. By increasing the range $r$ or the number of states $|S|$ the cardinality of the rule space is increased dramatically.

\begin{defn}
An Elementary Cellular Automaton at time step $t$, $A=(a(t),\{S_{A} \},f_{A})$ is composed of a lattice $a(t)$ of cells that can each assume a value from a finite alphabet ${S_{A}}$. A single cell is referenced as $a_{n}(t)$. The update rule $f_{A}$ for each time step is defined as $f_{A}: \{S^{2^{2 r+1}}\} \rightarrow \{S_{A}\}$ with $a_{n}(t+1)=f_{A}[a_{n-1}(t),a_{n}(t),a_{n+1}(t)]$. The entire lattice gets updated through the operation $f_{A}  a(t)$.~\cite{NavotGoldenfeld2006}
\end{defn}

For the so-called Primitive Cellular Automata (PCA) and General Cellular Automata (GCA) the neighbourhood range changes to $r=1/2$ and $r=3/2$ and the same definition than ECA holds. 

\subsection{Block emulation and Intrinsic universality}

The notion of computational universality used for cellular automata was an adaptation of classical Turing-universality~\cite{neumann}. A stronger notion called \textit{intrinsic universality} was later proposed (see~\cite{ollinger,ollinger2}).

The exploration of the computing capabilities of CA by \textit{block emulation} is a related mechanism by which the scale of space-time diagrams of CA are found and then coarse-grained.

\begin{defn}
Let $A$ and $B$ be two cellular automata. Then $A$ emulates $B$ if there exists a rescaling $P$ of $A$ such that $f^P_{A} = f_{B}$.
\end{defn}

An emulation consists thus in an embedding of the configuration space of the emulated cellular automaton into the configuration space of the emulating cellular automaton.

\begin{defn}
We will define a cellular automaton $A$ in rule space $R$ as intrinsically universal if, for each cellular automaton $B\in R$, there exists a rescaling $P$ (a reverse transformation of the block emulation) such that $f^P_{A} = f_{B}$.
\end{defn}

Effectively, if a cellular automata $A$ with compiler $C$ emulates $B$ according to $C$. Then given only $A$, $B$ and $C$, is it easy to deduce a formal proof that $A$ can fully emulate $B$ according to $C$, i.e. for any initial condition, this is because the CA rule (both local and global) can be deduced (see Sup. Mat. Section~\ref{causal} on Causal Decomposition).

The property of \textit{intrinsic universality} is strictly stronger that Turing-universality; the former implies the latter. Intrinsic universality requires $P$-completeness, i.e. a system is intrinsically universal if it simulates all others in polynomial time. While ECA rule 110 is Turing-universal~\cite{cook,nks}, no ECA is known to be intrinsically universal~\cite{ollinger}.

Here we show that most computer programs (by way of 1-dimensional CA) widen their emulation of other programs' capabilities as a function of compiler and rule space, which amounts to statistical evidence of pervasive intrinsic universality and hence Turing-universality. 

To prove that a particular 1-dimensional cellular automaton is intrinsically universal (and therefore Turing-universal), it is sufficient to prove that it can simulate any other 1-dimensional cellular automaton~\cite{durand}. 

Our emulations are related to an even stronger form of intrinsic universality, namely linear-time intrinsic universality, meaning that all emulations only carry a linear overhead as the result of our brute force exploration of the compiler and rule space. 

Because of the undecidability of the halting problem, there exists no effective algorithm capable of deciding whether a CA is complex in any definite way or Turing-universal~\cite{mazoyer}. One strategy to prove universality is, however, to find a procedure for setting up certain initial conditions in one system that would make that system emulate some other class of systems~\cite{StephenWolfram1984}. 

Following these ideas, one can try out different possible compilers/encodings and see what type of CA these encodings are able to emulate (see Sup. Mat. for more details). It would be highly interesting to find encodings which would allow a CA to emulate a CA which has been proven to be universal, but it would also be interesting to find CA with large reprogrammability capabilities that are not known to be universal that can simulate other CA characterized by qualitatively very different behaviour, thereby making them too natural candidates for intrinsic and Turing-universality. Even if this is impossible to do exhaustively, given that there is an infinite number of possible encodings/compilers for any CA, one can make an assumption of uniformity and derive conclusions from statistical behaviour.

Here we define a \textit{linear} block transformation~\cite{StephenWolfram1983,StephenWolfram1986}, i.e. a scheme where the original CA $A=(a(t) , SA , f_{A})$ is emulated by a CA $B=(b(t) , SB , f_{B})$ through the lattice transformation $b_k=P(a_{N k} ,a_{N k+1}, \ldots ,a_{N k+N-1})$. The rescaling function $P: \{SA\} \rightarrow\{SA\}^N$ projects a single cell of the initial condition of CA A to a block of $N$ cells, which we call a block. 
$P_a$ denotes the block-wise application of $P$ on the entire lattice $a$. 

More generally, one can consider a block transformation $P$ of block size $T$ acting on an initial condition and then running rule $f_A$ for $T t$ time steps. This emulates rule $f_{B}$. In fact, taking only every $T$ time step of the result of running $f_A$ for $T t$ time steps, one gets exactly the output one would if one were to run $f_{B}$ for $t$ time steps on the same initial condition. 

In order for CA $f_A$ and rescaling $P$ to generate a successful emulation (we use `simulation' and `emulation' interchangeably) of CA $f_B$, they must satisfy the commutativity condition:

\begin{equation}
\label{block_emulation_condition}
f^{t}_{B} a(0)=P^{-1} f^{T t}_{A} P a(0)
\end{equation}

The constant $T$ is the time scale of the emulation. In other words, running rule $f_{A}$ for $T$ time steps with the block projected initial condition $a(0)$ and projecting the output back with $P^{-1}$ is identical to running the emulated rule $f_{B}$ for $t$ time steps. 

\begin{figure}[htpb!]
\centering  \includegraphics[width=15cm]{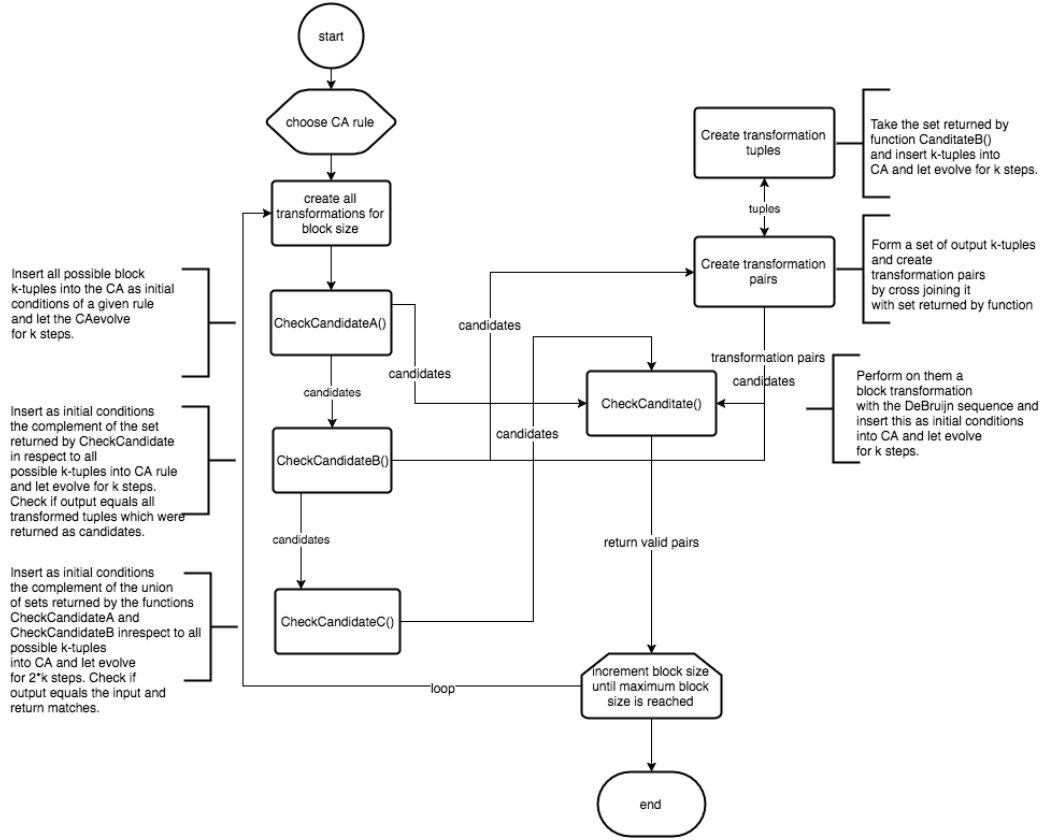}
\caption{
\label{fig_algorithm}Flow diagram of the exploration of the program (rule) space and compiler spaces.
}
\end{figure}

\begin{figure}[htpb!]
\centering  \includegraphics[width=13cm]{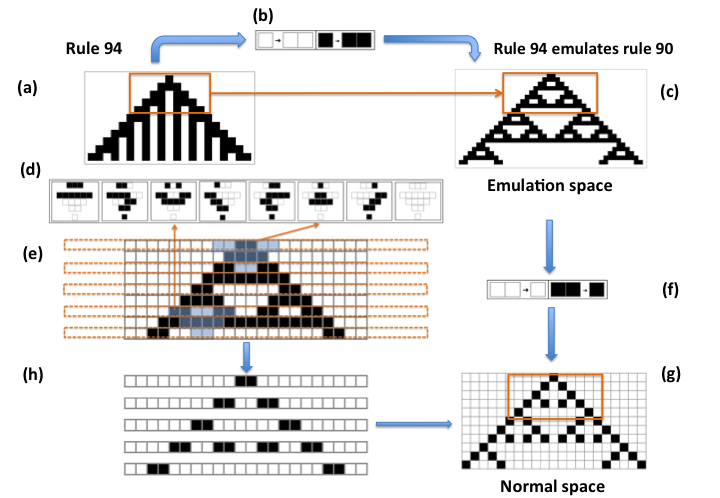}
\caption{
\label{fig_emul_diagram}Flow diagram of a CA emulation: Compiler (d) for rule 94 (a) to emulate rule 90 (c) for any initial condition (in the example only one is depicted, the simplest black cell). At the top are the rule icons (source code) of the 2 ECA. (f) Block transformation (h) coarse-graining (g) final emulation computes rule 94.
}
\end{figure}


Fig.~\ref{fig_algorithm} describes the algorithm in detail to reproduce all results for the exhaustive exploration of emulation capabilities of CAs for each increasing size CA rule space and for a quasi-lexicographic enumeration of compilers of increasing length. Fig.~\ref{fig_emul_diagram} demonstrates the emulation process for any initial condition. Here we start, on the one hand, with the most simple initial condition, without loss of generality, e.g. one black cell. By applying rule 94 we let the CA evolve for 10 steps (a). If, on the other hand, we apply the block transformation $\Box \rightarrow \Box \Box$ and $\blacksquare \rightarrow \blacksquare \blacksquare$ (b) to the above initial condition, we get two black cells. If we then let the transformed initial condition also evolve with rule 94, now for 20 steps, we get the space-time diagram of a different CA (c). In this particular case it is the same as ECA rule 90 taking one black cell as the initial condition and then letting the CA evolve for 10 steps (g). To see this better, we take the boxed area of (c) and highlight every second step (e). Extracting these steps leads to the space-time diagram given in (h). By applying the back transformation $\Box \Box \rightarrow \Box$ and $\blacksquare \blacksquare \rightarrow \blacksquare$ (f) to (h) we get ECA rule 90, i.e. the boxed area in (g). The back transformation is  the actual coarse-graining process after the compilers have created their output for each emulation step. 

In Fig.~\ref{fig_emul_diagram} (d) (top) we give an example of a shortest compiler (all compilers we found include the shortest possible compiler for emulating a specific CA) which allows ECA rule 94 to emulate ECA rule 90 by defining the rescaling $P=\Box \rightarrow \Box \Box, \blacksquare \rightarrow \blacksquare \blacksquare$. Each represents the encoding of all eight possible inputs of the compiler. The output comprises the bits for ECA rule 90. Each compilation displays a distinct pattern. We highlight in (e) two of the compiler components, showing how the compiler effectively translates cells from ECA rule 94 to ECA rule 90.

\begin{figure}[h!]
\centering  \includegraphics[width=10cm]{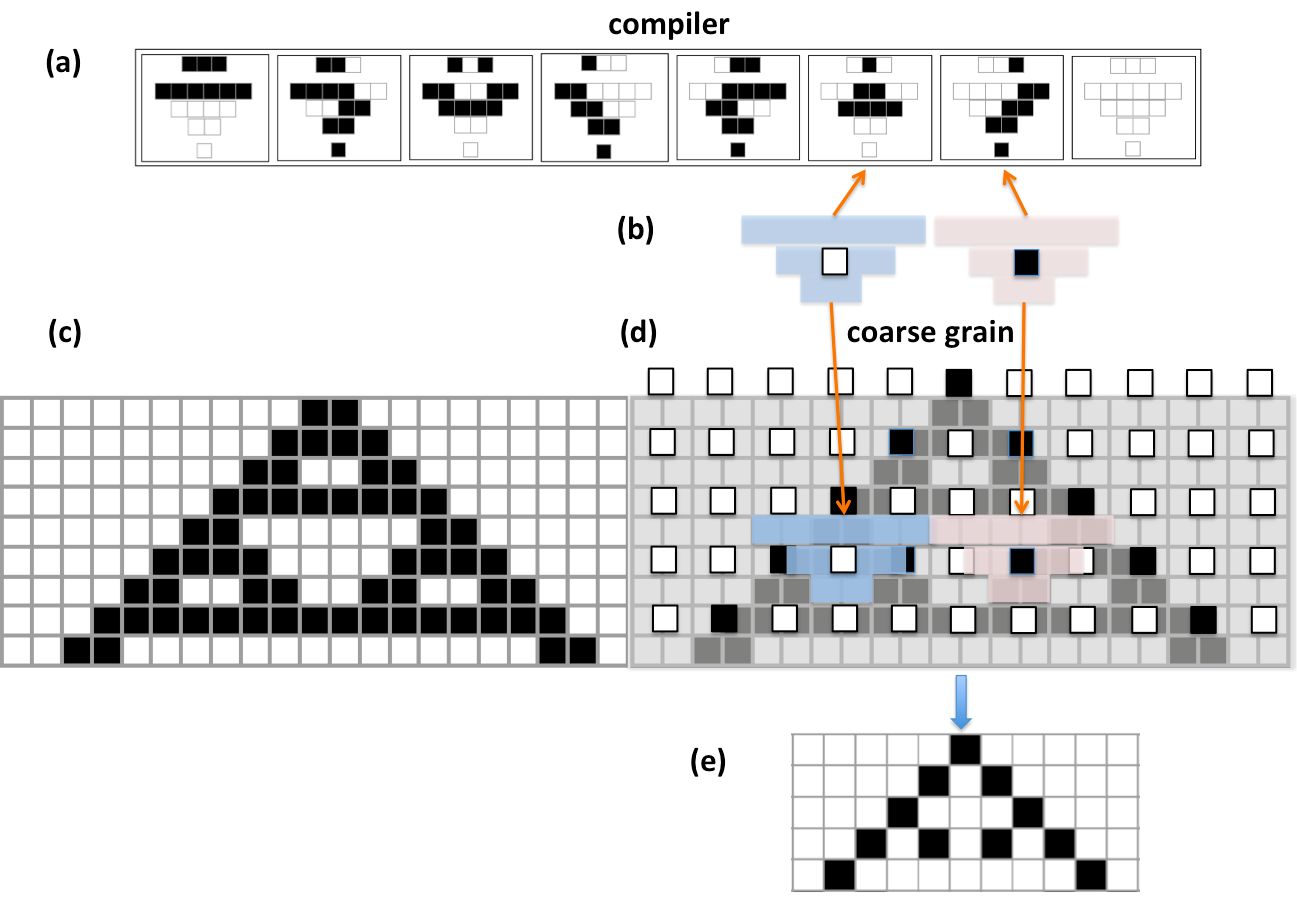}
\caption{
\label{fig_emul_diagram_coarse} Visualizing a CA coarse-graining using the space-time diagram of ECA rule 94 (c) for any initial condition (in the example only one is depicted, the simplest black cell) with compiler (a) found by exhaustive enumerating search with each local rule a `supercell' (b) which are coarse-grained (d) to yield space-time diagram of ECA rule 90 (f). Emulating and emulator are thus fine and coarse-grained versions of each other.
}
\end{figure}


In Fig.~\ref{fig_emul_diagram_coarse} we demonstrate the coarse-graining in more detail for the same initial condition as in Fig.~\ref{fig_emul_diagram}. We start with the space-time diagram of rule 94 (c) for the initial condition of one black cell as used in Fig.~\ref{fig_emul_diagram} and then block transform it with  $\Box \rightarrow \Box \Box$ and $\blacksquare \rightarrow \blacksquare \blacksquare$. Taking the compiler components (a) and identifying the encoding of all eight possible inputs of the compiler as supercells~\cite{NavotGoldenfeld2006}, one can understand the compiler to be a mapping of each supercell to a single cell, either black or white (b). Combining this with (c) one can depict the necessarily overlapping supercells as 'scales' on the space-time of the block transformed CA (d). By simply taking the coarse-grained cells we get the coarse-grained version of ECA rule 90.

Block emulation is a common technique in proofs for  Turing-universality. For example, ECA rule 110~\cite{nks,cook} and the 2,3 Wolfram Turing machine~\cite{smalltm} were both proven using compilers to translate a model of computation known to be universal (a cyclic tag system) into the proper initial conditions for the system to be proven universal. Here we proceed in a similar fashion by brute force (not fundamentally different from  the other author's techniques).

\subsection{Determining complexity classes}

An heuristic method to classify the behaviour of computer programs into four qualitatively different types was introduced by Wolfram~\cite{nks}. We will use the original Wolfram classification only for ECA and only to analyze the data and not for anything essential for the results.

For GCA and the general case other than ECA (unless otherwise specified), a candidate complexity class $W$ of a cellular automaton $CA$ can recursively be given by~\cite{zenilchaos} as follows:

\begin{defn}
$$W(CA)=\lim_{i,t\rightarrow\infty}\max{W(CA(i,t))}$$
\end{defn}

\noindent where $W(CA(i,t))$ is the maximum candidate class of $CA$ reached across all inputs $i$ (for an arbitrary enumeration) up to runtime $t$ according to a compressibility criterion. It is clear that $W(CA)$ is approachable from below. That is, if $W(CA(i,t))=C_n$ and $W(CA(i^\prime,t^\prime))=C_n^\prime$, then $CA$ belongs to class $C_{n\prime} < C_n$, where $n^\prime < n$. $W$ introduces a partition, given that a system $s$ cannot belong to two different Wolfram classes at the limit. That is, $\bigcap_{n=1}^4 C_n=\emptyset$ and $\sum_{n=1}^4 |C_n| = |\bigcup_{n=1}^4 C_n|$. Which does not mean one cannot misclassify a system for initial values $i$ and $t$. $W(CA(i,t))$ can then be formalized by using a suitable complexity measure, for example, the algorithmic complexity $K(CA(i,t))$ for initial conditions $i$ up to time $t$. Then one takes the maximum or tries to calculate a limit. In practice this is impossible because $K$ is semi-computable, but a lossless compression algorithm implementing Entropy can be used as a sufficient test for non-randomness, and therefore as a loose upper bound on $K$. Thresholds are then trained to divide $W$ into candidate classes $C_n$ motivated by Wolfram's original classes.

The measure, based on the change of the asymptotic size of the compressed evolutions for increasing $i$ and $t$, is calculated by following an enumeration of $i$ based on a Gray-code enumeration, as suggested in~\cite{zenilca,zenilchaos}, to establish a distance metric between initial configurations. The measure then gauges the resiliency or sensitivity of a system based on its initial conditions. The phase transition coefficient thus defined led to an interesting characterization and classification of systems, which when applied to elementary CA, yielded exactly Wolfram's four classes of system behaviour (except for one borderline case). The coefficient works by compressing the changes of the different evolutions through time, normalized by evolution space. It has been shown to be an interesting approach to CA behavioural classification questions~\cite{zenilca,zenilchaos}.

\begin{defn}
$$NC(CA) = \frac{K(CA)}{t}$$
\end{defn}

The criteria for classifying the asymptotic behaviour of the space-time evolution of a computer program, particularly a cellular automaton, are as follows:

\begin{myitemize}
\item Classes 1 and 2: The evolution of the compressed system diverges from its original size for any number of steps;
\item Classes 3 and 4: The lengths of the compressed evolutions asymptotically converge to the size of the uncompressed evolution. 
\end{myitemize}

The asymptotic behaviour of a cellular automaton is therefore 

$$\lim NC(CA(i,t))_{i,t\rightarrow \infty} = 1$$
\noindent for complex behaviour, and 
$$\lim NC(CA(i,t))_{i,t\rightarrow \infty} = 0$$
\noindent for simple behaviour~\cite{zenilchaos}. Here we will use $NC$ and entropy  to determine the class $W$ of a cellular automaton for the sole purpose of analyzing the emulation and emulating data. If $NC(CA) \rightarrow 1$, then $W(CA) = C_{3;4}$; otherwise $NC(s) \rightarrow 1$, then $W(CA) = C_{1;2}$. However, a numerical approximation of $NC(CA)$ is needed, which means evaluating $NC(CA(i,t))$ for a number of initial conditions $i$, following the Gray-code numbering scheme suggested in~\cite{zenilchaos}, and for a number of time steps. 

We believe that these measures introduced in~\cite{zenilchaos} provide a scalable formalization to Wolfram's original heuristic approach to quantify behavioural classes. Under no circumstance, however, these measures play any fundamental role and are only used to analyze the data and qualitatively differentiate cellular automata for which it is undecidable to tell if they are different or the same. 

In what follows, and only for data analysis and reproducibility purposes, we will identify 3 classes of complexity: low (L), medium (M) and high (H) complexity according to these measures using one of the most popular lossless compression algorithms (Compress based on LZW) with H the compressed size of the CA evolution reaching 90\% of its uncompressed original size for the same runtime, L compressed by at least half its size and M for the intermediate cases.

\subsection{Coarse-graining/back transformation}

In a dynamical system, coarse-graining is used to describe large scale structures in the evolution of a system. This process has been applied to CA before~\cite{mazoyer,NavotGoldenfeld2006}. In theory there are many ways one can define the coarse-graining of a dynamical system such as a CA. Here we follow a definition in~\cite{NavotGoldenfeld2006}. 

\begin{defn}
A renormalization scheme is defined where the original CA $f_A$ is coarse-grained by a lattice transformation $b_k=P(a_{N k} ,a_{N k+1}, \ldots, a_{N k+N-1})$. In this case a rescaling function $P: \{S_A\}^N \rightarrow \{S_B\}$ maps the value of a block of $N$ cells (also called a ``supercell''~\cite{mazoyer}) in CA $f_A$ to a single cell in CA $f_B$. 
\end{defn}

The coarse-graining emulation of CA $f_A$ by CA $f_B$ with the rescaling $P$ only works if the following commutative condition is imposed:

\begin{equation}
\label{coarse_grain_condition}
f^{t}_{B} P a(0)=P f^{T t}_{A} a(0)
\end{equation}

One can now apply this description of coarse-graining to the block emulation procedure described above. Applying the rescaling $P$ to (\ref{block_emulation_condition}) on both sides, one obtains 

\begin{equation}
\label{refcoarse_grain_equivalence}
f^{T t}_{A} P a(0)=P f^{t}_{B} a(0)
\end{equation}

Note, CA $f_{A}$, CA $f_{B}$, and $P$ are in principle different from the ones defined for eq. (\ref{coarse_grain_condition}). We can now compare  (\ref{refcoarse_grain_equivalence}) with (\ref{coarse_grain_condition}) and ascertain that it describes the coarse-graining emulation of CA $f_B$ by CA $f_A$ with the rescaling $P$. In other words, the block emulated CA $f_B$ is a coarse-grained description of the CA $f_A$  transformed with $P$, i.e. in Fig~\ref{fig_emul_diagram} e.) is the coarse-grained description of the block transformed CA in Fig.~\ref{fig_emul_diagram} b).


\section{Results}

\subsection{An example of emulation in the PCA rulespace}
We start with the PCA rule 13. By defining the rescaling $P=\Box \rightarrow \Box \blacksquare, \blacksquare \rightarrow \blacksquare \blacksquare$ to random initial conditions $a(0)$, i e. $a'(0)=P a(0)$, rule 13 can be made to emulate rule 12. In this particular case the block size is 2 and it takes 2 time steps for rule 13 to emulate one lattice entry of rule 12. The lattice entry $a(T t)$ is identical to the lattice entry of rule 12 at $a'(t)$ if one applies the back rescaling $P^{-1} a$.

\subsection{Reprogrammability of cellular automata}

We now show some representative cases showing how simple and complex behaviour cellular automata can emulate other simple and complex cellular automata without fundamental distinction or boundaries.

\subsubsection{A class 4 ECA to behave like a class 2 ECA}

\begin{figure}[h!]
\centering 
{\includegraphics[width=11cm,angle=0]{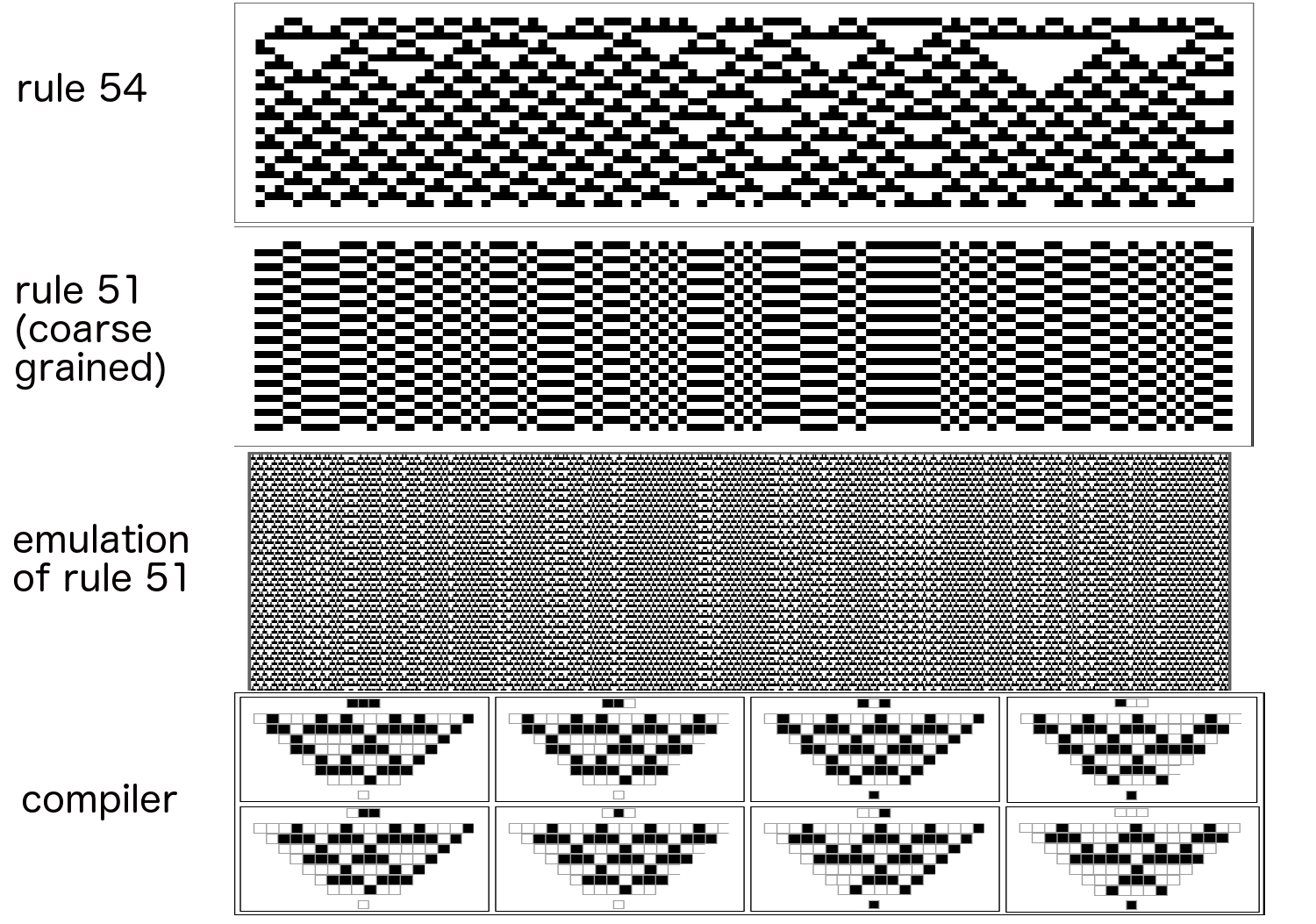}}
\caption{\label{fig_mass2}Reprogramming class 4 ECA rule 54 (top) to compute a class 2 ECA, rule 51 (middle top), for any initial condition (here only one is depicted for illustration purposes, in the middle bottom, which is also the coarse-grained version of the evolution in the middle bottom). The smallest compiler (bottom), of size 7, found in the compiler space (up to size 15) was used. The compiler only interacts with rule 54 at step $t=0$ but not afterwards, i.e., from $t=1$ on the computation is carried out by ECA rule 54 alone.
\label{fig_emul_54_51}
}
\end{figure}

If one applies the rescaling $P=\Box \rightarrow \Box \Box \Box \blacksquare \Box \Box, \blacksquare \rightarrow \Box \blacksquare \Box \Box \Box \blacksquare$ to random initial condition $a(0)$, rule 54 can be made to emulate rule 51 (see Fig.~\ref{fig_emul_54_51}). 

\subsubsection{A Wolfram ECA class 3 rule emulating a class 2 rule}

Using rescaling $P=\Box \rightarrow \blacksquare \blacksquare \blacksquare \blacksquare \Box \Box \blacksquare \Box \Box \blacksquare, \blacksquare \rightarrow \blacksquare \blacksquare \blacksquare \blacksquare \blacksquare \blacksquare \Box \Box \Box \blacksquare $ to random initial condition $a(0)$, ECA rule 45 can be made to emulate ECA rule 15. Rule 45 emulates rule 15, which is a high heat CA being programmed to emulate a simple rule. This was not shown in~\cite{StephenWolfram1986}. In~\cite{nks}, rule 45 is shown to emulate rule 90 with a time shifted block emulation, which is different from the block emulation used in this paper. 
The actual block emulation process of rule 15 by rule 45 shows that rule 45 takes $T=10$ steps in order to reproduce one line of rule 15.

\subsubsection{A Wolfram class 2 emulates a class 3 rule}

Another example of emulation is the emulation of ECA 90 by ECA 164. In this case the rescaling is $P = \Box \rightarrow \blacksquare \blacksquare \blacksquare \blacksquare, \blacksquare \rightarrow \blacksquare \Box \blacksquare \Box$. On applying the rescaling $P$ to random initial condition $a(0)$ using rule 164 to evolve the lattice, one obtains as a result an emulation of rule 90. This is especially interesting since rule 164 is a class 2 ECA and rule 90 a class 3 ECA.

\subsubsection{A random-looking rule emulates a simple-looking rule}

Cellular automata in larger rule spaces (GCA) provide a set of interesting class cross-boundary emulations over a much larger domain than ECA and PCA. 

An example is the emulation of GCA rule 782 by GCA rule 4086. In this case the rescaling is $P=\Box \rightarrow \Box \Box, \blacksquare \rightarrow \Box \blacksquare$. On applying the rescaling $P$ to random initial condition $a(0)$ using rule 4086 to evolve the lattice, one obtains as a  result an emulation of rule 782. Rule 4086 is more complex than rule 782, and the emulation is another example of a class emulating another, less complex one, which is very common.

Another example of an interesting emulation of a GCA rule emulating an apparently more complex rule is the emulation of GCA 4382 by GCA 17\,910. In this case the rescaling is $P=\Box \rightarrow \Box \Box, \blacksquare \rightarrow \blacksquare \Box$. On applying the rescaling $P$ to random initial condition $a(0)$ using rule 17\,910 to evolve the lattice, one obtains as a result an emulation of rule 4382. Rule 4382 is more complex than rule 17\,910. This is another example of a ``jumper'' rule, just like ECA rule 164.

\subsubsection{A random-looking rule emulates a simple rule}

An example of a simple GCA rule emulating a class 3 GCA is the emulation of GCA rule 27\,030 by GCA rule 13\,960. In this case the rescaling is $P=\Box \rightarrow \Box \blacksquare \Box, \blacksquare \rightarrow \Box \blacksquare \blacksquare$. On applying the rescaling $P$ to random initial condition $a(0)$ using rule 13\,960 to evolve the lattice, one obtains an emulation of rule 27\,030. Rule 27\,030 is more complex than rule 13\,960, and the emulation is another example of one class emulating another, less complex one, which is very common. 

Despite the disorganized nature of class 3 rules, they also display a wide range of reprogramming capabilities, being able to compute and behave in an orderly fashion, as, for example, class 2 or 4 rules. Fig.~\ref{fig_cross_class} (in the Supplemental Material section), shows how class 3 rules can be reprogrammed to emulate other class 3 rules, but also class 4 rules. Hence they cannot be overlooked as candidates for Turing-universality on the grounds that they seem difficult to control, as this uncontrollability is only apparent.

\subsubsection{A simple rule emulates a random-looking rule}

An example of a Wolfram class 1 emulating a Wolfram class 3 GCA is the emulation of GCA rule 6696 by GCA rule 27\,030. In this case the rescaling is again $P=\Box \rightarrow \Box \Box \blacksquare, \blacksquare \rightarrow \Box \blacksquare \blacksquare$. Applying the rescaling $P$ to random initial condition $a(0)$ using rule 6696 to evolve the lattice, one obtains an emulation of rule 27\,030. Rule 6696 is much less complex than rule 13\,960, and the emulation is an example of one class emulating another, more complex rule.

\subsubsection{A complex rule emulates another complex rule}

An example of an intra class 4 emulation is the emulation of GCA 2966 by GCA 25\,542. In this case the rescaling is $P=\Box \rightarrow \Box \Box, \blacksquare \rightarrow \Box \blacksquare$. Applying the rescaling $P$ to random initial condition $a(0)$ using rule 2966 to evolve the lattice, one obtains an emulation of rule 25\,542. This is interesting, since both rules 2966 and 25\,542 are complex GCA rules.

\subsection{Efficient simulation time complexity}

The depth of the ``compiler" (or length of block encoding) provides the time complexity overhead of the simulation. For example, in Fig.~\ref{fig_mass2}, rule 54 requires 7 time steps to emulate every time step of rule 51, and the time complexity overhead of the simulation with respect to the simulated is $O(n+7)=O(n)$. This also means that the block transformation to get the actual computation of the simulated from the simulating one has to take the coarse-grained version of the latter, only taking every other 7 steps to get the exact output of the simulated rule.

\begin{figure}
\begin{tabular}{cc}
  \includegraphics[width=12.2cm]{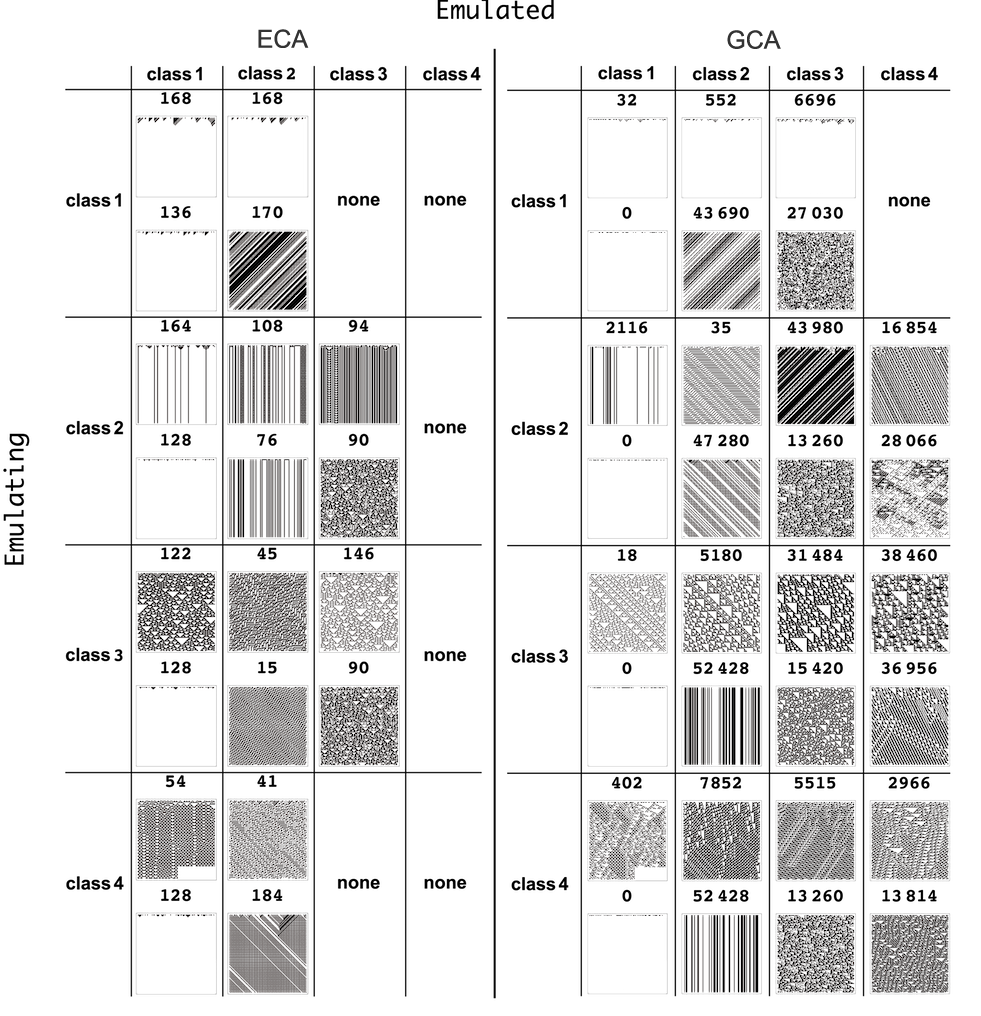}
\end{tabular}
\caption{\label{fig_cross_class} Cross-boundary ECA and GCA rules. For ECA we used the `official' Wolfram class according to~\cite{nks} while for PCA and GCA a compression mechanism based on Compress (LZW) was used. The grid shows cross-boundary ECA and GCA rule pairs (the top rule is the emulating rule, the bottom one is the emulated rule) for different Wolfram classes (rows are emulating- and columns emulated rules). GCA show a wide range of reprogrammability, with almost every rule capable of being reprogrammed. Notable cases are classes 2 and 3, which are capable of emulating classes 3 and 4, with class 3 in fact being able to emulate any other class (from GCA on) up to the compiler space size explored. Hence all these classes can be reprogrammed to behave like any other computer program, displaying alternative qualitative behaviour (complexity).}
\end{figure}

\subsection{A complexity measure based on node degree}

We construct emulation graphs. A rule is connected to another if it emulates (outgoing edge) or is emulated (incoming edge).

\begin{figure}[h!]
\centering
\textbf{Ingoing (emulated frequency):}\\
\includegraphics[width=5.7cm,angle=0]{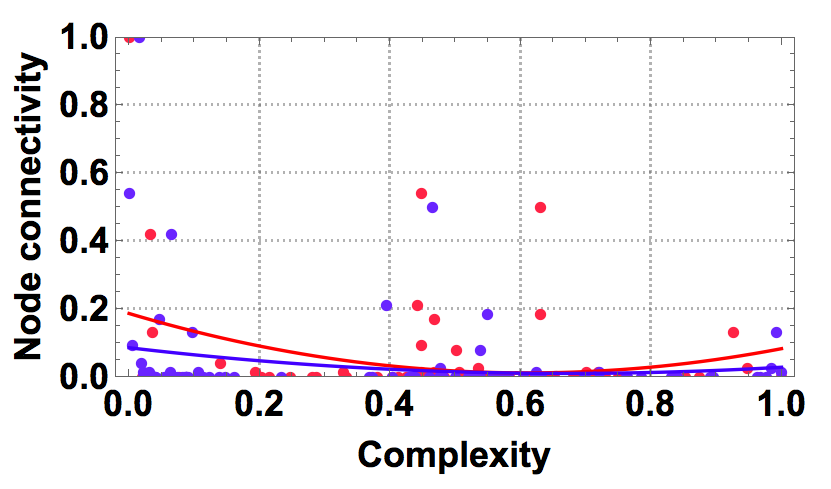} \hspace{.2cm}\includegraphics[width=6.0cm,angle=0]{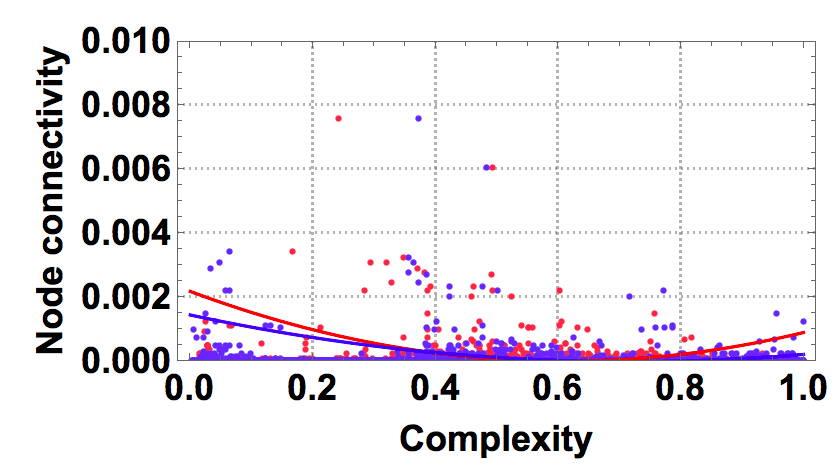}\\
\centering 
\textbf{Outgoing (emulating frequency):}\\
\includegraphics[width=6cm,angle=0]{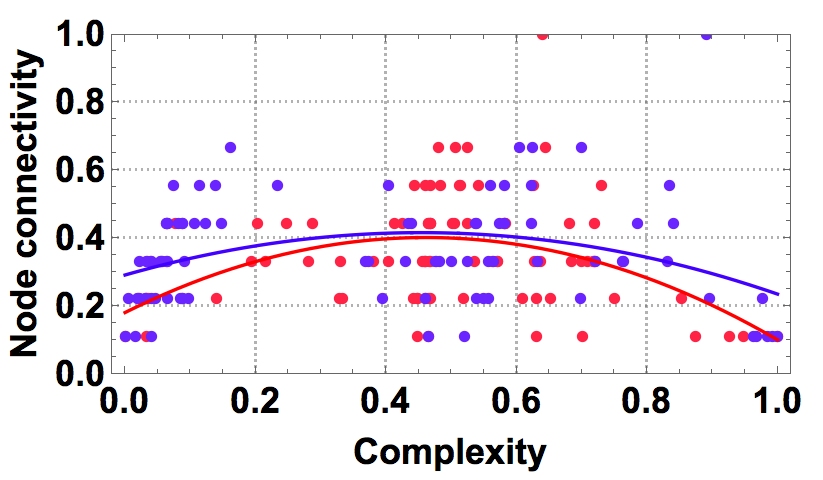} \includegraphics[width=6.0cm,angle=0]{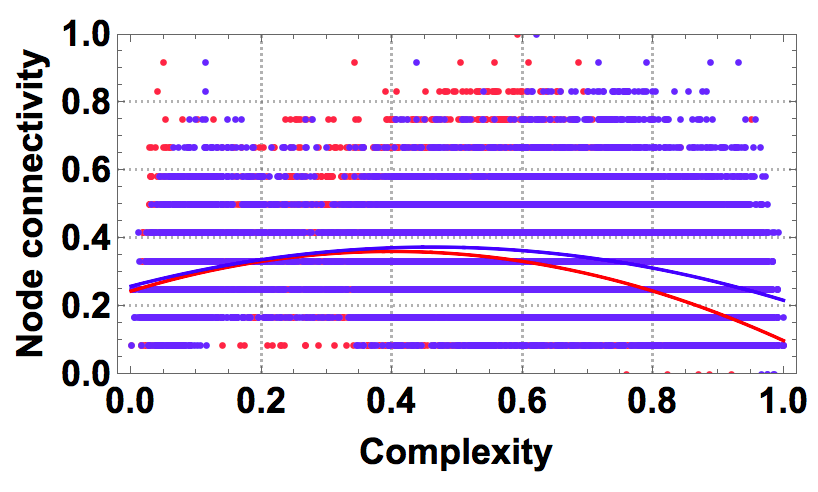}\\
\caption{\label{fig_GCAEmulationTopologicalComplexity}Topological-computational rule classification: Connectivity distribution ($y$-axis) vs. rule complexity (Normalized Entropy and Normalized lossless compression or $NC$) ($x$-axis) for ECA (left) and GCA (right) rule spaces for both ingoing (left) and outgoing (right) degree distributions. Ingoing plots show how many times different complexity classes can emulate other rules. Outgoing plots show how many times a rule is emulated by rules of different complexity classes. Class 1 followed by class 2 are the least complex. Class 3 is the most complex. Fitting lines are polynomials of second degree.}
\end{figure}

Fig.~\ref{complexityECAs} shows how rules belonging to Wolfram class 3 and particularly class 4 are on average less frequently emulated, while rules of Wolfram classes 1 and 2 tend to have a higher node degree, i.e., many more rules are able to emulate them for small compiler programs (block encodings of short lengths). This points to an interesting connection between complexity and graph topology that we confirmed in the GCA rulespace by classifying rules by their complexity estimations, and finding that high complexity rules are of low degree, as shown in Fig.~\ref{fig_GCAEmulationTopologicalComplexity}.

Fig.~\ref{fig_GCAEmulationTopologicalComplexity}(top) provides a topological measure of computational capacity and of class complexity. It suggests that class 4 rules can emulate more rules, which is compatible with the conjecture that class 4-type rules are likely Turing-universal~\cite{nks}, and thereby constitutes statistical evidence of computational capabilities from a topological perspective, based on the information from the emulating network. Additionally, Fig.~\ref{fig_GCAEmulationTopologicalComplexity}(bottom) suggests that class 4-type rules are less frequently emulated, thereby providing a fine and novel index of computational complexity, pinpointing that class 4-type rules (to the immediate right of the middle valley) are amongst the most difficult to emulate~\footnote{Results were generated with data from an exhaustive search of the full compiler space up to block size 15 for ECA and 12 for GCA. The exploration took about a week with a 10 CPU (2-core at 2.2GHz each) server and up to 64Gb of RAM.}. The plots are in agreement with the fact that class 4 computer programs are believed to be more programmable to reach computational universality, while at the same time being more difficult to emulate. In other words they make better emulators than emulated. On the other hand, GCA suggests that medium low complexity (equivalent to class 2 in Wolfram’s ECA classification) is equally amenable to being reprogrammed.

\begin{table}[ht]
\caption{Rule complexity classification table based on emulated rule count (ECA with ECA for all compilers up to size 15). All other rules not in the table can be considered more complex than those included because they were not simulated by any other rule up to the compiler size explored.}
\centering
\begin{tabular}{rrr}
  \hline
rank & ECA rule & emul. count \\ 
  \hline
1 & 170 & 35130 \\ 
  2 & 204 & 35051 \\ 
  3 & 150 & 32766 \\ 
  4 &   0 & 576 \\ 
  5 & 128 & 177 \\ 
  6 & 136 & 67 \\ 
  7 &  51 & 53 \\ 
  8 &  15 & 52 \\ 
  9 &  34 & 33 \\ 
  10 &  90 & 26 \\ 
  11 & 184 & 20 \\ 
  12 & 200 & 7 \\ 
  13 &   4 & 4 \\ 
  14 & 162 & 2 \\ 
  15 &  12 & 1 \\ 
  16 &  50 & 1 \\ 
  17 &  76 & 1 \\ 
  18 & 132 & 1 \\ 
  19 & 134 & 1 \\ 
  20 & 146 & 1 \\ 
   \hline
\end{tabular}
\end{table}

\subsection{Apparent Pervasiveness of Reprogrammability in CA}

While we consider how many times a rule is emulated a measure of its complexity--–which is deeply related to its Kolmogorov complexity by way of algorithmic probability---we also consider how many other programs a given computer program can emulate up to a certain compiler size, treating this number as a measure indicating its likelihood of Turing-universality.

 There are three conditions that if met point towards pervasive reprogrammability as evidence of ubiquitous intrinsic universality and therefore of Turing-universality:

\begin{myenumerate}
\item that the number of different non-trivial emulated rules per emulating rule increases (as a function of compiler length),
\item that the phenomenon replicates in larger rule spaces, and
\item that the emulations are not all characterized by a particular (trivial) behaviour.
\end{myenumerate}

The results reported here provide strong evidence on all counts, even if for item 2 only a relatively small set of computer program and compiler pairs can be explored  (despite being huge in absolute numbers). 

On the one hand, in ECA almost all rules emulate at least one other rule, almost all of them emulate at least 2 Wolfram classes, and 3 rules (26, 94, 164) emulate 3 different Wolfram classes. It is worth noting that the number of native low complexity rules exponentially decays~\cite{zenilchaos} and is therefore one of the reasons for the lack of emulation in the first panel. In other words, the fact that native high complexity rules dominate higher rule spaces strengthens the pervasive programmability-- and thereby pervasive universality-- hypothesis. 

\begin{figure}[h!]
\centering
  \includegraphics[width=10cm]{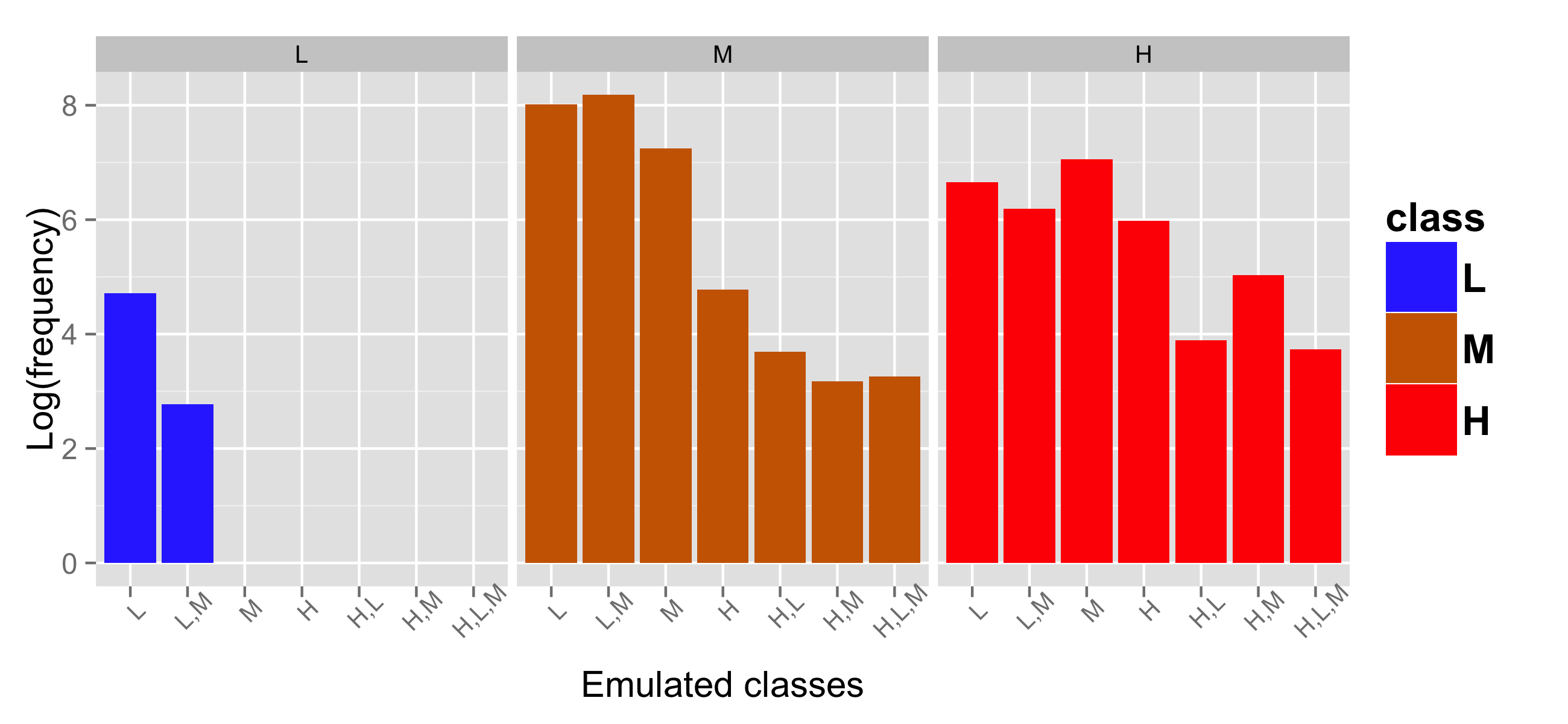}\\
    \includegraphics[width=10cm]{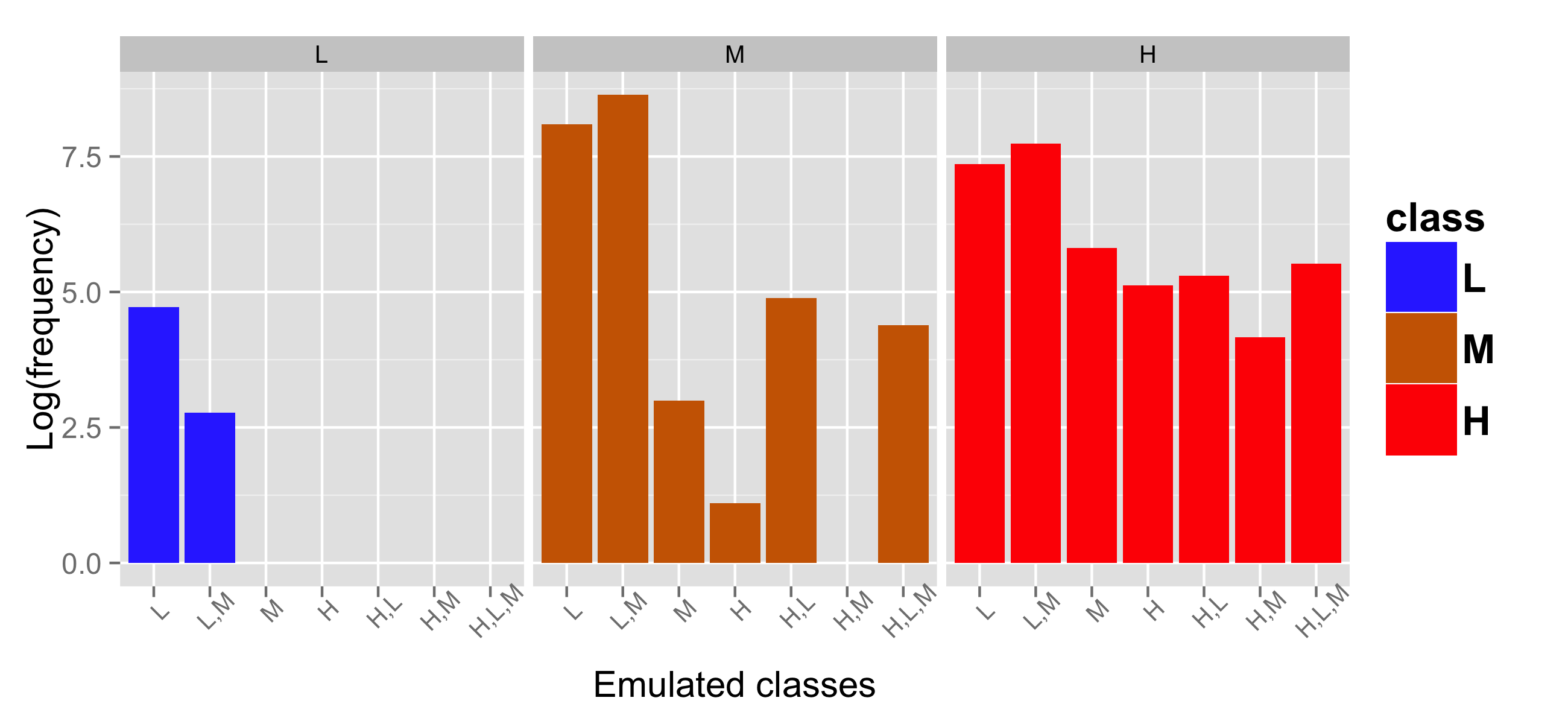} \caption{\label{mainb} Distributions of emulating and emulated rules of different complexity among all GCA emulations (L stands for low complexity, M for medium and H for high), showing that (1) emulating rules emulate non-trivial (in this case, M and H) rules, including the full range of all possible behaviours (rules that emulate all classes of complexity in which the GCA was divided by compressibility and entropy) and (2) the more complex the rule the greater the programmability (there is a short compiler to carry the emulations). Top: for all compilers of size 4. Bottom: for all compilers of size 8.}
\end{figure}

On the other hand, Fig.~\ref{mainb}, shows that in GCA, almost all rules emulate at least one other complexity class, and that among high complexity rules the number of multiemulating rules emulating 2 different complexity classes numbers around 500 (more than 3\%). 

\begin{figure}[ht!]
\centering
  \includegraphics[width=94mm]{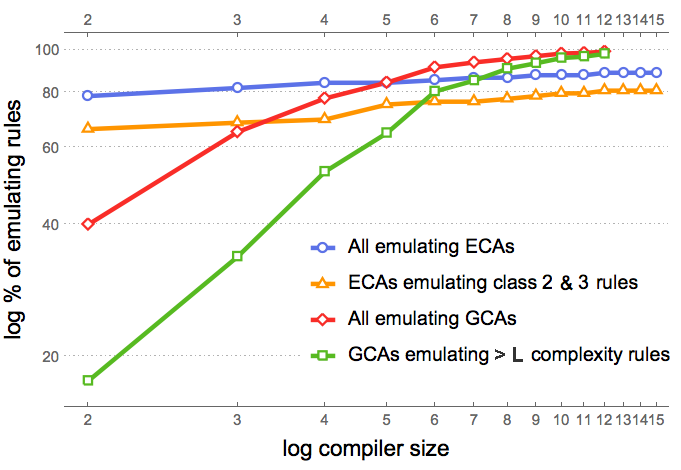}
\caption{\label{main} Accumulated rules that can be reprogrammed to behave as at least one other non trivially equivalent rule in the same rule space with a compiler up to size 15 (for ECA) and size 12 (for GCA). The number of non-trivially equivalent emulating GCAs emulating other GCAs of medium complexity grows asymptotically, just as in all other cases, strongly indicating reprogramming capabilities rapidly saturating its rule space, even for the smallest and simplest rule and compiler combinations.}
\end{figure}

Fig.~\ref{main} shows that the rule space together with the compiler space saturates the number of (non-trivial) emulations in the rule space and that the larger the rule space the greater the number of emulating and emulated rules, thereby suggesting a path towards ubiquitous intrinsic universality and therefore Turing-universal computation even for the smallest rule spaces and for the simplest rules, with the exception of the set of most trivial rules (e.g. rule 0 and rule 255 in ECA), which is shown to remain constant as a function of growing rule spaces and therefore asymptotically negligible and effectively of size 0. Other rules, such as the linear ones, can be proven to not be intrinsically universal~\cite{ollinger}, yet they seem to remain low in number as compared to both the growing number of emulators and the growing number of emulating rules per emulator in larger spaces.

As seen in Fig.~\ref{main}, almost all ECA rules are capable of being reprogrammed to behave like some other rule, showing the wide range of computing capabilities of the rule space up to the explored compiler size 15, with this range asymptotically reaching 100\% of each CA space emulating some other non-trivial and apparent high complexity rule. Evidently a small number of trivial rules like ECA rule 0, 255, the shift, the identity and the `XOR' (rule 150) will never be reprogrammable and while the number of emulations emulating trivial rules dominate, we show results for rules that do not show any of this trivial behaviour.

\begin{figure}[ht!]
\begin{tabular}{cc}
  \includegraphics[width=60mm]{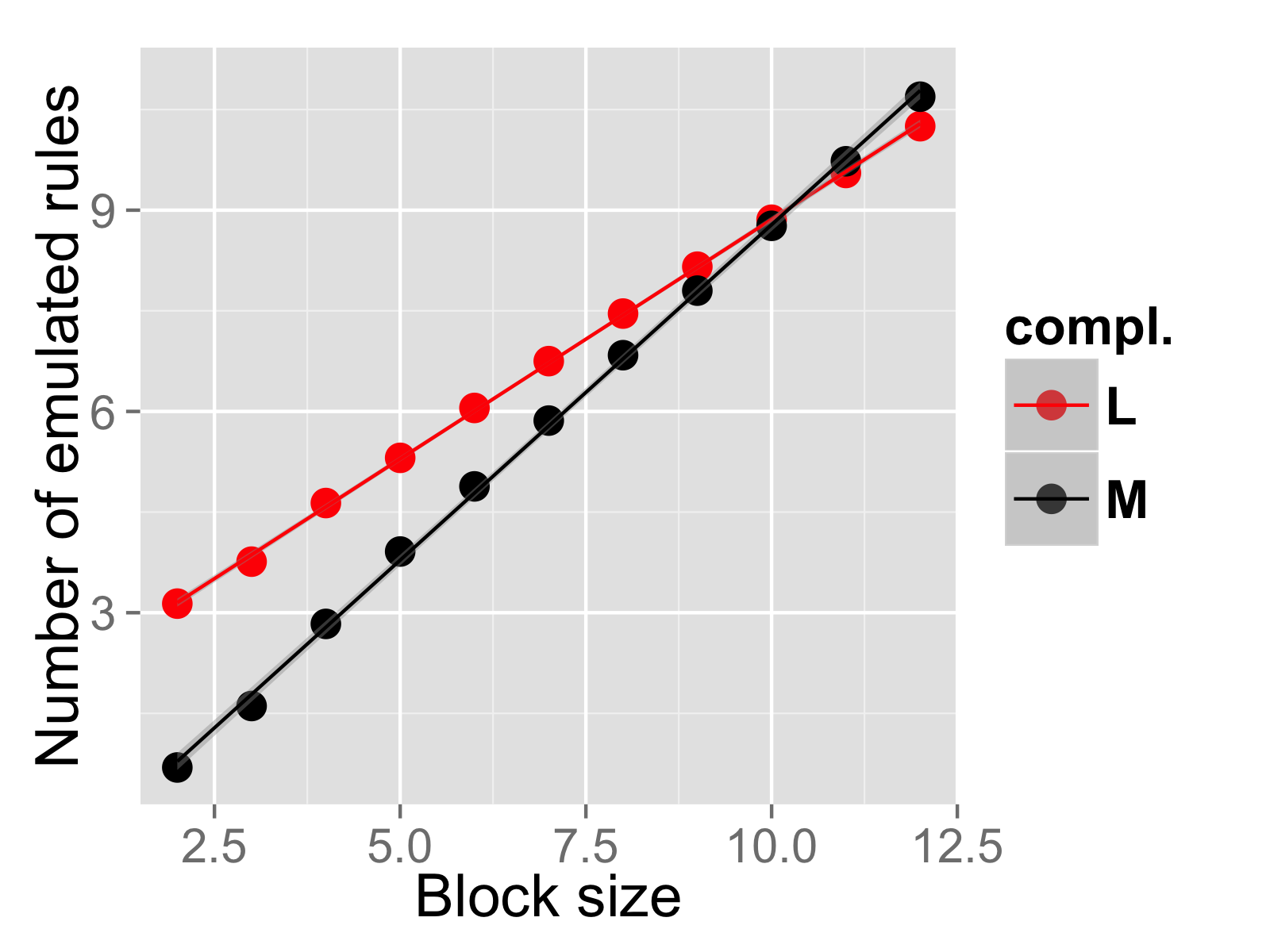} &   \includegraphics[width=60mm]{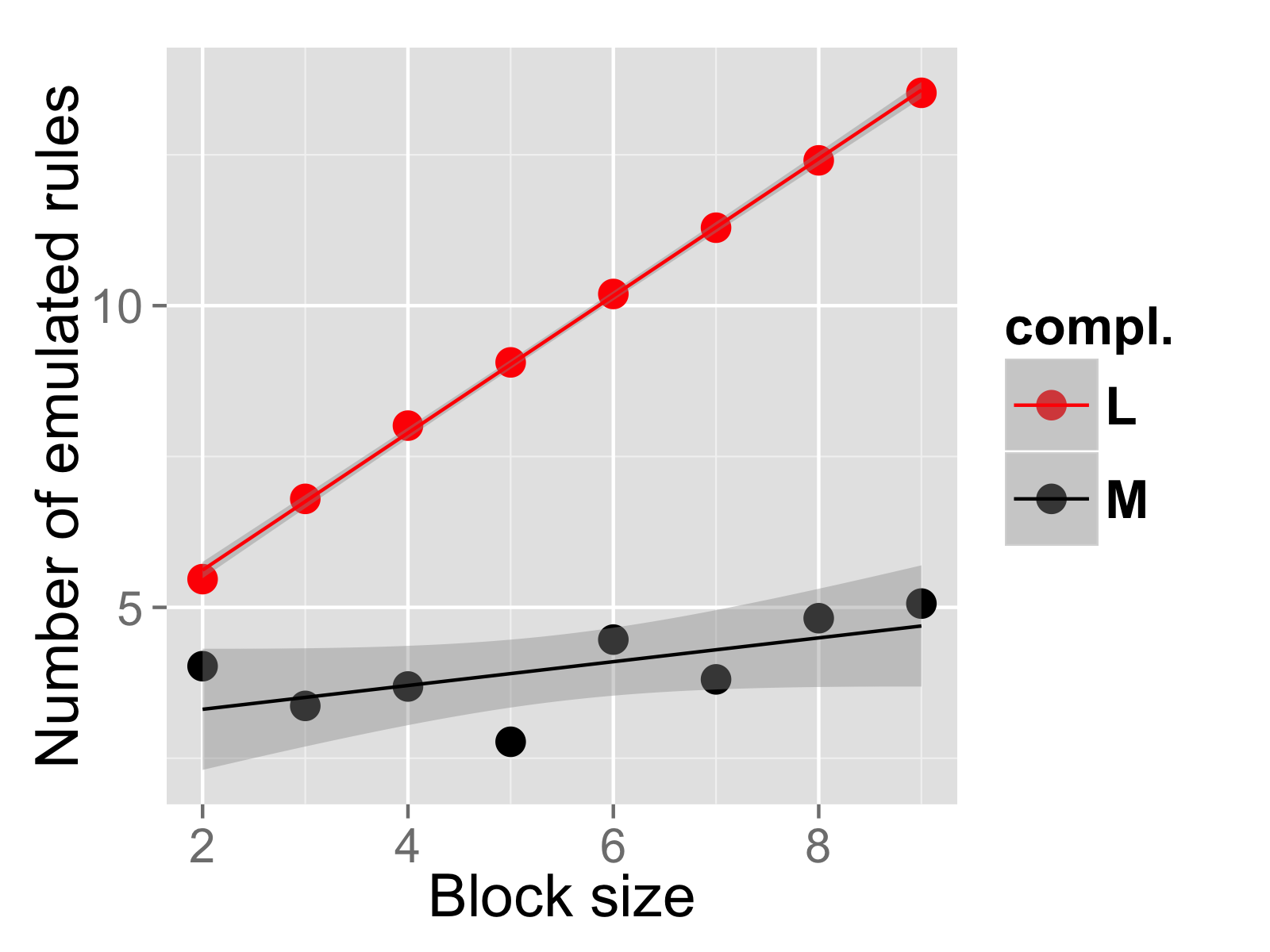} \\
(a) Low complexity ECA rules & (d) Low complexity GCA rules \\[5pt]
 \includegraphics[width=60mm]{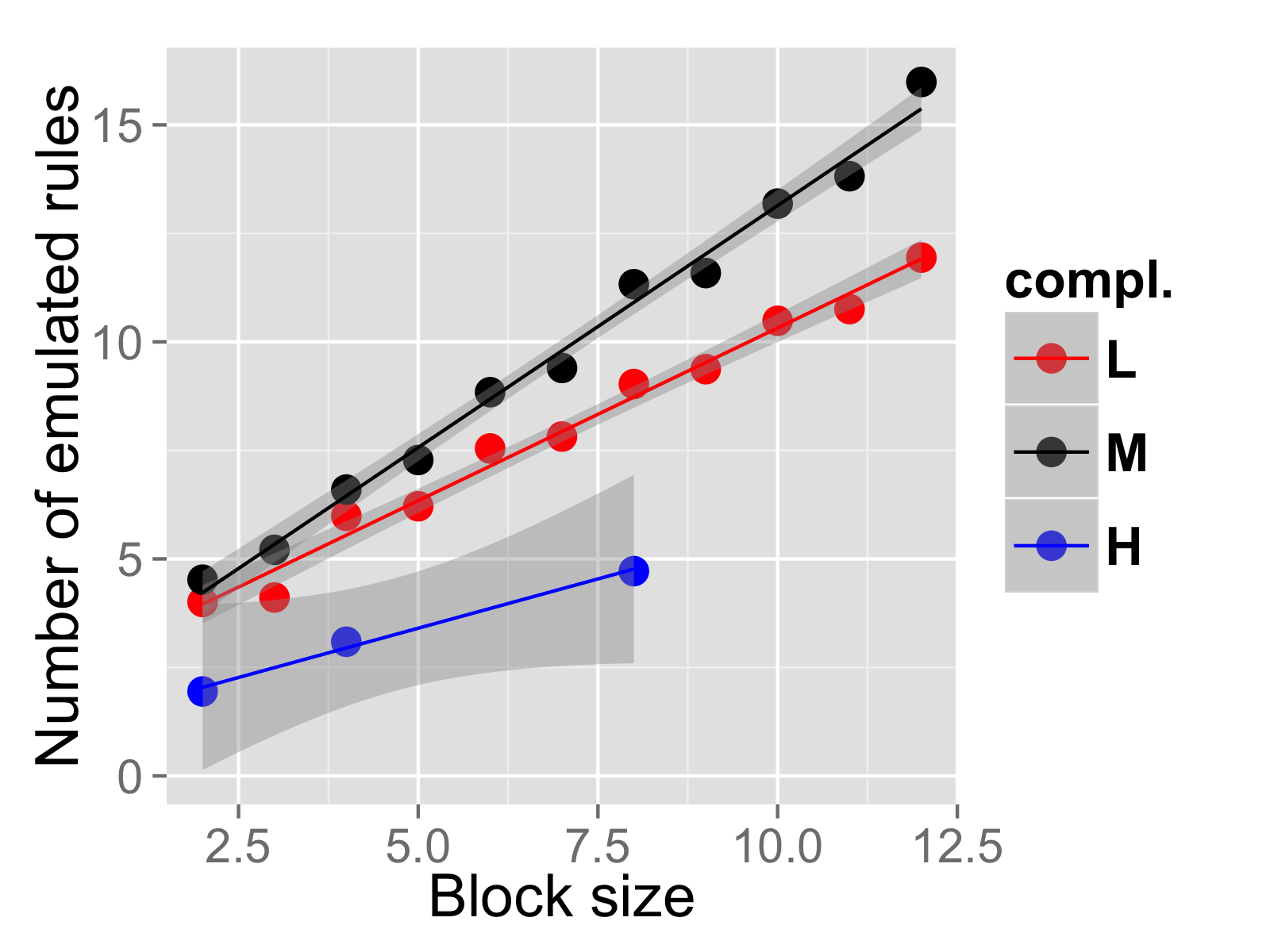} &   \includegraphics[width=60mm]{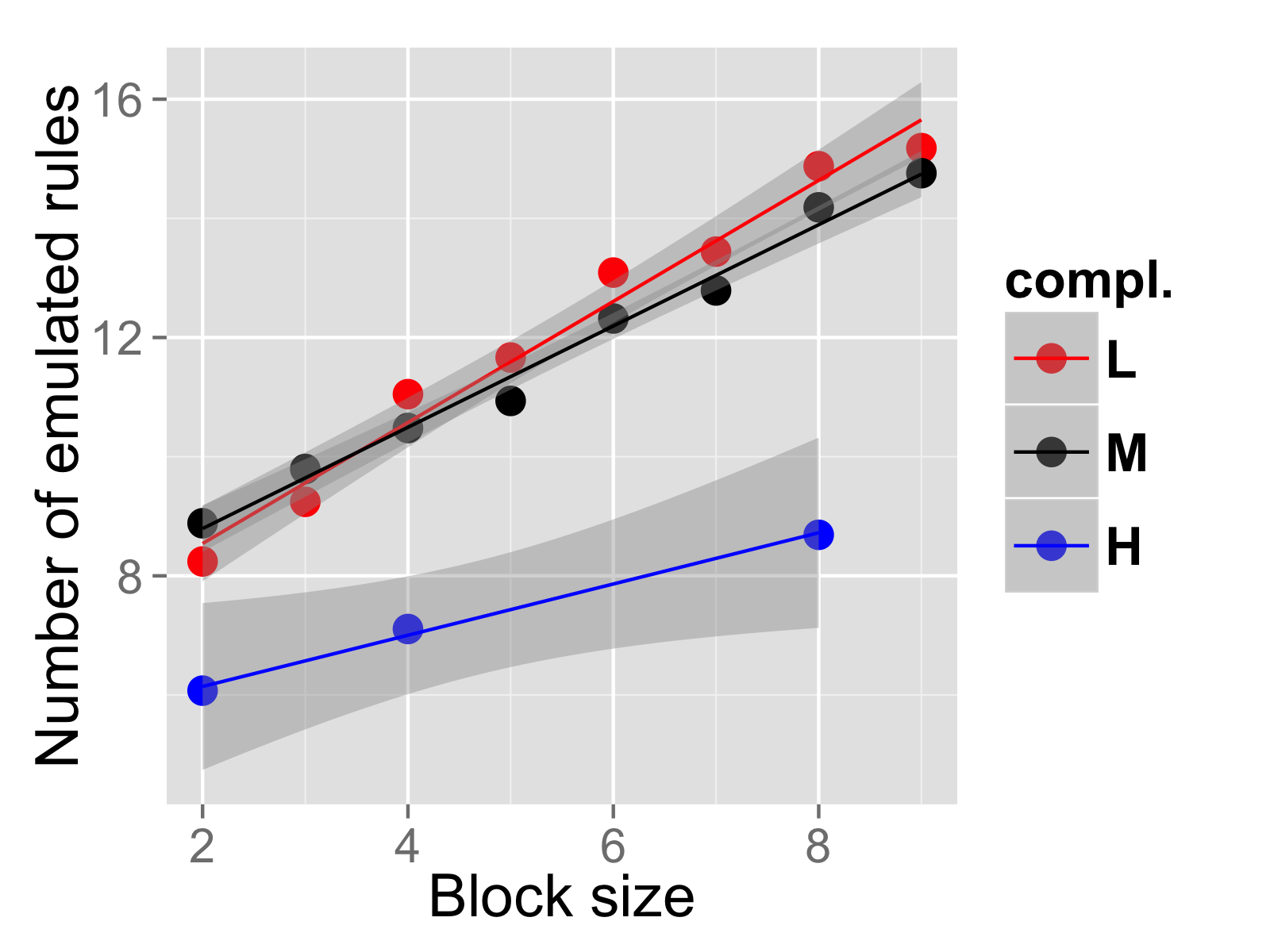} \\
(b) Medium complexity ECA rules & (e) Medium complexity GCA rules \\[5pt]
 \includegraphics[width=60mm]{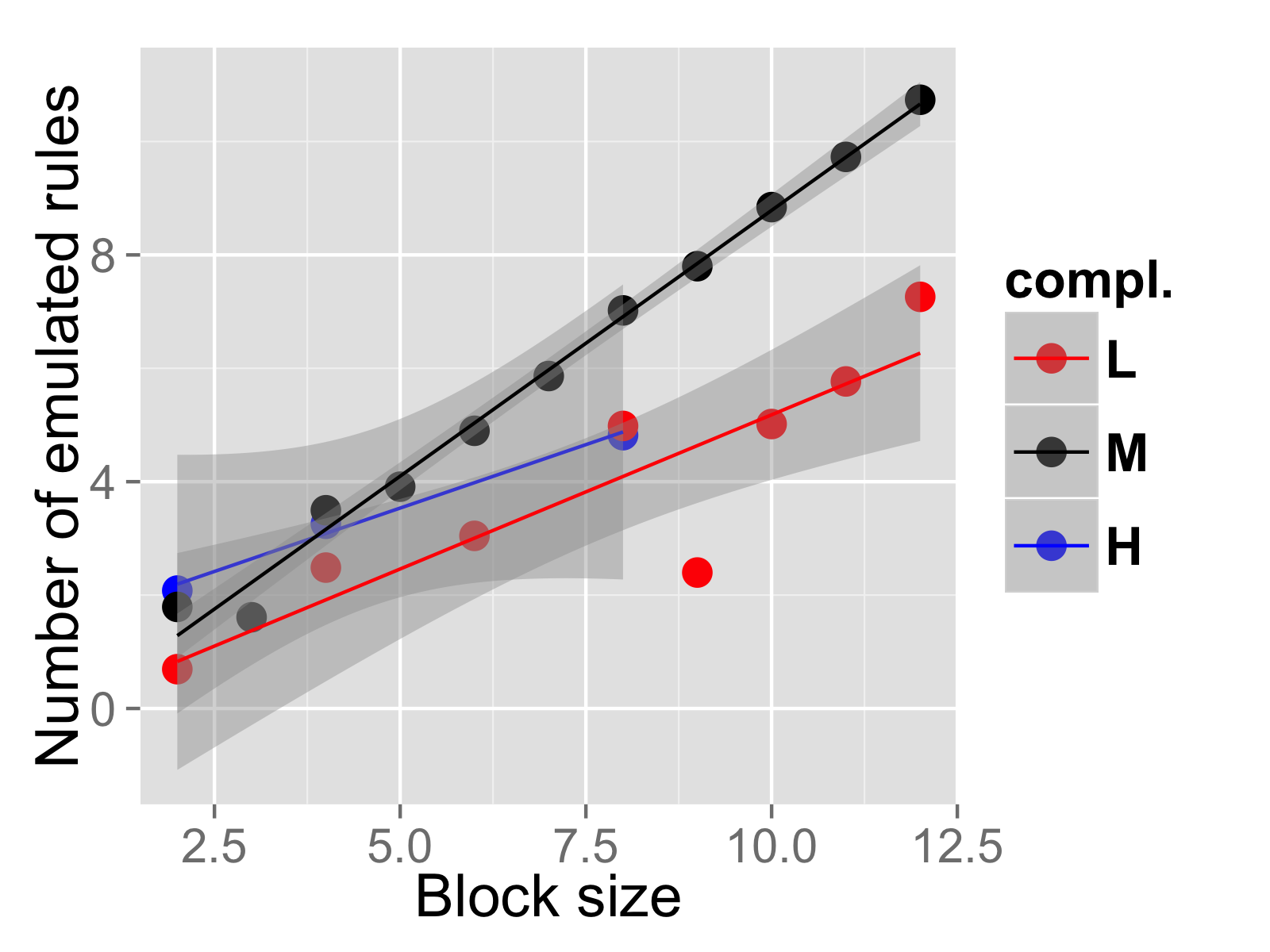} &   \includegraphics[width=60mm]{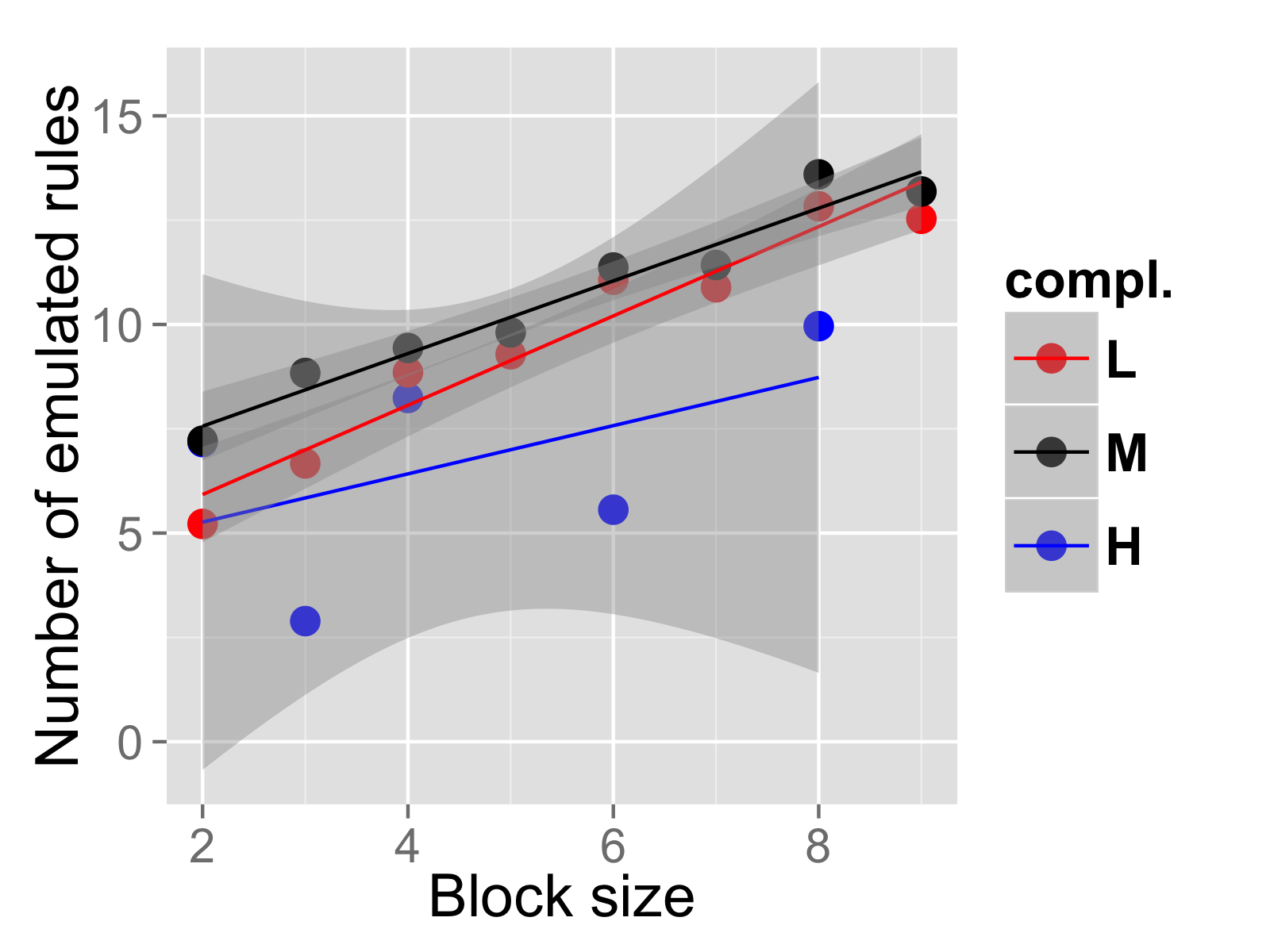} \\
(c) High complexity ECA rules & (f) High complexity GCA rules \\[5pt]
\end{tabular}
\caption{\label{fig_EmulFreq} Emulation frequency of high, medium, and low complexity ECA and GCA rules. The top three graphs (a)-(c) show the emulation frequency of ECA rules in relation to the emulation block size. Graph (a) shows how frequently high complexity ECA (Wolfram classes 3 \& 4) emulate other ECA of high complexity, medium complexity (Wolfram class 2), and low complexity (Wolfram class 1). Similarly, graph (b) shows this for medium complexity ECA and graph (c) for low complexity ECA. Graphs (d) - (f) show the same for GCA rules.}
\end{figure}

Moreover, emulating rules emulate more than one other rule and with more than one compiler. GCA rules continue generating emulating rules after compiler size 12. While the fraction of possible emulations is small when considering the huge rule $+$ compiler pair combinatorial space (on $y$ axis, right plot), the plot on the right provides statistical evidence that the number of emulations (including all compilers that emulate rules that were emulated before) grows, and that the larger the rule space (GCA) the faster the potential convergence. A more fine-grained version of these results is shown in Fig.~\ref{fig_EmulFreq}.

In fact, while almost all rules in ECA emulated some other non-trivial ECA (including class 2 emulating class 3 rules), no class 4 ECA was emulated. However, in GCA, both medium high (equivalent to class 4) and lowest highest complexity rules emulated all other rule complexity cases, including a class 1 emulating a class 3, a class 2 emulating a class 3 and 4, a class 3 emulating 3 and 4, and class 4 emulating all others, including class 3 and other class 4 rules. i.e., all GCA emulated all other complexity classes, except for class 1, which did not emulate class 4.

\begin{figure}[htpb!]
\centering
Collapsed\\
\centering
\textbf{ECA}\hspace{4.5cm}\textbf{GCA}\\
  \includegraphics[width=55mm]{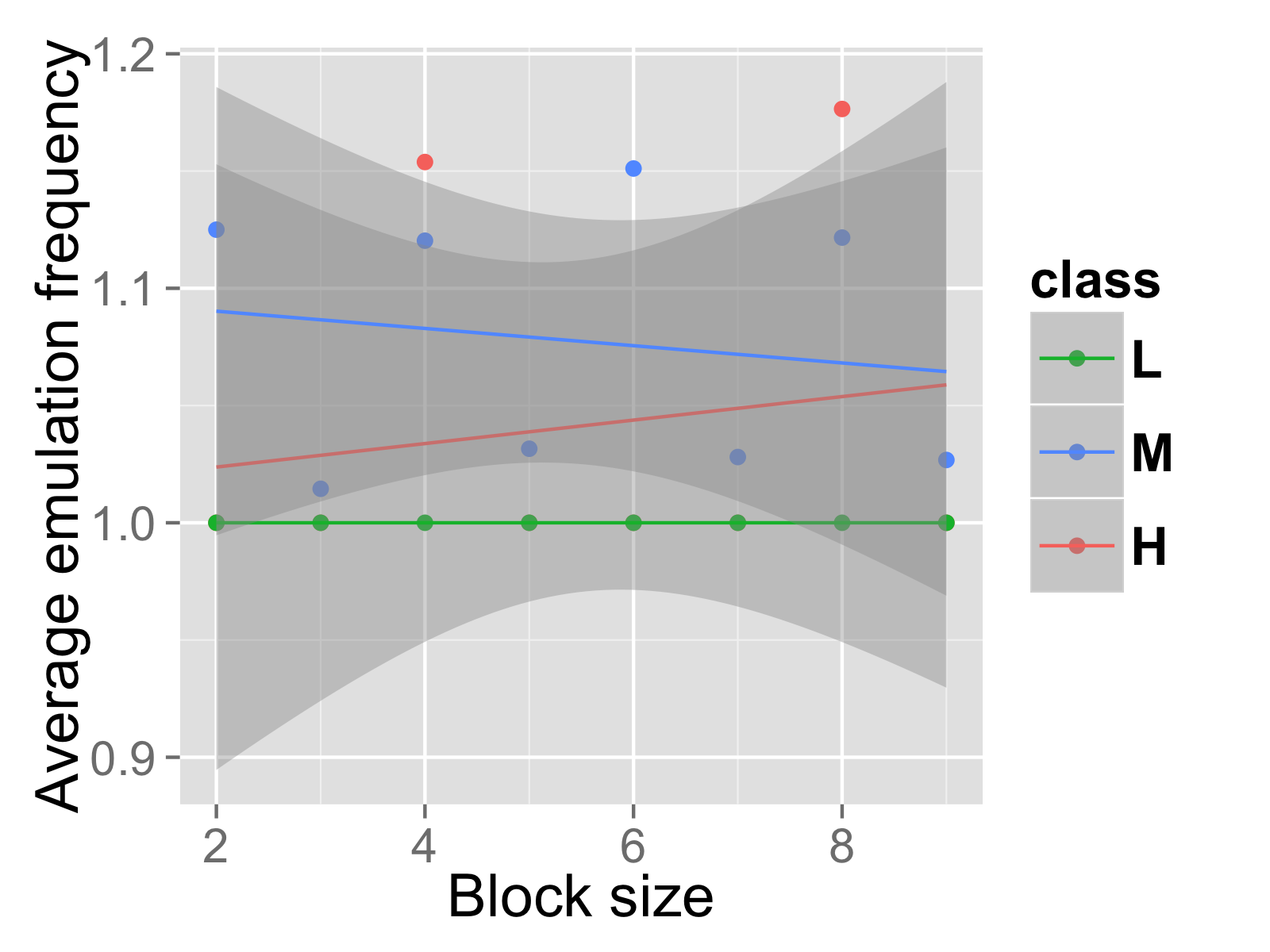}  \includegraphics[width=55mm]{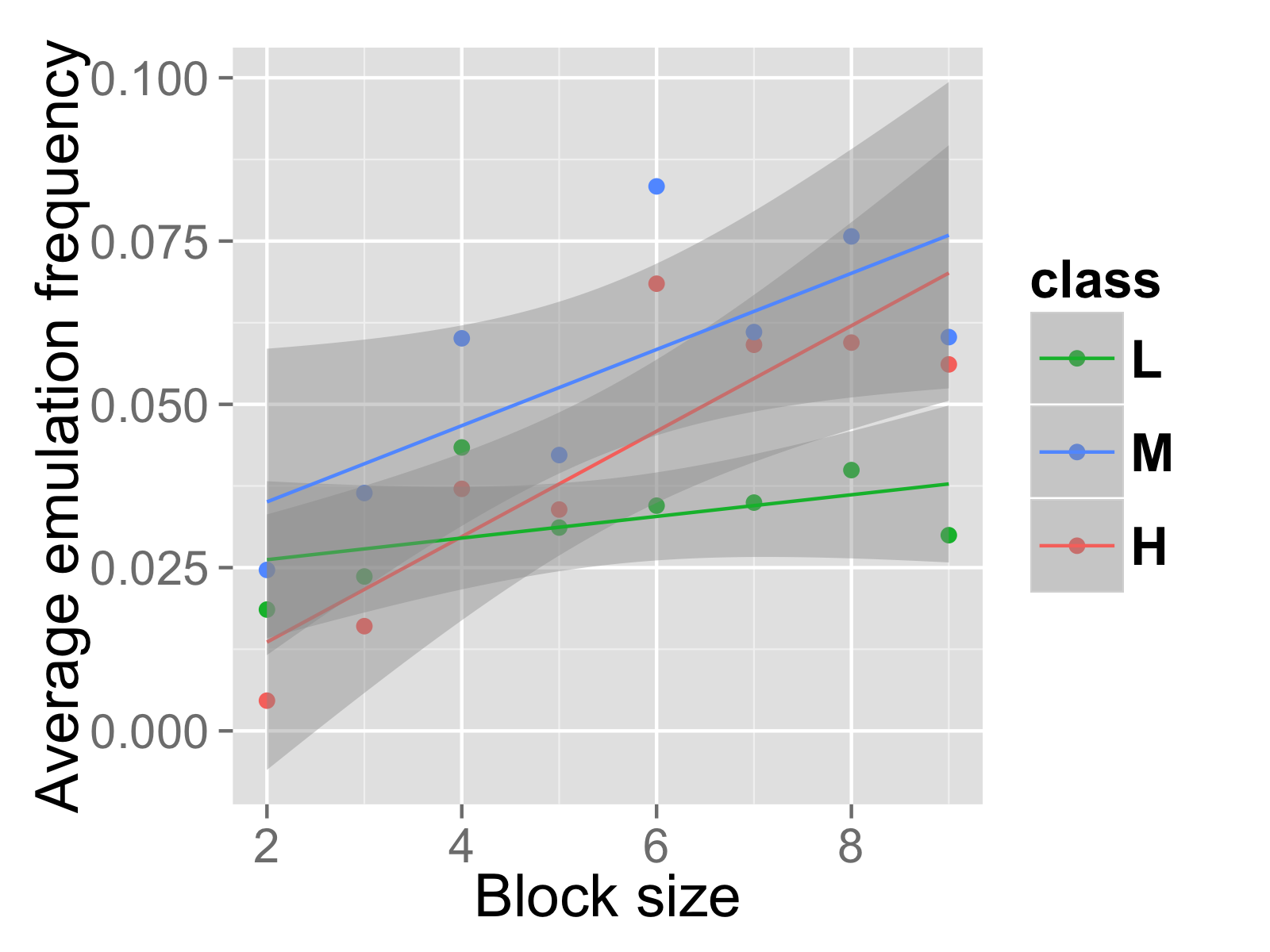}\\
  \medskip
  Non-collapsed\\
  \centering
 \textbf{ECA}\hspace{4.5cm}\textbf{GCA}\\
    \includegraphics[width=57mm]{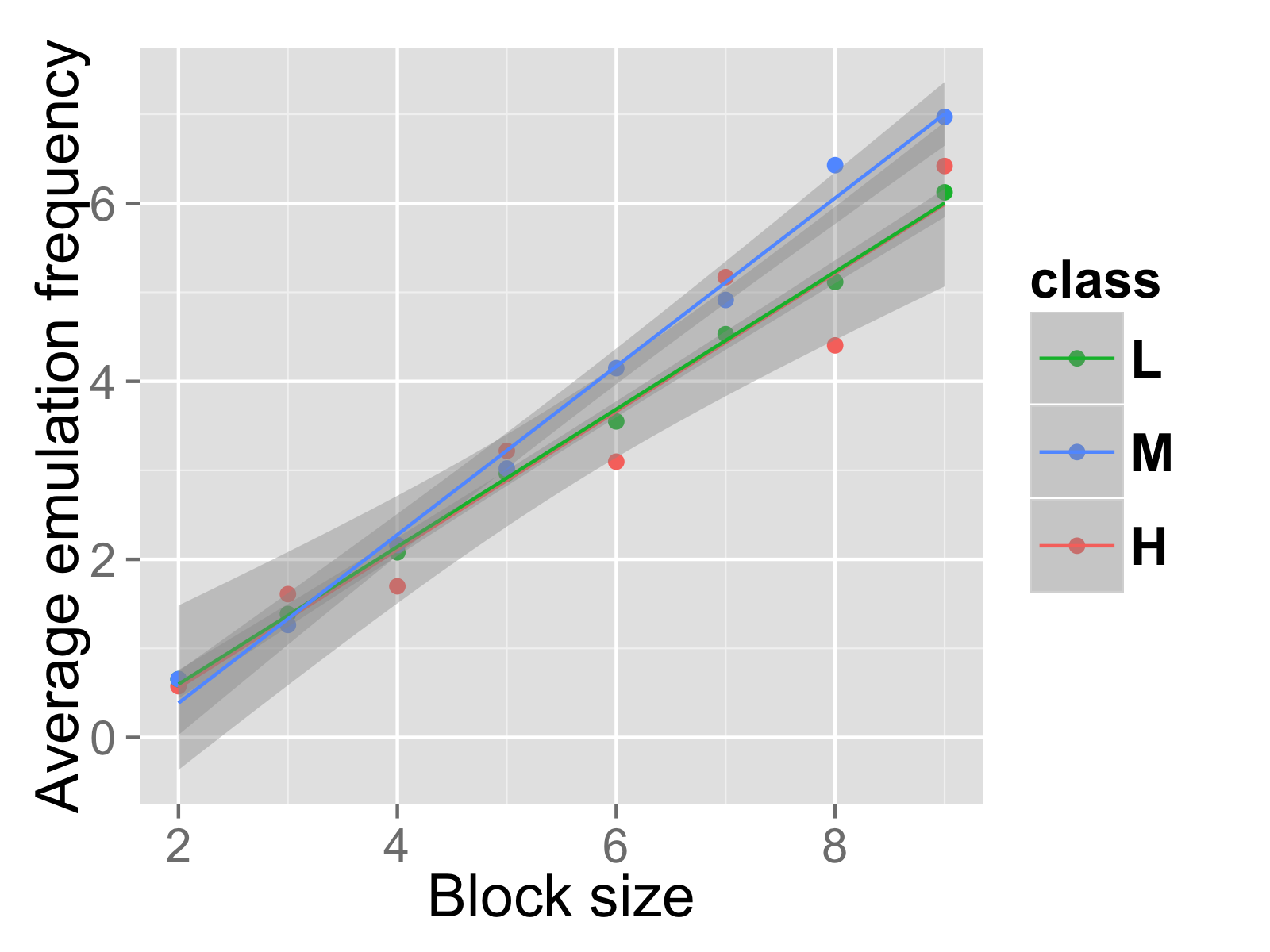}  \includegraphics[width=53mm]{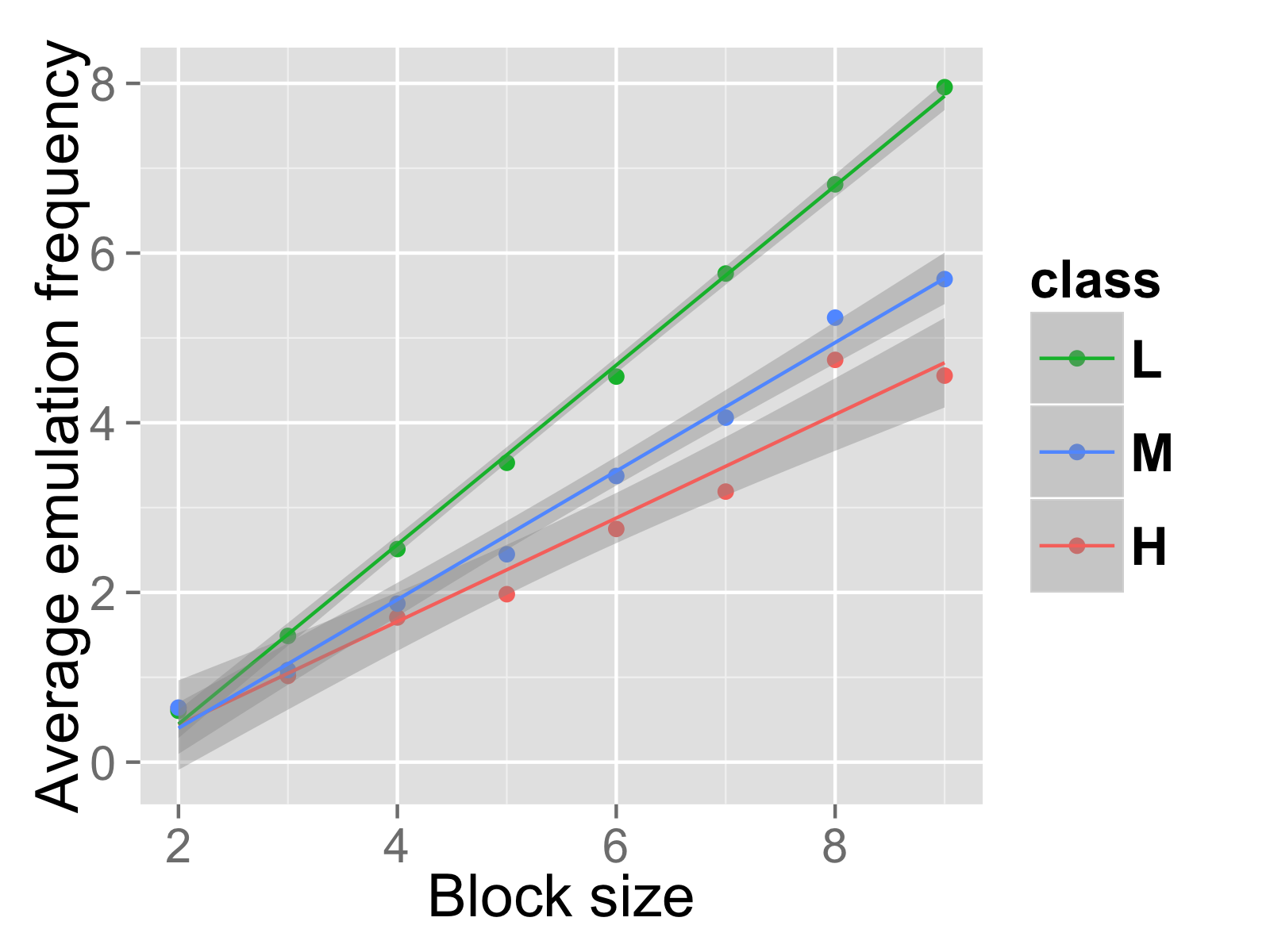}
\caption{\label{maina} Frequency counts for collapsed and non-collapsed cases. ($x$-axis block size and y-axis $\log$ average of emulation frequency per CA class (ECA  for Wolfram classes and GCA divided into Low and High complexity). For the collapsed case (i.e. distinct frequencies not considering different encodings per block size) the data is not enough. However, for non-collapsed ECA and both collapsed and non-collapsed GCA, the average increase of non-trivial reprogrammability per rule is confirmed.}
\end{figure}

Figs.~\ref{maina} demonstrates that among the emulators in ECA and GCA, each of them emulates on average (for all classes) a larger fraction of other rules as a function of compiler size. Plots for ECA and GCA for collapsed (only differently emulated) and non-collapsed cases (counting repetitions of emulations) show that all frequencies do increase in linear fashion. 

As shown in Fig.~\ref{maina}, the reprogrammability phenomenon in the emulation space is not dominated by trivial emulation. Taken together, the results suggest that reprogramming capabilities increase for higher rule and compiler size, and that every rule in every rule space is effectively non-trivially reprogrammable. While the growth of the plot on the right could be driven by trivial emulations, e.g., emulations of rule 0, the plot on the left only counts each rule emulation once and cannot therefore be driven by trivial cases, in light of the fact that most rules actually eventually become of the highest complexity class according to the asymptotic behaviour investigated in~\cite{zenilchaos}.

\subsection{The compiler space}

\begin{figure}[h!]
\begin{tabular}{cc}
  \includegraphics[width=60mm]{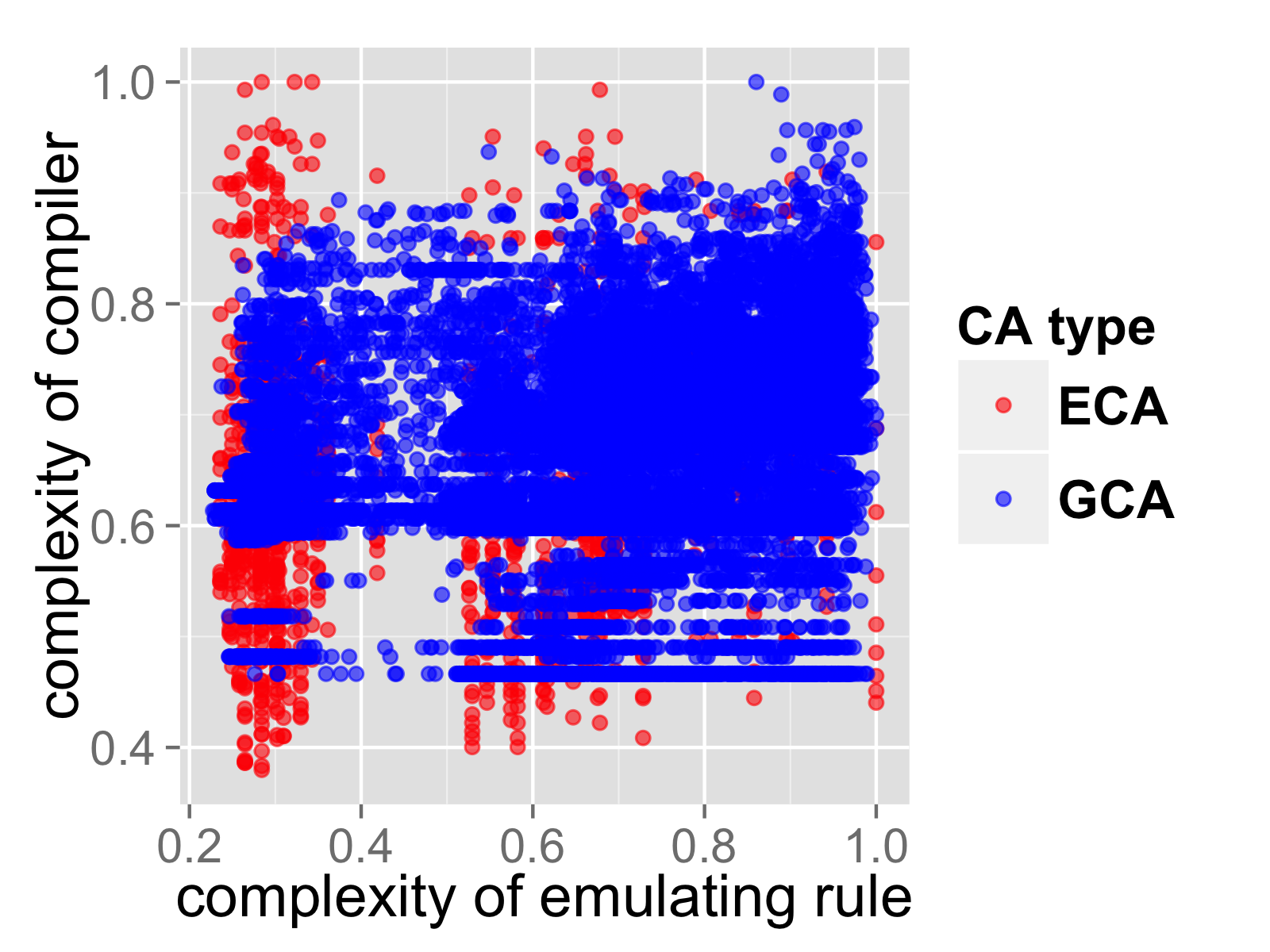} &   \includegraphics[width=60mm]{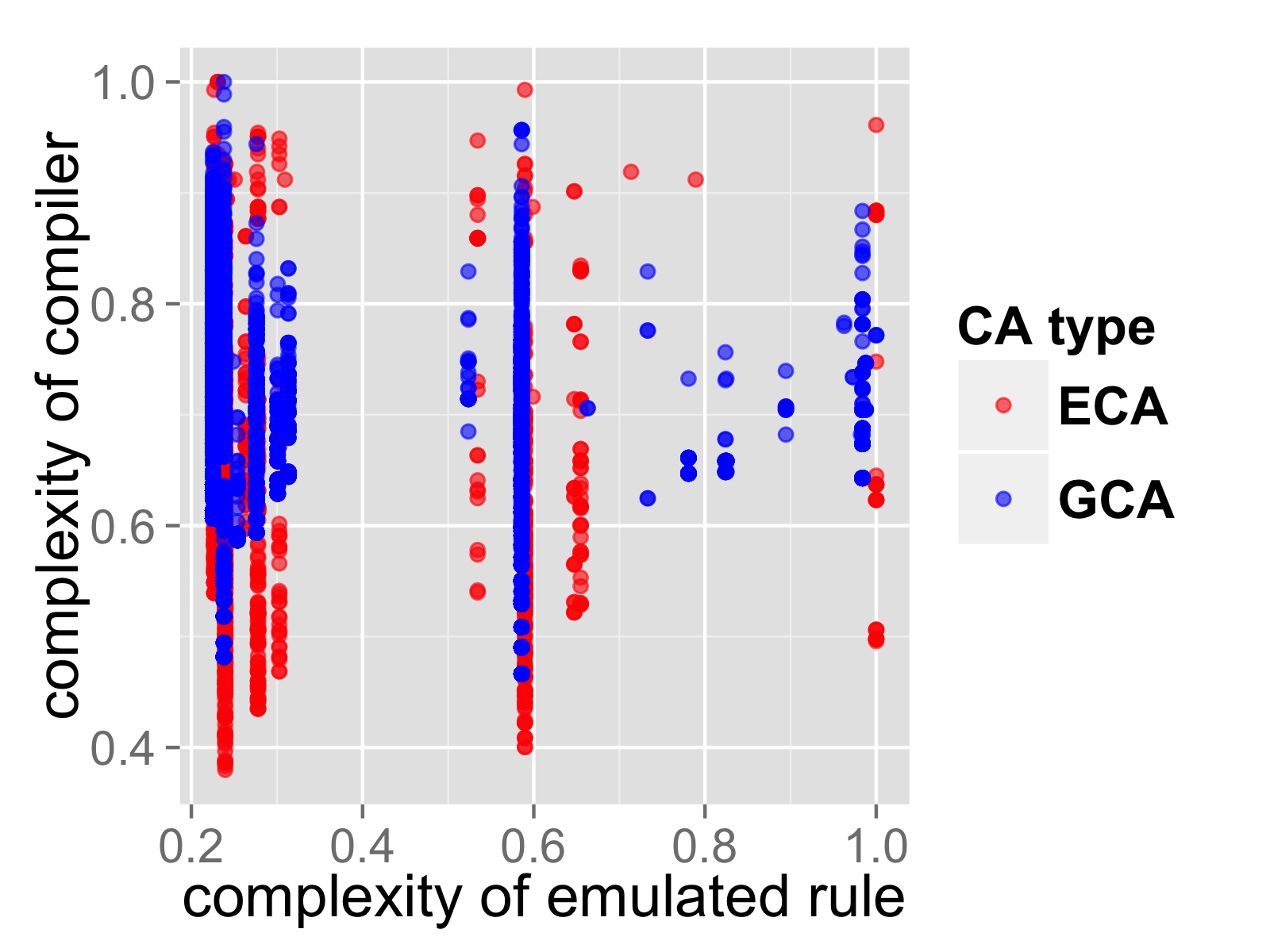} \\
(a)  & (b)  \\[6pt]
 \includegraphics[width=60mm]{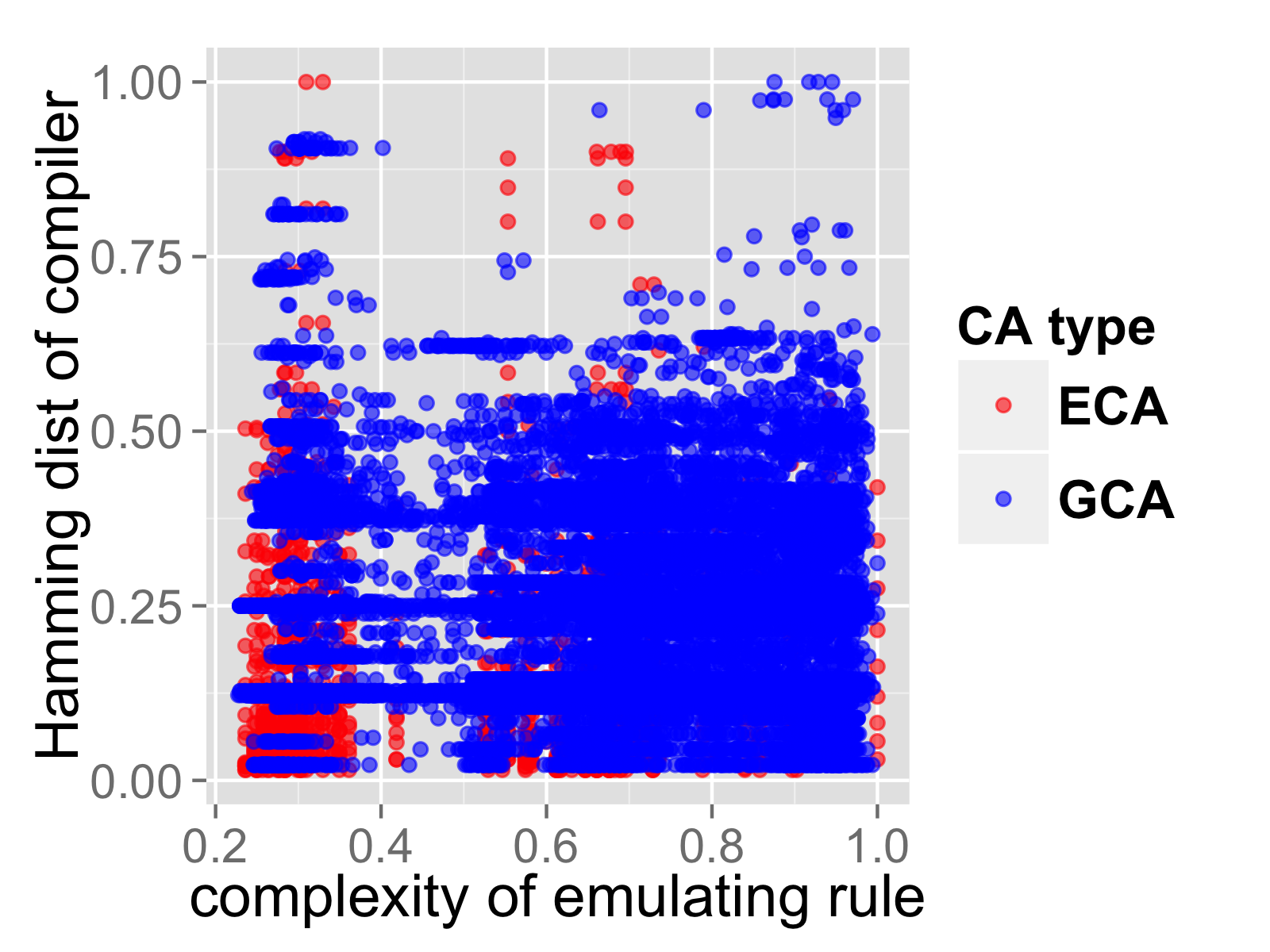} &   \includegraphics[width=60mm]{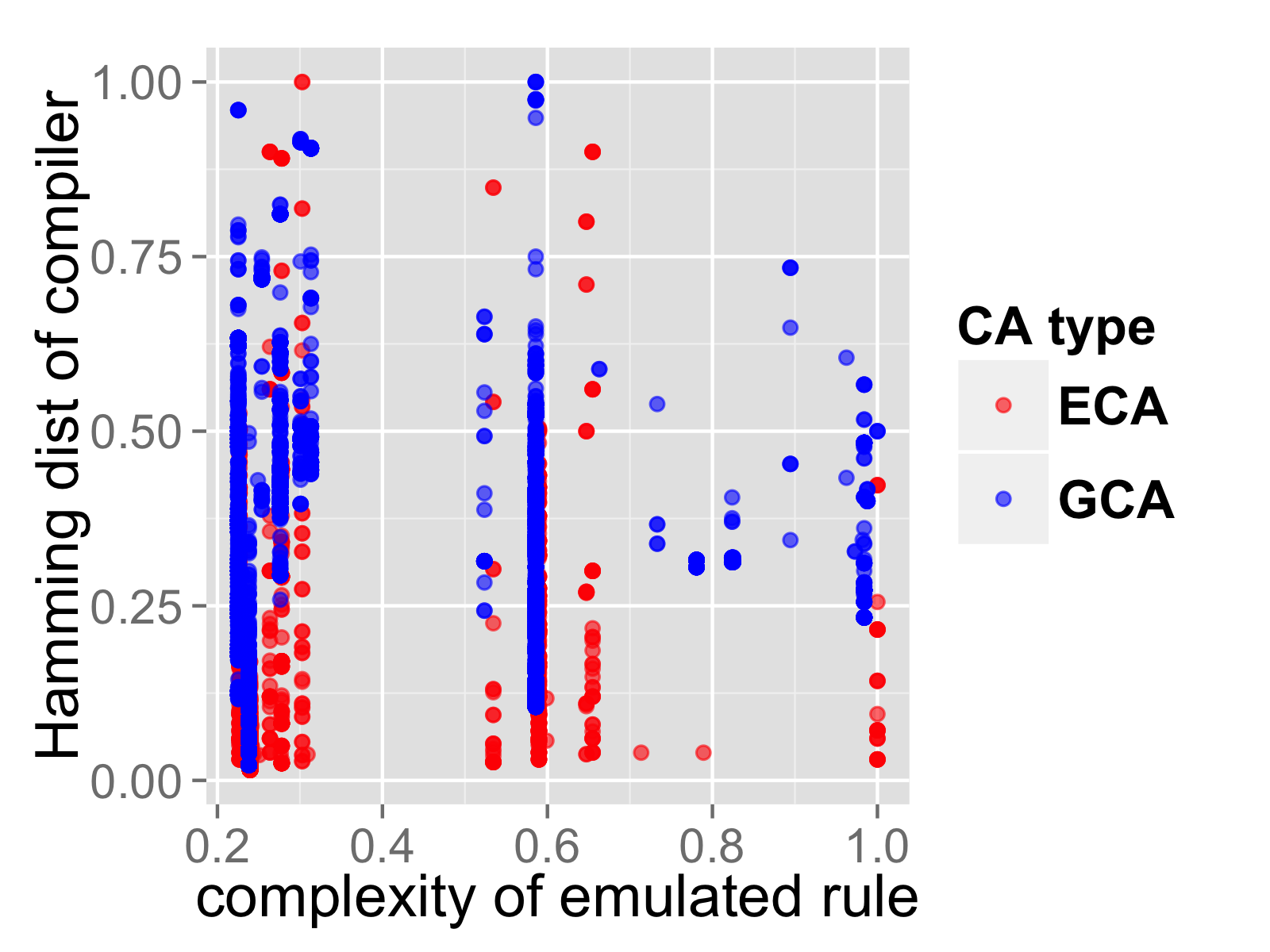} \\
(c)  & (d)  \\[6pt]
\end{tabular}
\caption{\label{fig_ECA_Compiler_Complexity} Exploring the space of compilers. Complexity comparison of ECA and GCA emulation rule pairs and the associated compilers for block size 12 (ECA) and block size 9 (GCA). Figure (a) shows the complexity of emulating rule vs.compiler complexity. Figure (b) shows the complexity of emulated rule vs.compiler complexity. Figure (c) shows the complexity of emulating rule vs. compiler Hamming distance. Figure (d) shows the complexity of emulated rule vs. compiler Hamming distance. These plots suggest that there is no hacking strategy based on complexity, i.e. compiler complexity  or compiler similarity provides no clue to the computational capabilities of the resulting system. In other words, the emulating rule and compiler tuple is non-additive.}
\end{figure}

Fig.~\ref{fig_ECA_Compiler_Complexity} shows that there is no correlation between the compiler complexity (entropy and compression) and emulated or emulating rules. That is, more complex compilers do not necessarily do better or worse when it comes to allowing more or less complex rules to behave in one way or another. In other words, if one assumed that using a high complexity compiler to make a simple rule behave like a complex one amounted to a clever ruse, this is not the case, because even simple compilers can make simple rules behave like complex ones, while both simple and complex compilers can make complex rules behave like simple ones. 

Furthermore, the compilers have no similarity, as measured by their Hamming distance as well as by their complexity, indicating that in general only brute force exploration is possible when attempting to reprogram computing programs. Which suggests that there is no cleverer hacking strategy that might be resorted to. 

While it has been known that universality can be achieved by coupling two simple systems-- e.g. a 2-state 3-symbol Turing machine that uses only 5 instructions is universal when coupled with a finite automaton~\cite{maurice1,maurice2}--the block emulation approach merely requires an initial input translation that only interacts with the computer program at step $t=0$, hence effectively rewriting the initial condition only, and not intervening in any way afterwards, leaving the actual original computer program to carry out all the computation.

 Additionally, Fig.~\ref{fig_ECA_Compiler_Complexity} shows that compiling produces a mapping between emulating and emulated rules that emerges from a uniform complexity behaviour but produces 3 apparent clusters of emulating complexity rules, which we found interesting to report and for which we have no hypothesis, given that the same complexity measures (Shannon entropy and $NC$) are used for both the emulating and emulated rules. In other words, the compiling process seems to result in groups of rules with similar complexity. 

\section{Conclusions}

One negative reading of the results is a collapse of the qualitative complexity classes. However, the number of simple rules not emulating other simple and complex rules is of measure 0 with respect to the rule/program space size (which also determines the maximum program size), in contrast to those able to emulate other rules which saturate every rule/program space. Thus, while there are no essential differences separating classes by their computational capabilities (other than trivial versus the rest), the results represent a definite alternative to fundamental classification schemes that attempt to provide information about the computing capabilities of the programs classified. 

The results reported here imply that if computer programs are taken as toy models of digital ``universes'', the initial conditions are as fundamental or more so than the underlying rules or instructions. Rules in these computer programs can be seen as physical laws that cannot be broken, but as we have seen, even the simplest ``physical laws'' are able to emulate other computer programs, including complex ones, disregarding their otherwise shallow behaviour for simple or random initial conditions, thereby occluding much and revealing little of the nature of the laws or even the initial conditions, which are only revealed by the number of compilers which a rule can use to behave in a particular, e.g. complex, fashion. 

Because of the undecidability of the halting problem related to block emulation and intrinsic universality~\cite{mazoyer}, analytical results relevant to this paper seem difficult if not impossible to obtain. This implies that ultimately these results could be deceiving because emulations can
occur between rules of very different apparent complexity because block encoded configurations in the emulator CA have measure 0 (general case). For instance, by just adding a spreading state to an intrinsically universal CA, one gets
a universal simulator with the simplest apparent complexity (any cell turns into the spreading state in expected constant time on random configurations) for that time scale.

The results suggest, however, that at the studied time scale the more complex a system the more sensitive it is to its initial conditions, thereby becoming fundamentally less predictable and that even the most complex and random looking systems can effectively and qualitatively be reprogrammed. Indeed, the common belief was that CA of Wolfram's class 3 were in some sense ``too wild" (of highest entropy) to be controlled and made to perform useful computations (i.e. to be practically reprogrammable), but we have shown that this is not the case in any fundamental sense.

Furthermore, examples of simple cellular automata (e.g. classes 1 and 2) that were believed to be of limited computational power turned out to be highly programmable too, able to carry out even the most complex qualitative computations, thereby strongly suggesting that even the simplest rules are strong candidates for Turing universality. The results shown bear out a generalized phenomenon of pervasive reprogrammability.





\newpage
\appendix

\section{Supplemental Material}

\subsection{Reducing a program (CA) rule space}
Local rules define the dynamical behaviour of CA. However, not all rules show essentially different dynamical behaviour. To focus on the number of rules in a rule space which show essentially different dynamical properties, one can introduce the following symmetry transformations: \\

{\bf Reflection}: 
\begin{equation}
 f_r(x_1,x_2,...,x_n)=f(x_n,...,x_1,x_2)
\end{equation}

{\bf Conjugation}: 
\begin{equation}
 f_c(x_1,x_2,...,x_n)=q-1-f(q-1-x_1,q-1-x_2,...,q-1-x_n)
\end{equation}

{\bf Joint transformation}, i.e., conjugation and reflection:
\begin{equation}
 f_c\circ f_r (x_1,x_2,...,x_n)=q-1-f(q-1-x_n,...,q-1-x_1,q-1-x_2)
\end{equation}

Under these transformations two CA rules are equivalent and they induce equivalence classes in the rule space. Taking from each equivalence class a single representative (by convention the one with the smallest rule number), one gets a set which contains essentially different rules, i.e., rules which show different global behaviour.

 Let $G(\chi(f_r),\chi(f_c),\chi(f_cr))$ be a group under the operation $\circ$ acting on a set $X$ of all possible neighbourhood templates. Using the orbit counting theorem one can state the following theorem:

\begin{equation}
 \frac{1}{|G|}\sum_{g\in G}\chi(g)=\frac{\chi(f_I)+\chi(f_r)+\chi(f_c)+\chi(f_{cr})}{4}
\end{equation}
with $\chi(g)$ being the number of elements of $X$ fixed by $g$.

For the PCA one finds:
\begin{equation}
 \chi(g)=\frac{2^{4}+2^{3}+2^{2}+0}{4}=7 ,
\end{equation}
For the ECA one finds:
\begin{equation}
 \chi(g)=\frac{2^{8}+2^{4}+2^{6}+2^{4}}{4}=88,
\end{equation}
and 
for the GCA one finds:
\begin{equation}
 \chi(g)=\frac{2^{16}+2^{10}+2^{8}+0}{4}=16\,704
\end{equation}
essentially different rules. We will therefore only subject these essentially different rules to analysis throughout the paper.

\subsection{Linear (Additive) Rules}
\label{additive}

A \textit{linear} rule is defined by the additivity condition: 

\begin{equation}
\label{additivity_rule}
 f_{a b c} [x_{n-1},x_n,...,x_{n+1}]= a x_{n-1} \otimes b x_{n} \otimes  c x_{n+1}, a,b,c \in \{0,1\}
\end{equation}

If a CA rule satisfies the additivity condition (\ref{additivity_rule}), the factors a, b, c are integer constants 0 or 1. Linear rules are usually simple rules. However, there are linear rules which exhibit complex behaviour, such as Wolfram class 3 rules. Looking at the different rule spaces one finds:

\begin{enumerate}
\item PCA rule space: There are 4 essentially different linear rules: (0, 3, 6, 10). Besides the trivial rule 0, rule 3 and 10 are Wolfram class 2 rules. The only complex rule is 6, which is a Wolfram class 3 rule.
\item ECA rule space: There are 8 essentially different linear ECA rules: (0, 15, 51, 60, 90, 105, 150, 170, 204). Rules 0 (a=b=c=0), 15 (right shift with toggle), 170 (left shift rule), and 204 (identity rule) are Wolfram class 2 rules. The rules 60,90,105  (sum rule with toggle), and 150 (sum rule) are Wolfram class 3 rules.
\item GCA rule space: There are 14 essentially different rules (0, 255, 3855, 4080, 13260, 15420, 15555, 21930, 23205, 27030, 38550, \
39270, 43690, 52428). Rules 255, 3855, 43690, and 5242 are  Wolfram class 2 rules. However, rules 4080, 13260, 15420, 15555, 21930, 23205, 27030, 38550, and 39270 are Wolfram class 3 rules.

Linear CA have the advantage that one can perform an algebraic analysis in order to determine certain of their global properties~\cite{wolframmartin,bruyn}. We will refer to linear CA throughout the paper in the context of their role in block emulation, and point out some of the special roles they play.
\end{enumerate}

\section{Cellular automata rulespaces}

\subsection{PCA rulespace}

We first studied the emulation network of the small PCA rulespace of which there are only 16 rules. The number of essentially different PCA rules is only 7 and they are 0, 1, 2, 3, 6, 8 and 10. Of these, the following PCA rules are linear (also known as additive~\cite{nks,bruyn} c.f.~\ref{additive}): 0, 3, 6, 10. The network graph for PCA is given in Fig. \ref{fig_PCAClasses}. Also included in the Figure are the non-essentially different PCA rules, to show the full symmetry of the emulation graph.The PCA rules can be classified according to Wolfram's classification, as follows (the numbers in brackets are equivalent rule numbers): 

\begin{myitemize}
\item Class 1 PCA rules are 0 (15), 8 (14)
\item Class 2 PCA rules are 1 (7), 2 (4, 11, 13), 3 (5), 10 (12)
\item Class 3 PCA rules are 6 (9)
\item There are no Class 4 rules.
\end{myitemize}

Self-emulating rules are 8, 14, 10, 5, 12, 3, 6, 9. The minimal emulation depth, i.e., emulation block size, is usually 2 but is maximally 3 for rules 2, 11 and 13, for rule 4 emulating rule 0, as well as for rules 5 and 3 emulating themselves.

\subsection{ECA rulespace}

The ECA rule space contains 88 essentially different rules and 9 linear (additive) rules (0, 15, 51, 60, 90, 105, 150, 170, 204). The Wolfram classification groups the ECA rules as follows:

\begin{myitemize}
\item 8 Class 1 ECA rules (0, 8, 32, 40, 128, 136, 160, 168)
\item 65 Class 2 ECA rules (1, 2, 3, 4, 5, 6, 7, 9, 10, 11, 12, 13, 14, 15, 19, 23, 24, 25, 26, \
27, 28, 29, 33, 34, 35, 36, 37, 38, 42, 43, 44, 46, 50, 51, 56, 57, \
58, 62, 72, 73, 74, 76, 77, 78, 94, 104, 108, 130, 132, 134, 138, \
140, 142, 152, 154, 156, 162, 164, 170, 172, 178, 184, 200, 204, 232)
\item 11 Class 3 ECA rules (18, 22, 30, 45, 60, 90, 105, 122, 126, 146, 150)
\item 4 class 4 rules (41, 54, 106, 110)
\end{myitemize}

\begin{figure}
\centering \subfigure[\ Class 4 emulations of Class 4 rules]{\includegraphics[width=12.0cm,angle=0]{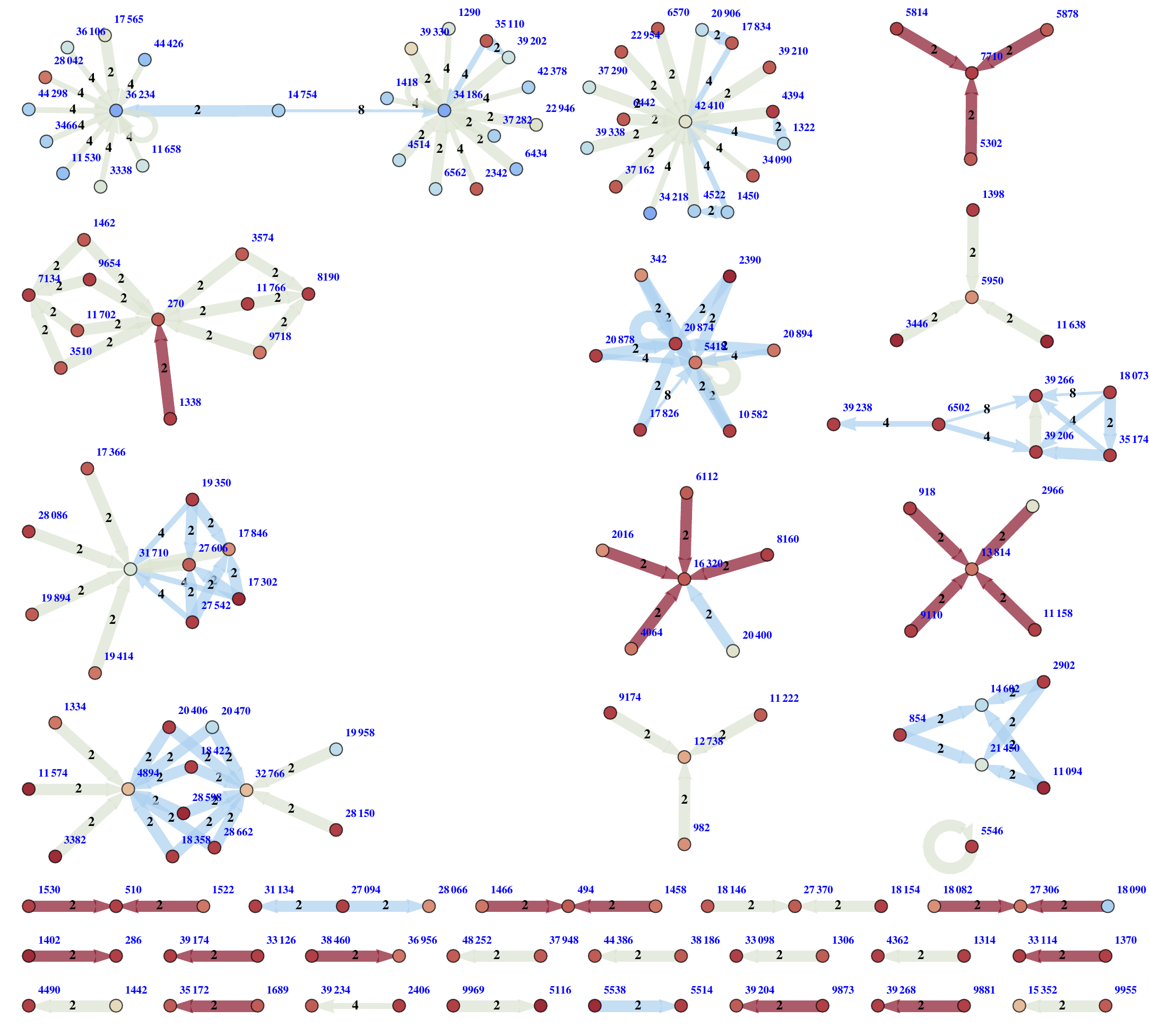}}
\caption{\label{fig_GCAClass4}Rule emulation graph of GCA class 4 emulating class 4 rules. The full graph is too large to be included in a single page.
}
\end{figure}

Fig.~\ref{fig_GCAClass4} shows all rule emulations for class 4 GCA. It is interesting  that no class 4 rule seems to be able to emulate another class 4 rule up to block size 20 in this rulespace.

\begin{figure}
\centering
\includegraphics[width=5.5cm,angle=0]{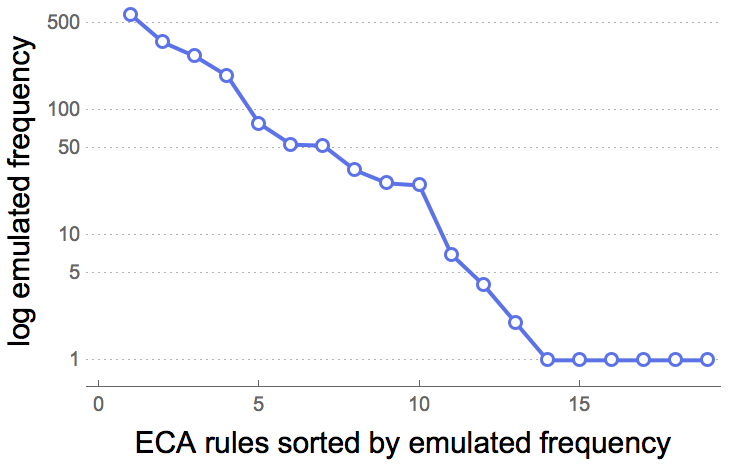}\vspace{.1cm} \includegraphics[width=5.5cm,angle=0]{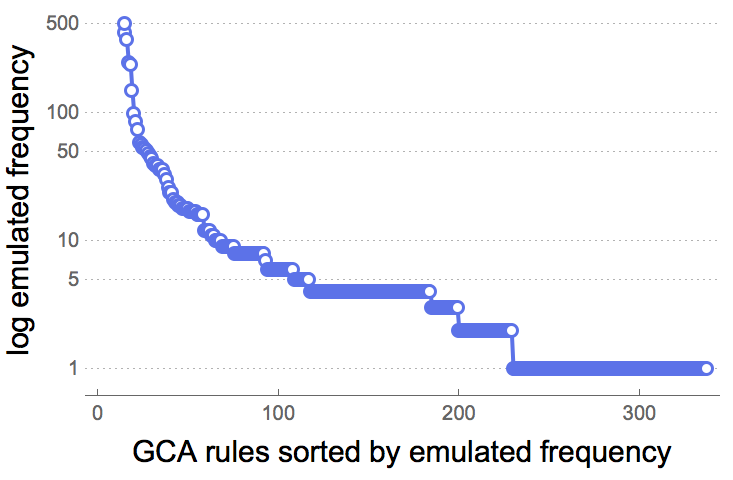}\\
\medskip
\includegraphics[width=5.7cm,angle=0]{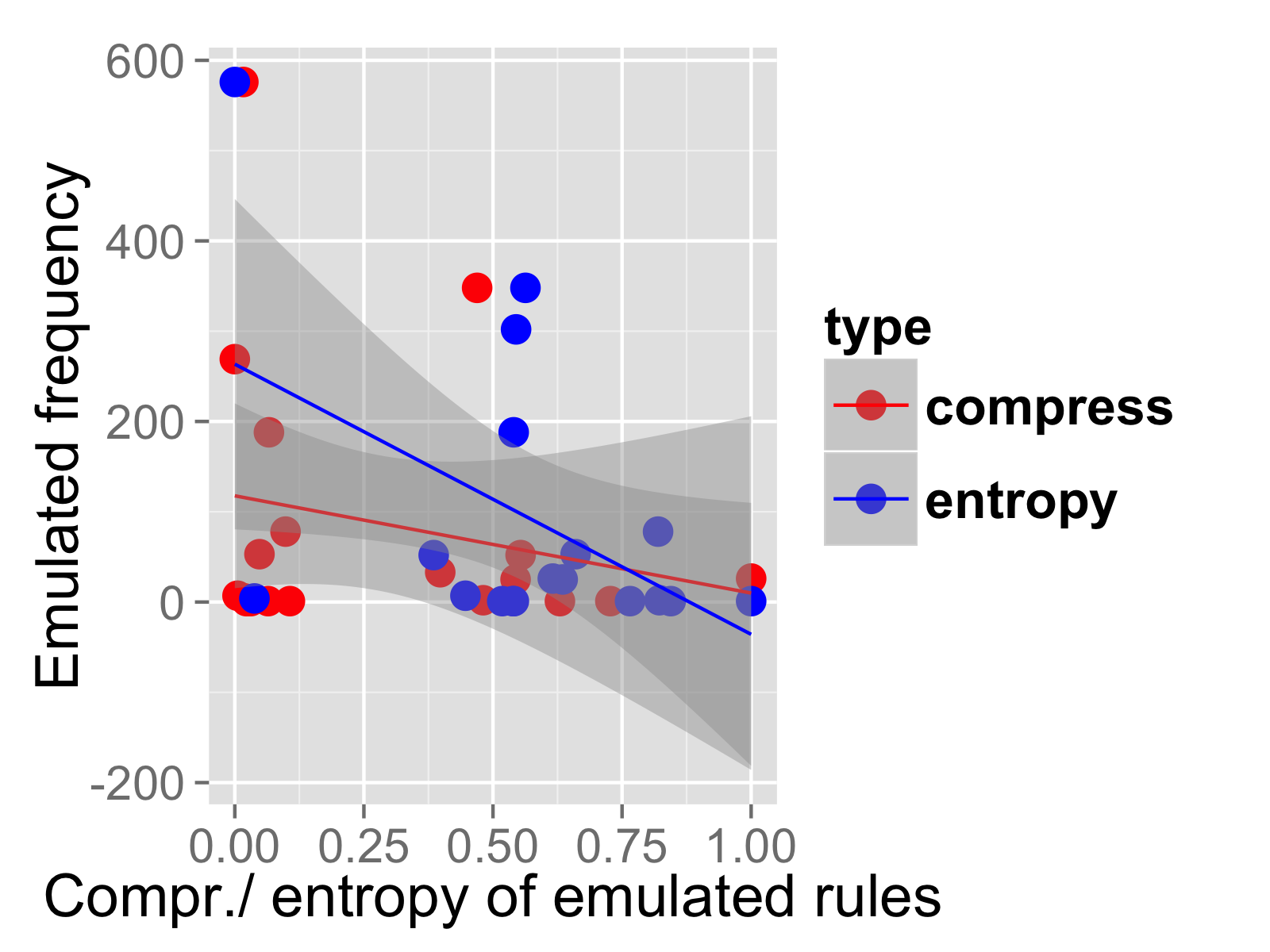}\includegraphics[width=5.7cm,angle=0]{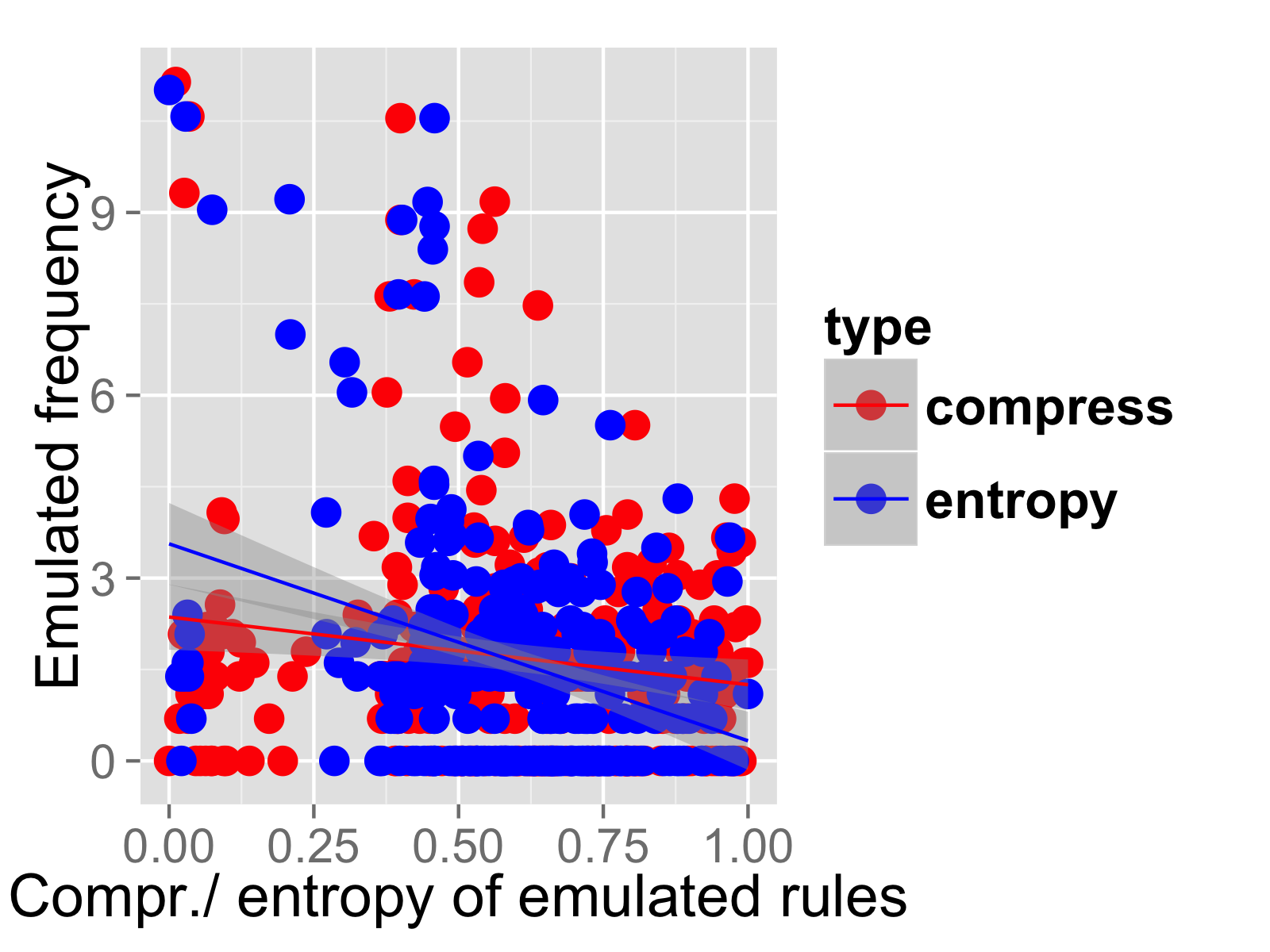}
\caption{\label{ap}Top: Algorithmic probability of ECA and GCA from emulation frequency. Emulation frequency provides an indication of each rule complexity by means of the \textit{algorithmic coding theorem} relating frequency of production and Kolmogorov complexity. Bottom: When comparing frequency of emulation versus Entropy rate and Compress that we used to classify GCA, we find an expected correlation (linear fitting lines are shown), thereby validating both directions: compression as an index for rule complexity to classify GCA to approximate its ``Wolfram class'', and the frequency-based complexity in agreement with both Compress (a form of entropy rate for fixed window length) and intuitive complexity location of ECA along the distribution.}
\end{figure}

Fig.~\ref{ap} (top) classifies ECA and GCA rules by their emulation frequency, which is related to the number of times a computer program is produced from another computer program and is therefore close to the concept of algorithmic probability. The induced order is in agreement with the intuition of the complexity of the distributed rules and also correlates (Fig.~\ref{ap} (bottom)) with the lossless compression index used througought this paper to classify GCA in a systematic formal approach equivalent to Wolfram's ECA classes.

The distributions in Fig.~\ref{ap} (top) follow the characteristic exponential decay~\cite{miracle} and the locations of the rules correspond to their expected complexity. For example, trivial rules, such as ECA rule 0, are amongst the most frequently emulated, but more random-looking ones such as rule 30 and class 4 ECA are in the tail. 

\subsection{GCA rulespace}

In order to classify the much larger rule space of GCA, we adopt the classification scheme described in~\cite{zenilca,zenilchaos} based on the measure $NC$, as described above. 

The so-called GCA rule space~\cite{nks} has 16\,704 essentially different rules. We investigate rule emulation up to block size 10. We divide the GCA rule space according to our measure $NC$, described above. 

We are taking a closer look at the emulation capacity of ECA and GCA under block transformation. For instance, in the GCA rule space, out of 60\,737 distinct GCA rule pairs, there are 42\,996 rule pairs which have a linear GCA as the emulated rule (only 17 rule pairs have a linear GCA as the emulating CA). The number of remaining rule pairs which do not contain any linear GCA rules is 17\,741. 

In order to look for a general pattern, we separate the CA emulating other CA into groups of varying complexity. For the ECA rule space we use Wolfram's classes for the grouping. Here we identify the group of high complexity ECA rules with Wolfram classes 3 and 4, the medium complexity ECA with Wolfram class 2, and finally the low complexity rules with Wolfram class 1. In the case of the GCA rule space, we use the compression distance $NC$, described above, in order to differentiate high, medium, and low complexity classes beyond the ECA rule space. 

For each of the complexity groups we count the distinct number of emulation rule pairs for different emulation block sizes. For both ECA and GCA we go up to emulation block size 10. The result of the analysis is shown in Fig.~\ref{fig_EmulFreq}.\\

We find that:

\begin{myitemize}
\item Low complexity ECA for the given block size range are capable of emulating only other low and medium complexity ECA. The emulation frequency for both complexity classes is clearly linear, increasing by block size. Here, the slope for the medium complexity rules is greater. In the case of GCA, the picture is similar. However, the slope of low complexity rule emulation is greater than the one for medium complexity GCA.
\item Medium complexity ECA are capable of emulating all three complexity classes (low, medium and high). Again, the medium complexity classes are more frequently emulated than low complexity ECA. The emulation frequency of high complexity rules appears to be linear as well, with an increasing slope. The picture is very similar for GCA. However, the emulation frequencies for low complexity GCA are larger than the ones for medium complexity GCA.
\item High complexity ECA are capable of emulating all three ECA complexity classes (low, medium and high). 
\item Overall, the high complexity rules tend to emulate other high complexity rules with high frequency. Further analysis shows that most such high complexity to high complexity rule emulations are performed by class 3 linear rules. This is also the case with the high complexity GCA emulated by medium complexity GCA.
\item In summary, simple rules can be reprogrammed to behave like low and medium complexity rules but less often as high complexity rules (as seen in ECA), but medium and high complexity rules can emulate low and medium complexity rules almost equally well, though they are more difficult to reprogram to emulate other high complexity rules, even when they can in fact emulate them. 

\end{myitemize}

\section{Emulation by input translation}

\subsection{Block transformation}

To find possible block transformation candidates for a given CA rule and block size $k$ one can adopt the following algorithm:

\begin{enumerate}
\item For ECA set the CA range to 1 and for GCA set range to $3/2$.
\item Given the block size one insert all possible block $k$-tuples into the CA as initial conditions of a given rule and let the CA evolve for $k$ steps. Check whether output equals input. Return matches. This is implemented in function \texttt{CandidateA[]}.
\item Insert as initial conditions the complement of the set returned by function \texttt{CandidateA[]} in respect to all possible $k$-tuples into CA rule and let evolve for k steps. Check whether output equals all transform tuples which were returned as candidates by function \texttt{CandidateA[]}. Return matches. This is implemented in function \texttt{CandidateB[]}.
\item Insert as initial conditions the complement of the union of sets returned by the functions \texttt{CandidatesA[]} and \texttt{CandidateB[]} in respect to all possible $k$-tuples into CA and let evolve for $2\times k$ steps. Check whether output equals the input and return matches. This is implemented in function \texttt{CandidateC[]}.
\item Use $k$-tuple pairs created from sets returned by functions \texttt{CandidateA[]} and \texttt{CandidateC[]} and with them perform a block transformation on the De Bruijn sequence. i.e. for ECA use \texttt{\{0, 0, 0, 0, 1, 0, 1, 1, 1, 1, 0, 1\}} and for GCA use \texttt{\{0, 0, 0, 0, 1, 0, 0, 1, 1, 0, 1, 0, 1, 1, 1, 1\}}.  Insert these as initial conditions into CA and let evolve for $k$ steps.  The output is checked if it contains all the input $k$-tuples. If so, return true, otherwise false. This is implemented in the function \texttt{CheckCandidate[]}.
\item Take the set returned by function \texttt{CandidateB[]} and insert $k$-tuples into CA and let evolve for $k$ steps. Form a set of output $k$-tuples and create transformation pairs by cross joining it with set returned by function \texttt{CandidateB[]}, i.e. the input. First run the elements of set returned by function \texttt{CandidateB[]} through the CA with given rule for k steps and combine the output with the input elements of set returned by function \texttt{CandidateB[]} to form pairs. Check whether these pairs are real candidates by passing them through the function \texttt{CheckCandidate[]}.
\item Collect all $k$-tuple pairs for which the output from the function \texttt{CheckCandidate[]} returns true. The resulting set contains valid block transformations for given rule and block size. 
\end{enumerate}

Steps 5 to 7 are implemented in the function \texttt{Candidates[]} written in the Wolfram Language and uploaded as a package of Wolfram Mathematica notebooks to GitHub (available at \url{https://github.com/algorithmicnature}).

\begin{figure}
\centering \subfigure[\ Class 4 emulations of Class 3 rules]{\includegraphics[width=13cm,angle=0]{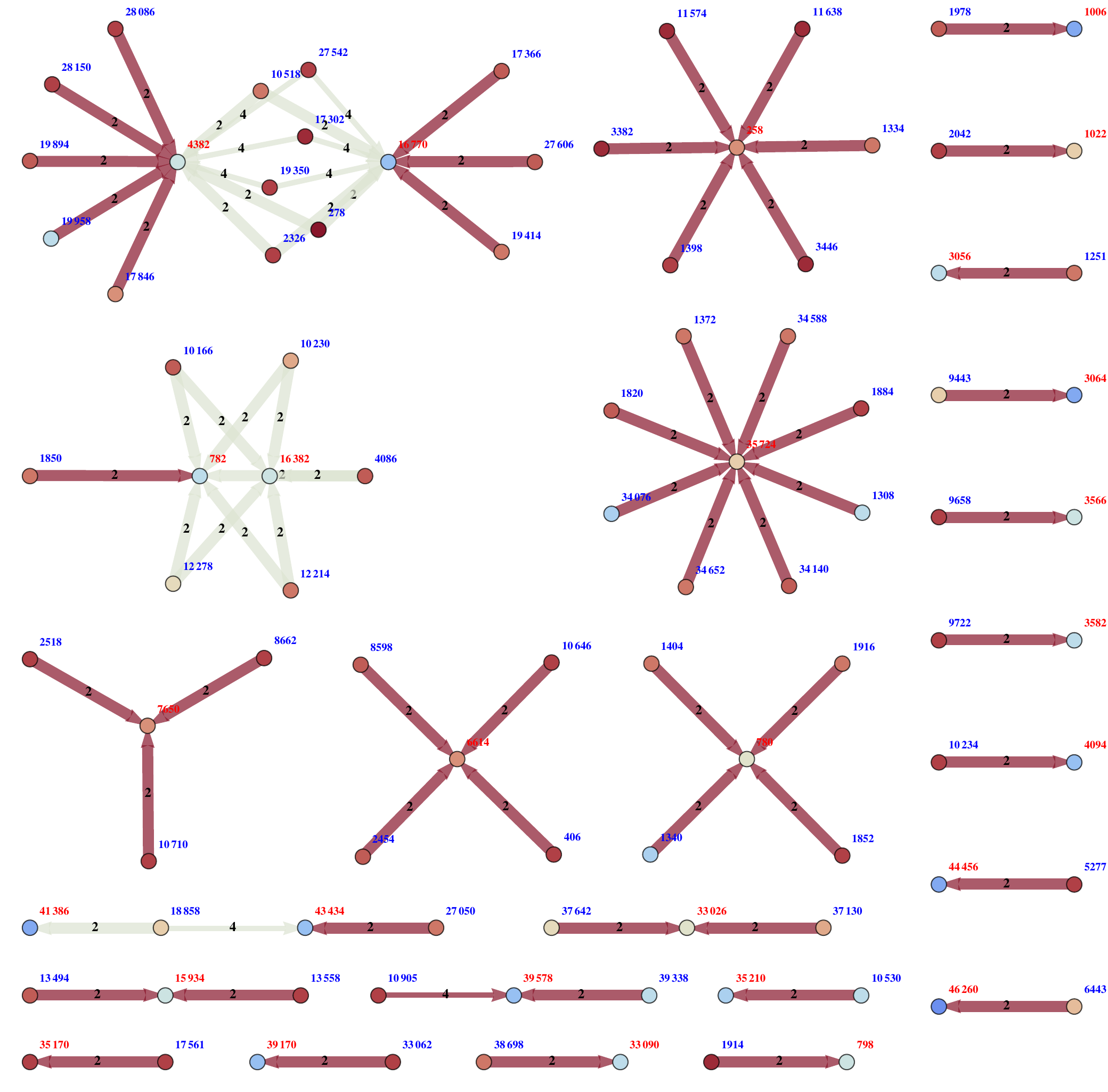}}
\caption{\label{fig_GCAClass4b}Rule emulation graph of GCA class 4 emulating class 3 rules. The full graph is too large to be included in a single page.}
\end{figure}

For a set of CAs one can construct a directed graph (see Figs.~\ref{fig_PCAClasses}, \ref{fig_GCAClass4b}, \ref{fig_GCAClass4c} and \ref{fig_GCAClass4d}) of rule emulations by taking rule numbers (both the emulating and emulated) as the vertices and connecting the vertices (rules) when 2 CA are emulated with at least one compiler. Hereby we set the direction of the edges to point from emulator rule to emulated rule, and we call these graphs \textit{emulation networks}.

\begin{figure}
\centering
\includegraphics[width=11cm,angle=0]{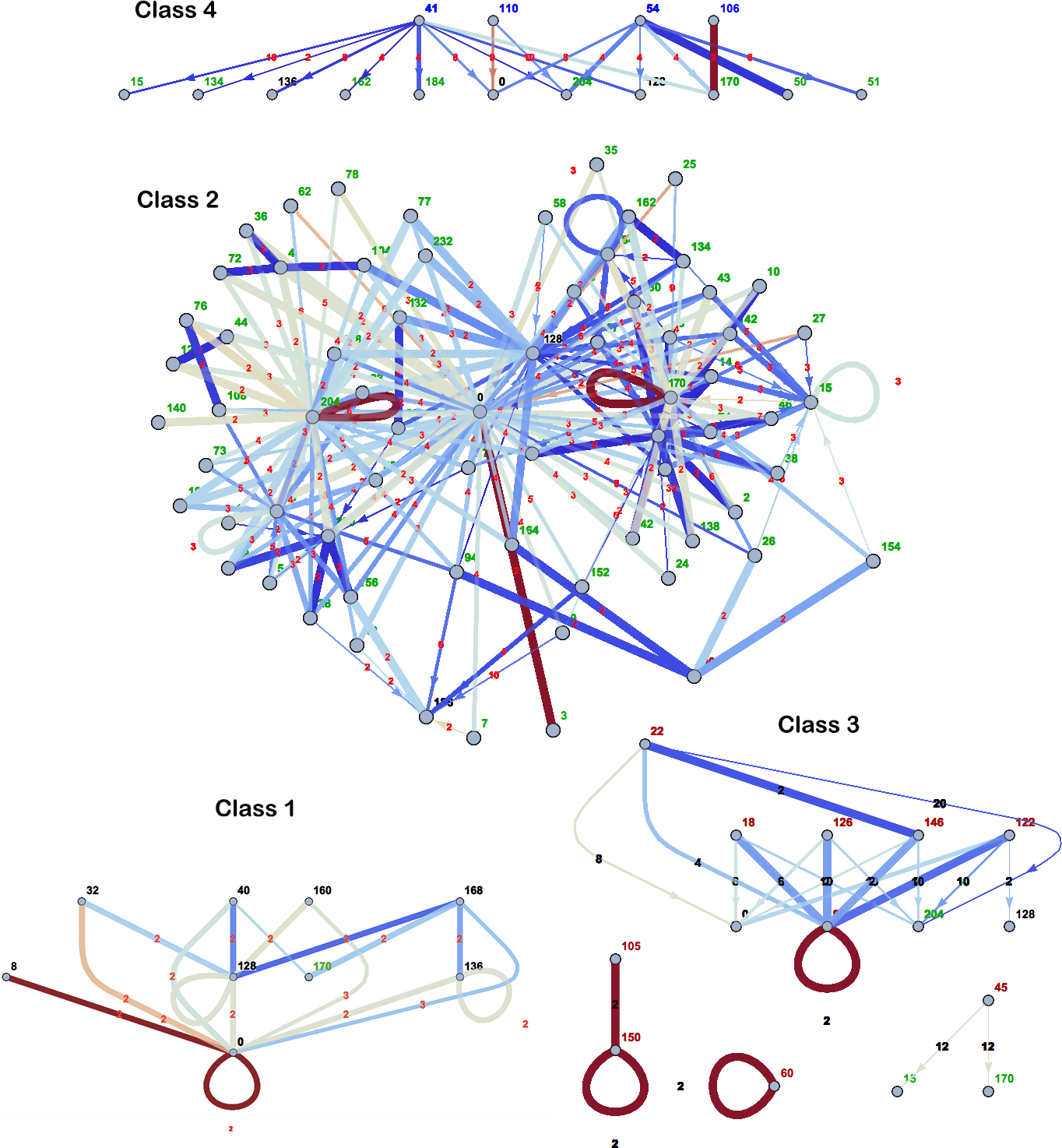}
\caption{\label{emulationnetwork}Rule emulation network for ECA. Network constructed with block size up to 20 for Wolfram classes 1, 2, 3 and 4. Loops represent self-emulations with linear time overhead as a result of using compilers of respective block lengths. In reality this emulation network is all connected, because as we have demonstrated, there are ``jumper'' rules that trespass class boundaries, i.e. emulate a class different from their own, whether of higher or lower complexity. However, for clarity we have segmented the network, and ``jumpers" are shown in a font of a different colour than the colour of their native class. For visualization purposes we only show some interesting cases with different maximum lengths for different classes. The colour of the edges is determined by the number of emulations. The more emulations, the closer to red the edge colour; the fewer possible emulations, the closer to blue (traditional heat colours). The thickness of the edge is inversely related to the minimum block length of the emulation. If the minimal block length is 2, the edge is maximally thick and decreases if the minimal block length increases. In addition, the minimal encoding block length is displayed at each edge. The rule numbers are colour coded. The codes are: blue$=$class 4, red$=$class 3, green$=$class 2, and black$=$class 1. Hub nodes confirm that the attractors are mostly linear--also known as additive CA rules.}
\end{figure}

\begin{figure}
\centering \includegraphics[width=11cm]{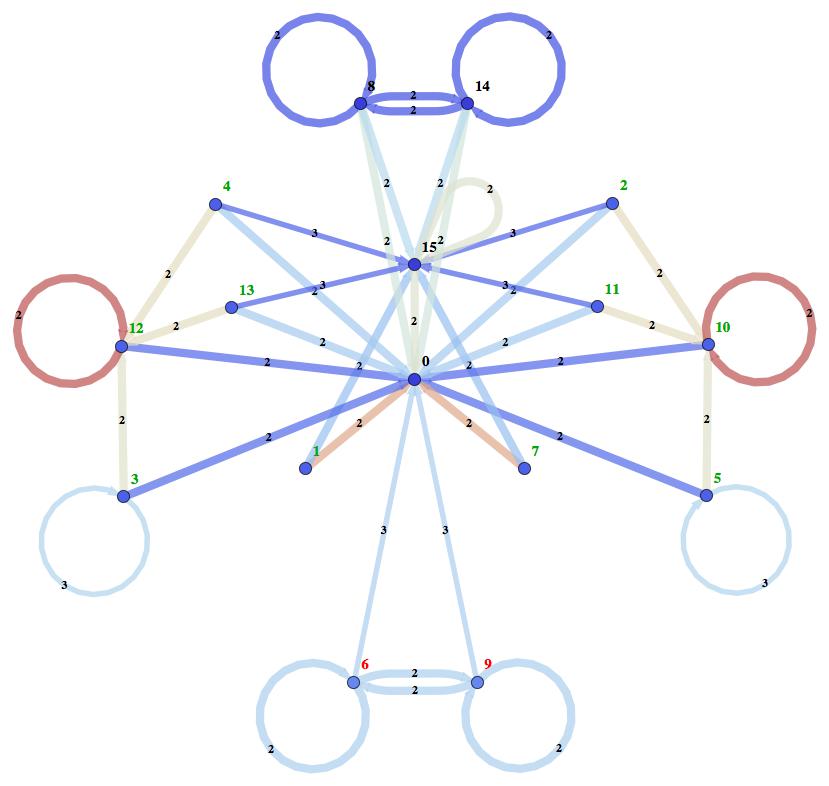}
\caption{\label{fig_PCAClasses}Rule emulation graph of the PCA rule space for block emulators (compilers) up to size 10. Edge weights give the number of emulations; the colour and the width are as in Fig.~\ref{emulationnetwork}. In this emulation network no reduction to essential rules was performed, and it can be seen that the network reflects this by producing a symmetric graph.}
\end{figure}

\begin{figure}
\centering \includegraphics[width=12.0cm,angle=0]{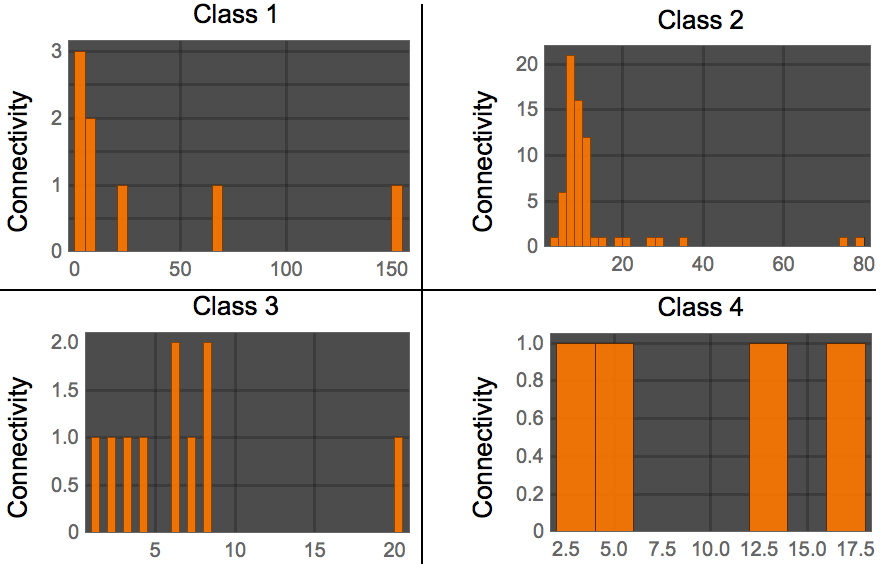}
\caption{\label{complexityECAs} Degree distribution histograms of emulating networks per Wolfram ECA class. In the ECA rule space, the more complex (Wolfram classes 3 and 4) the lower the degree, i.e., they are less frequently emulated up to a fixed compiler size. Rules in simpler Classes 1 and 2, however, are emulated by more ECA $+$ compiler pairs.
}
\end{figure}

\subsection{Algorithmic complexity from emulation frequency}

There are very interesting connections with Algorithmic Probability~\cite{levin,solomonoff}. By the so-called \textit{algorithmic coding theorem}~\cite{levin}, we know that the Kolmogorov (or Kolmogorov-Chaitin) complexity~\cite{kolmo,chaitin} $K$ of an object $x$, defined as the length of the shortest computer program that produces $x$, is inversely proportional to the logarithm (in base 2) of the algorithmic probability of $x$, that is, the probability that $x$ will be produced by a random (\textit{prefix-free}~\footnote{A \textit{prefix-free} Turing machine is a technical requirement imposed on its domain of valid computer programs such that the sum of all possible outcomes is not larger than 1, using the so-called Kraft's inequality. In other words, \textit{prefix-freeness} implies that no valid computer program is the initial segment of any other valid computer program.}) universal Turing machine (a machine whose instruction rule table in binary was chosen by chance with each bit equiprobable).

\begin{myobservation} From emulation results one can devise analytical approximations to the conditional Kolmogorov complexity $K$ of a rule emulated by another with the same conditional complexity. Which illustrates how one can take advantage of emulation data to explore rule complexity.
\end{myobservation}

\begin{mytheorem}
Let $K(f_A)$ and $K(f_B)$ be the Kolmogorov complexity of cellular automata $f_A$ and $f_B$ respectively. $|K(f_A) - K(f_B)| < c$ if $f_A$ is emulated by $B$ with a compiler of size/block length $c$. Therefore $K(f_A) = K(f_B) + c$.
\end{mytheorem}

This implies that $K$ for the emulated rules is roughly the same as that for the emulating rules if nothing else is taken into consideration, such as the complexity of the initial conditions and the difficulty of finding a proper compiler (because the compiler's own complexity is irrelevant, both for the asymptotic behaviour accounting for $K$ of the CA, and in practice, as suggested in Fig.~\ref{fig_ECA_Compiler_Complexity}).

\section{Causality and (re)programmability}

The natural time scale for a CA is the discrete time $t$, i.e. one step per time unit. The CA rules driving the evolution of a CA in time are local rules. This means that the new state of a cell of the next time step $t+$ is only determined by rule $f_{A}$ and the neighbourhood template of the current time step. The cells of the templates are independent of each other and they do not have a memory of the states of cells from the prior time step. However, the set of possible states that influenced the system in some past are called 'light cones'. The name is an analogue of the idea of light cones in special relativity, where space-time is causally decided by a light cone, thus defining regions which are causally connected or causally disconnected. There is also a notion of a `speed of light' in a CA, since information cannot propagate instantaneously from a cell to a far-away cell. In that sense, every cell has its own light cone. It is common to think of cellular automata as physics-like models of computation. As such the 'laws of physics' in a CA are, by homogeneity, the same everywhere, and by locality there is no action at a distance. Thus the whole system is deterministic since the local rules are.

The block transformation of the initial condition of a CA $f_A$ introduces a program. This program is being executed $T$ times per time unit $t$, where $T$ is the block length of the encoding. For $T$ time steps a compiler specific to the chosen block encoding is being executed and leads to an output of the program which defines the new states of cells for time step $T t$. The CA rules are the basic components of the program and the program or compiler is determined by different initial conditions, or in this case by a different block transformation. This is similar to the evolution of the sequence of symbols on the tape of a Turing machine \cite{StephenWolfram1983}.

\subsection{Causal Decomposition}
\label{causal}

Two CA are the same according to the following criteria:

\begin{enumerate}
\item we have a valid transformation showing that the block transformed rule tuples (8 for ECA), i.e. the 'pasts' of the light cones, have only two possible 'futures', which must be the same as the two defined by the block transformation
\item if the 'futures' of the light cones map back to zeros and ones, given by the back transformation, which are equal to the rule encoding of the emulated rule.
\end{enumerate} 

(1) and (2) amount to the statement that the light cones are causally disconnected. Notice we are only able to do this because we know the domain of the ECA and thus the size of the generating rule.

\begin{figure}[h!]
\centering 
  \includegraphics[width=120mm]{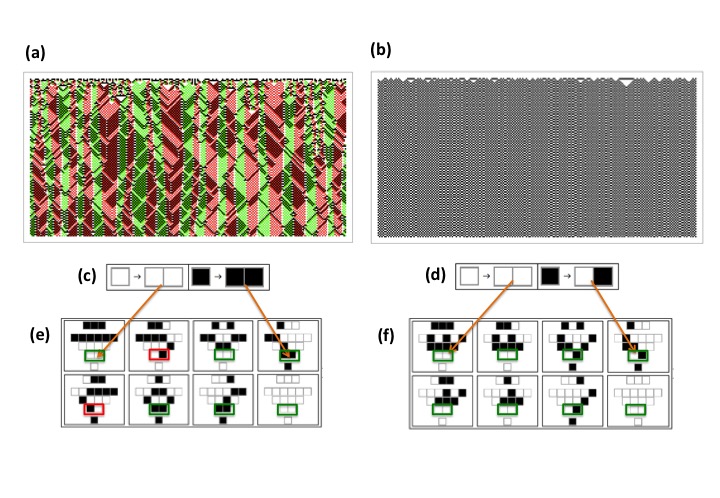}
\caption{\label{Fig_1} Here is shown the decomposition of ECA rule 54 into causally disconnected `light cones' and the condition when a decomposition is mutually exclusive. Subfigures (a), (c) and (e) show an example for a block transformation $P=\Box \rightarrow \Box \Box, \blacksquare \rightarrow \blacksquare \blacksquare$ which does not lead to a valid CA emulation within the same rule space. The red squares in (e) indicate that the 'futures' of the block transformation induced light cones do not all map back to zeroes  $P=\Box$ or ones  $P= \blacksquare$. To demonstrate that the block transformation is leaving the ECA rule space we colour code the tuples in the following way: $\Box \Box \rightarrow \Box$, $\blacksquare \blacksquare \rightarrow \blacksquare$, $\Box \blacksquare \rightarrow \mathcolor{red}{\blacksquare}$ and $\blacksquare \Box \rightarrow \mathcolor{green}{\blacksquare}$, i.e. the new CA actually has 4 colours. In contrast subfigures (b), (d) and (f) show a successful emulation of ECA rule 50 by ECA rule 54 via the block transformation $P=\Box \rightarrow \Box \Box, \blacksquare \rightarrow \Box \blacksquare$. In this case the 'futures' of all the block transformation induced 'light cones' map back to zeroes  $P=\Box$ or ones  $P= \blacksquare$ and therefore the emulated CA is within the same ECA rule space, i.e. maps back to two colours with the same colour coding as above.
\label{fig_emul_visual}
}
\end{figure}

 To visualize the situation we take the $2^{3}=8$ possible compiler components which are induced by a block transformation of an ECA. Each compiler component represents a `light cone' which has a `past', represented by the first row. This row is induced by the block transformation on all 8 tuples of the ECA rule encoding. The last row of the `light cone' is its `future'. In order to produce a valid decomposition which produces mutually exclusive `light cones', i.e. a valid ECA emulation, the `past' must evolve into at most 2 distinct `futures'. Only then will there exist a valid back transformation. Fig.~\ref{Fig_1} shows an example of a block transformation of ECA rule 54 which leaves the ECA rule space (see sub Figures (a), (c) and (e)) and a valid emulation of ECA rule 50 with ECA rule 54  (see subfigures (b), (d) and (f)). We can generalize this to any CA of $k$ states and range $r$:

\begin{cond}
A decomposition of mutually exclusive `light cones' for a given block transformation of a $k$-state and range $r$ CA is possible only if the 'past' of the light cones evolves into at most $k$ distinct `futures' given by the block transformation.
\end{cond}

\subsection{Mapping to a higher colour rule space}

We now expand on the idea of causal decomposition of light cones as described above. The colour coding of the transformation tuples (for block size 2 see Fig.~\ref{fig_emul_visual}) can be viewed as a transformation from a 2-colour rule supace to a 4-colour rule space. Let's now consider such a 4-colour CA with random 4-colour initial condition, and further consider block transformations of the transform $\Box \rightarrow \Box \Box$, $\mathcolor{blue}{\blacksquare} \rightarrow \blacksquare \blacksquare$, $\mathcolor{red}{\blacksquare} \rightarrow \Box \blacksquare$ and $\mathcolor{green}{\blacksquare} \rightarrow \blacksquare \Box$ and all 256 permutations. In addition to the block transformation we also define a one-to-one colour coding of the tuples as a mapping from the 2-colour space to the 4-colour space in the following way: $\Box \Box \rightarrow \Box$, $\blacksquare \blacksquare \rightarrow \mathcolor{blue}{\blacksquare}$, $\Box \blacksquare \rightarrow \mathcolor{red}{\blacksquare}$ and $\blacksquare \Box \rightarrow \mathcolor{green}{\blacksquare}$. Note that the block transformation as a mapping from the 4-colour space to the 2-colour space is not necessarily one-to-one. 

To construct the actual block emulation of a 2-colour CA (ECA) we adopt the following procedure:

\begin{itemize}
\item We start with a random initial 4-colour condition and perform the above block transformation of block size 2.
\item With the now 2-colour initial condition we let the ECA evolve for n steps.
\item From the space-time of the emulated CA we take every 2nd row or lattice and perform the mapping from the 2-colour space to the 4-colour space.
\item The result is a 4-colour emulated CA.
\end{itemize}

The resulting emulated CA generated by all of the 256 block transformations of block size 2 as described above can be classified into two sets. The first (ECA4) is of most interest as it contains all CA with a 2-colour space-time. In general, the second group has a 4-colour space-time. This means that the block transformation also splits into two sets. The block transformation leading to ECA4 transforms an ECA into a 4-colour representation of itself. 

\subsubsection{Mapping 4-colour space-time to 2-colour space-time}

Further insight into which conditions result in a 2-colour CA within the 4-colour rule space can be gained by looking at the 4-colour rule tuples. There are 6 subsets of the 64 rules tuples which are 2-colour. A list of the 6 basis vectors is provided in table \ref{table:2colorbasis}.

\begin{table}[!ht]
  \small
  \centering
\begin{tabular}{p{1cm}|p{6cm}}
 \multicolumn{2}{c}{2-colour basis vectors} \\
 \hline
 colours& basis vector\\
 \hline
 $\Box \mathcolor{blue}{\blacksquare}$& $b_{0,1} = \irow{4^0&4^1&4^4&4^5&4^{16}&4^{17}&4^{20}&4^{21}}$    \\
 $\Box \mathcolor{red}{\blacksquare}$& $b_{0,2} = \irow{4^0&4^2&4^8&4^{10}&4^{32}&4^{34}&4^{40}&4^{42}}$   \\
 $\Box \mathcolor{green}{\blacksquare}$& $b_{0,3} = \irow{4^0&4^3&4^{12}&4^{15}&4^{48}&4^{51}&4^{60}&4^{63}}$  \\
 $\mathcolor{blue}{\blacksquare} \mathcolor{red}{\blacksquare}$& $b_{1,2} = \irow{4^{21}&4^{22}&4^{25}&4^{26}&4^{37}&4^{38}&4^{41}&4^{42}}$  \\
 $\mathcolor{blue}{\blacksquare} \mathcolor{green}{\blacksquare}$& $b_{1,2} = \irow{4^{21}&4^{23}&4^{29}&4^{31}&4^{53}&4^{55}&4^{61}&4^{63}} $   \\
 $\mathcolor{red}{\blacksquare} \mathcolor{green}{\blacksquare} $ & $b_{2,3} = \irow{4^{42}&4^{43}&4^{46}&4^{47}&4^{58}&4^{59}&4^{62}&4^{63}}$\\
 \hline
\end{tabular} 
\caption{\label{table:2colorbasis}List of all 6 basis vectors that divide the 4-colour rule space into a 2-colour rule space equivalent to the ECA rule space.}
\end{table}

\begin{table}[!ht]
  \small
  \centering
\begin{tabular}{ p{2cm}|p{8.5cm} }
 \multicolumn{2}{c}{2-colour basis vectors} \\
 \hline
 basis & basis vector\\
 \hline
 $b_{0,1}$ ($\Box \mathcolor{blue}{\blacksquare}$ )& $\iroww{\irow{0&0&0}&\rightarrow~4^0}, \iroww{\irow{0&0&1}&\rightarrow ~4^1}, \iroww{\irow{0&1&0}&\rightarrow ~4^4}, \iroww{\irow{0&1&1}&\rightarrow ~4^5}$, $\iroww{\irow{1&0&0}&\rightarrow ~4^{16}}, \iroww{\irow{1&0&1}&\rightarrow ~4^{17}}, \iroww{\irow{1&1&0}&\rightarrow ~4^{20}}, \iroww{\irow{1&1&1}&\rightarrow ~4^{21}}$     \\
$b_{0,2}$ ($\Box \mathcolor{red}{\blacksquare}$) & $\iroww{\irow{0&0&0}&\rightarrow ~4^0}, \iroww{\irow{0&0&2}&\rightarrow ~4^2}, \iroww{\irow{0&2&0}&\rightarrow ~4^8}, \iroww{\irow{0&2&2}&\rightarrow ~4^10}$, $\iroww{\irow{2&0&0}&\rightarrow ~4^{32}}, \iroww{\irow{2&0&1}&\rightarrow ~4^{34}}, \iroww{\irow{2&2&0}&\rightarrow ~4^{40}}, \iroww{\irow{2&2&2}&\rightarrow ~4^{42}}$    \\
$b_{0,3}$ ($\Box \mathcolor{green}{\blacksquare}$)& $\iroww{\irow{0&0&0}&\rightarrow ~4^0}, \iroww{\irow{0&0&3}&\rightarrow ~4^3}, \iroww{\irow{0&3&0}&\rightarrow ~4^{12}}, \iroww{\irow{0&3&3}&\rightarrow ~4^{15}}$, $\iroww{\irow{3&0&0}&\rightarrow ~4^{48}}, \iroww{\irow{3&0&3}&\rightarrow ~4^{51}}, \iroww{\irow{3&3&0}&\rightarrow ~4^{60}}, \iroww{\irow{3&3&3}&\rightarrow ~4^{21}}$   \\
$b_{1,2}$ ($\mathcolor{blue}{\blacksquare} \mathcolor{red}{\blacksquare})$ &$\iroww{\irow{1&1&1}&\rightarrow ~4^{21}}, \iroww{\irow{1&1&2}&\rightarrow ~4^{22}}, \iroww{\irow{1&2&1}&\rightarrow ~4^{25}}, \iroww{\irow{1&2&2}&\rightarrow ~4^26}$, $\iroww{\irow{2&1&1}&\rightarrow ~4^{37}}, \iroww{\irow{2&1&2}&\rightarrow ~4^{38}}, \iroww{\irow{2&2&1}&\rightarrow ~4^{42}}, \iroww{\irow{2&2&2}&\rightarrow ~4^{42}}$  \\
$b_{1,3}$ ($\mathcolor{blue}{\blacksquare} \mathcolor{green}{\blacksquare}$)& $\iroww{\irow{1&1&1}&\rightarrow ~4^{21}}, \iroww{\irow{1&1&3}&\rightarrow ~4^{23}}, \iroww{\irow{1&3&1}&\rightarrow ~4^{29}}, \iroww{\irow{1&3&3}&\rightarrow ~4^{31}}$, $\iroww{\irow{3&1&1}&\rightarrow ~4^{53}}, \iroww{\irow{3&1&3}&\rightarrow ~4^{55}}, \iroww{\irow{3&3&1}&\rightarrow ~4^{61}}, \iroww{\irow{3&3&3}&\rightarrow ~4^{63}}$    \\
$b_{2,3}$ ($\mathcolor{red}{\blacksquare} \mathcolor{green}{\blacksquare}$)  & $\iroww{\irow{2&2&2}&\rightarrow ~4^{42}}, \iroww{\irow{2&2&3}&\rightarrow ~4^{43}}, \iroww{\irow{2&3&2}&\rightarrow ~4^{46}}, \iroww{\irow{2&3&3}&\rightarrow ~4^{47}}$, $\iroww{\irow{3&2&2}&\rightarrow ~4^{58}}, \iroww{\irow{3&2&3}&\rightarrow ~4^{59}}, \iroww{\irow{3&3&2}&\rightarrow ~4^{62}}, \iroww{\irow{3&3&3}&\rightarrow ~4^{63}}$ \\
 \hline
\end{tabular}
\caption{\label{table:2colorbasisrep}Alternative representation 1.}
\end{table}

Each of the 6 bases map to tuples of size 3 containing 2 colours. Given the specific encoded $\Box \rightarrow 0, \mathcolor{blue}{\blacksquare} \rightarrow 1, \mathcolor{red}{\blacksquare} \rightarrow 2$ and $\mathcolor{green}{\blacksquare} \rightarrow$ there are 6 distinct sets of tuples of size 3. For example, the tuples with colours $\Box \mathcolor{blue}{\blacksquare}$ map to the powers of 4 as  $\{0, 0, 0\} \rightarrow 4^0, \{0, 0, 1\} \rightarrow 4^1, \{0, 1, 0\} \rightarrow 4^4, \{0, 1, 1\} \rightarrow 4^5, \{1, 0, 0\} \rightarrow 4^{16}, \{1, 0, 1\} \rightarrow 4^{17}, \{1, 1, 0\} \rightarrow 4^{20}, \{1, 1, 1\} \rightarrow 4^{21}$. A complete list of all 6 basis vectors is given in table \ref{table:2colorbasisrep}. 
Each of the basis vectors divides the 4-colour rule space into a 2-colour rule space equivalent to the ECA rule space. All ECA rules can be recovered with the respective colour coding of the basis. The transformation from the ECA rule space to the the 2 colour subspace is defined as follows:

\begin{defn}
	Let $\Re[\oplus_{2}]$ be the binary representation in vector form of an ECA rule number, $\Re[\oplus_{4}]$ the quarternary representation in vector form of a 4-colour CA rule number, having the same range $r=1$,  and let $b_{x,y}^2$ be the basis vectors of the 2-colour subset of the 4-colour CA rule transition function with colours $x, y \in \Sigma$, where $\Sigma$ is an alphabet composed of all integers modulo $s$: $\Sigma = \mathbb{Z}_s = \{0,1,2,3\}$. Then the mapping from an ECA rule to an equivalent 4-colour CA rule is defined as:
\begin{equation}
 	\Re[\oplus_{4}]= \Re[\oplus_{2}] \bullet  \vec{b_{x,y}^2}
\end{equation}
\end{defn}

This means that each ECA rule has 6 distinct representations. For example, the ECA rule 50 can be transformed as: 
\begin{equation}
	\vec {50_{4}} = \vec{50_{2}} \bullet b_{0,1}^2=\icol{0\\1\\0\\0\\1\\1\\0\\0} \bullet \icol{4^0\\4^1\\4^4\\4^5\\4^{16}\\4^{17}\\4^{20}\\4^{21}} = 4 + 4^{16} + 4^{17} = 21474836484
\end{equation}

All 6 4-colour representations of ECA rule 50 are listed in table \ref{table:2colorto4color}.

\begin{table}[!ht]
  \tiny
\begin{tabular}{ p{1cm}|p{12cm} }
 \multicolumn{2}{c}{2-colour to 4-colour mapping} \\
 \hline
 colors& basis vector\\ 
 \hline
 \\
 $\Box \mathcolor{blue}{\blacksquare}$& $\vec {54_{4}} = \vec{54_{2}} \bullet b_{0,1}=\icol{0\\1\\0\\0\\1\\1\\0\\0} \bullet \icol{4^0\\4^1\\4^4\\4^5\\4^{16}\\4^{17}\\4^{20}\\4^{21}} = 4 + 4^{16} + 4^{17} $    \\[10ex]
 $\Box \mathcolor{red}{\blacksquare}$& $\vec {54_{4}} = \vec{54_{2}} \bullet b_{0,1}=\icol{0\\1\\0\\0\\1\\1\\0\\0} \bullet 2 \icol{4^0\\4^2\\4^8\\4^{10}\\4^{32}\\4^{34}\\4^{40}\\4^{42}} = 4^2 + 4^8 + 4^{32} + 4^{34} $   \\[10ex]
 $\Box \mathcolor{green}{\blacksquare}$& $\vec {54_{4}} = \vec{54_{2}} \bullet b_{0,1}=\icol{0\\1\\0\\0\\1\\1\\0\\0} \bullet 3 \icol{4^0\\4^3\\4^12\\4^{15}\\4^{48}\\4^{51}\\4^{60}\\4^{63}} = 4 + 4^4 + 4^{16} + 4^{17} $  \\ [10ex]
 $\mathcolor{blue}{\blacksquare} \mathcolor{red}{\blacksquare}$& $\vec {54_{4}} = \vec{54_{2}} \bullet b_{0,1}=\icol{0\\1\\0\\0\\1\\1\\0\\0} \bullet \icol{4^{21}\\2\cdot4^{22}\\4^{25}\\4^{26}\\2\cdot4^{37}\\ 2\cdot4^{38}\\4^{41}\\4^{42}} = 4^{21} + 2\cdot4^{22} + 4^{25} + 4^{26} + 2\cdot4^{37} +  2\cdot4^{38} + 4^{41}  + 4^{42}$  \\ [10ex]
 $\mathcolor{blue}{\blacksquare} \mathcolor{green}{\blacksquare}$& $\vec {54_{4}} = \vec{54_{2}} \bullet b_{0,1}=\icol{0\\1\\0\\0\\1\\1\\0\\0} \bullet \icol{4^{21}\\3\cdot4^{23}\\4^{29}\\4^{31}\\3\cdot4^{53}\\3\cdot4^{55}\\4^{61}\\4^{63}} =  4^{21} + 3\cdot4^{23} + 4^{29} + 4^{31} + 3\cdot4^{53} + 3\cdot4^{55} + 4^{61} + 4^{63} $   \\ [10ex]
 $\mathcolor{red}{\blacksquare} \mathcolor{green}{\blacksquare} $ & $\vec {54_{4}} = \vec{54_{2}} \bullet b_{0,1}=\icol{0\\1\\0\\0\\1\\1\\0\\0} \bullet \icol{2\cdot4^{42}\\3\cdot4^{43}\\2\cdot4^{46}\\ 2\cdot4^{47}\\3\cdot4^{58}\\3\cdot4^{59}\\ 2\cdot4^{62}\\2\cdot4^{63}} = 2\cdot4^{42} + 3\cdot4^{43} + 2\cdot4^{46} + 2\cdot4^{47} + 3\cdot4^{58} + 3\cdot4^{59} + 2\cdot4^{62} + 2\cdot4^{63}$ \\ [10ex]
 \hline
\end{tabular} 
\caption{\label{table:2colorto4color}Alternative representation 2.}
\end{table}

\subsubsection{4-colour representations of ECA rule emulation}
\label{rule50}

By applying all 256 possible block transformations of block size 2 as described, we find that block transformations which map a 4-colour CA to a 2-colour CA, i.e. map to either of the 6 possible subsets of the rule space which map to 2-colour tuples, are representations of ECA rules. In fact there are just the allowed emulations of an ECA rule under a block transformation of block size 2. In the case of ECA rule 54 there are 2 distinct 2-colour mappings which contain the transformation tuples $\Box \Box$, $\Box \blacksquare$ and $\blacksquare \Box$. In Section~\ref{rule50}, we showed that the block transformations $\Box \rightarrow \Box \Box$, $\blacksquare \rightarrow \Box \blacksquare $ and $\Box \rightarrow \Box \Box$, $\blacksquare \rightarrow \blacksquare \Box$ allow ECA rule 54 to emulate ECA rule 50. These two distinct block transformations in the 2-colour space have their equivalents in the 4-colour space, i.e. \\

\begin{center}
$\Box  \rightarrow \Box \Box, \mathcolor{blue}{\blacksquare}  \rightarrow \Box \Box, \mathcolor{red}{\blacksquare}  \rightarrow \Box \Box,  \mathcolor{green}{\blacksquare}  \rightarrow \Box \blacksquare$ \\
$\Box  \rightarrow \Box \Box, \mathcolor{blue}{\blacksquare}  \rightarrow \Box \Box, \mathcolor{red}{\blacksquare}  \rightarrow \Box \blacksquare,  \mathcolor{green}{\blacksquare}  \rightarrow \Box \blacksquare$ \\
$\Box  \rightarrow \Box \Box, \mathcolor{blue}{\blacksquare}  \rightarrow \Box \blacksquare, \mathcolor{red}{\blacksquare}  \rightarrow \Box \blacksquare,  \mathcolor{green}{\blacksquare}  \rightarrow \Box \blacksquare$\\
$\Box  \rightarrow \Box \Box, \mathcolor{blue}{\blacksquare}  \rightarrow \Box \Box, \mathcolor{red}{\blacksquare}  \rightarrow \Box \Box,  \mathcolor{green}{\blacksquare}  \rightarrow \blacksquare \Box$ \\
$\Box  \rightarrow \Box \Box, \mathcolor{blue}{\blacksquare}  \rightarrow \Box \Box, \mathcolor{red}{\blacksquare}  \rightarrow \blacksquare \Box,  \mathcolor{green}{\blacksquare}  \rightarrow \blacksquare \Box$ \\
$\Box  \rightarrow \Box \Box, \mathcolor{blue}{\blacksquare}  \rightarrow \blacksquare \Box, \mathcolor{red}{\blacksquare}  \rightarrow  \blacksquare \Box,  \mathcolor{green}{\blacksquare}  \rightarrow \blacksquare \Box$
\end{center}

\noindent and their permutations. 

\begin{figure}[h!]
\centering 
  \includegraphics[width=120mm]{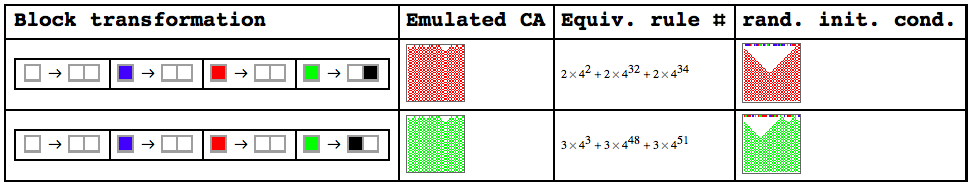}
\caption{\label{fig_2} 
\label{fig_emul_4color}
Showing two among 28 possible block transformations of ECA rule 54 to the 4-colour rule space leading to 4-colour equivalent CA rules of the emulated ECA rule 50. Column 1 contains the emulated CA's space-time, column 2 the space-time diagram of column 1 for a given initial condition, column 3 the quarternary representation and column 4 its space-time with random initial conditions.
}
\end{figure}

In total one gets 28 valid block transformations that lead from a 4-colour rule space to a 2-colour subspace. Fig.~\ref{fig_emul_4color} shows two of 28 possible block transformations with the emulated space-time, the quarternary representation and its space-time with random initial conditions. 

\subsubsection{4-colour representation and causal decomposition}

\begin{figure}[h!]
\centering 
  \includegraphics[width=120mm]{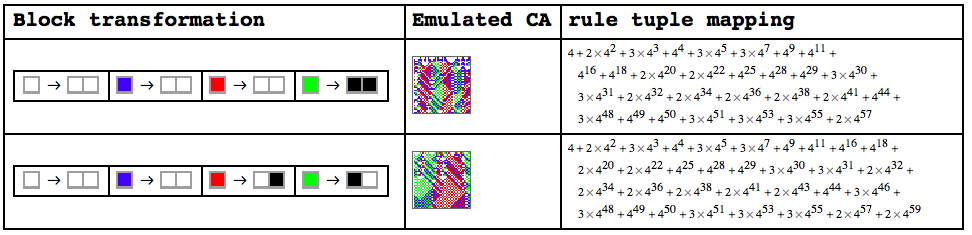}
\caption{\label{fig_3} 
\label{fig_emul_4color2}
Showing two of 228 possible block transformations of ECA rule 54 to the 4-colour rule space leading to 4-colour CA rules which are not equivalent emulations of ECA rule 54. Column 1 contains the space-time of the emulated CA, column 2 the space-time diagram of column 1 for given initial conditions, and column 3 the rule tuple mappings of the space-time in column 2.
}
\end{figure}

On the other hand, the remaining 228 (256-28) block transformations lead to CA which do not map exclusively to one or all 2-colour 6 basis vectors, i.e subsets of the 4-colour rule space. These transformations do not amount to representations of ECA emulations. In the case of ECA rule 54 Fig.~\ref{fig_emul_4color2} shows two such block transformations. Note that the rule tuple mapping contains a pure 2-colour basis and is directly sensitive to the choice of an initial condition. In contrast, this is not the case for the block transformation discussed above, meaning that it is not sensitive to initial conditions. 

\begin{figure}[h!]
\centering 
  \includegraphics[width=120mm]{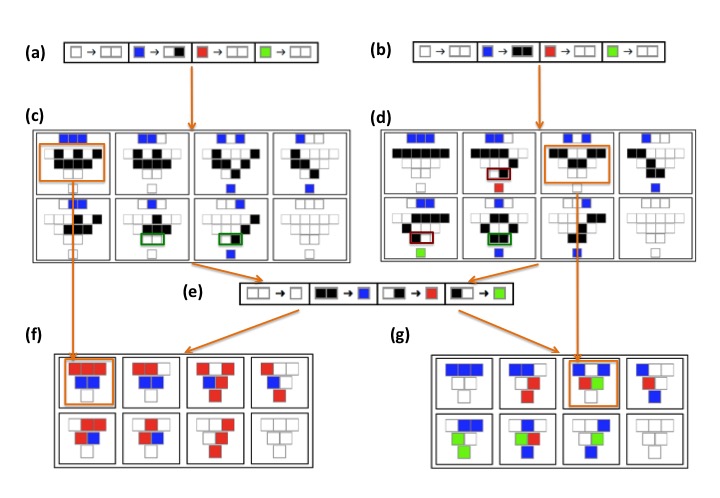}
\caption{\label{fig_4} 
\label{fig_emul_4color2Compiler}
In the case of ECA rule 54, two examples of block transformations which lead to mapping from the 4-colour space to the 2-colour subspace (left) and another which doesn't (right). Starting with the 8 rule tuples (here we have chosen only tuples of the 2-colour basis with colours $\Box \mathcolor{blue}{\blacksquare}$, but others are possible as well), applying the block transformations (a) or (b) one induces a compiler (c) or (d). Only (a) leads to an emulation which maps back to a 2-colour basis. In contrast, (b) does not. This can be best seen in the compiler space-times (f) and (g), which are mapped back via the colour mapping (e). 
}
\end{figure}

To visualize this situation better we look at the complier which is induced by the block transformations discussed above. In Fig.~\ref{fig_emul_4color2Compiler} we show two examples of block transformations which lead to mapping from the 4-colour space to the 2-colour subspace (left) and another which doesn't (right). Starting with the 8 rule tuples (here we have chosen only tuples of the 2-colour basis with colours $\Box \mathcolor{blue}{\blacksquare}$, but others are possible as well), applying the block transformations (a) or (b) one induces a compiler (c) or (d). Only (a) leads to an emulation which maps back to a 2-colour basis. In contrast, (b) does not. This can be best seen in the compiler space times (f) and (g), which are mapped back via the colour mapping (e).

It is precisely this robustness which indicates that the causal decomposition is induced by a certain block transformation which leads to a valid block transformation of an ECA. However, the sensitivity of the rule tuple mapping in the 4-colour space due to initial conditions indicates that a block transformation doesn't induce a causal decomposition. This means that this block transformation is not a valid emulation of an ECA.

\subsubsection{Generalization to block sizes $n$}

We now generalize the above discussion to block transformations of block size $n$ of $2$-colour CA with range $r$. There are $\frac{2^{n^n}-2^n}{2}$ possible $k$-colour basis vectors.

\begin{defn}
	Let $\Re[\oplus_{2}]$ be the binary representation in vector form of a CA rule number, $\Re[\oplus_{2^n}]$ the $2^n$-ary representation in vector form of a $2^n$-colour CA rule number ($n$ is the block size), having the same range $r$,  and $b_{2}$ be the basis vectors of the $k$-colour subset of the $2^n$-colour CA rule transition function with colours $2^n \in \Sigma$, where $\Sigma$ is an alphabet composed of all integers modulo $s$: $\Sigma = \mathbb{Z}_s = \{0,...,2^n\}$. Then the mapping from a $2$-colour rule to an equivalent $k^n$-colour CA rule is defined as:
\begin{equation}
 	\Re[\oplus_{2^n}]= \Re[\oplus_{2}] \bullet \vec{b^{2}}
\end{equation}
\end{defn}

We now show how to construct the $2^n$-colour basis. First we start with a function $\lambda_{\ell,m}$ calculating the difference between base tuples of size $n$, block size, and colour $2^n$ which only contain a single $1$ at position $i \ge 1$ to $i \le n$. With $\ell = 2^n$ we have:

\begin{equation}
 	\lambda_{\ell,m}= \frac{(\ell-2) \ell^m+1}{\ell-1}; \ell=2^n
\end{equation}

Another function $\delta(\ell,m)$ which we will use to encode values on a vector returns its first argument only if its second argument is $0$. Otherwise it returns the value $0$.
\begin{equation}
        \delta(\ell,m)  =  \left\{\begin{array}{l@{\quad}l}
                        		m=0 \rightarrow \ell \\
        				m \neq 0 \rightarrow 0
\end{array}\right\}
\end{equation}

Based on the above two functions we define a vector $\vec{\Delta}$ of size $2^{(n-1)}$ which contains the differences between all basis tuples containing only binary digits. Here $r$ again is the range:
\begin{equation}
\vec{\Delta} = \sum_{m=1}^{r}  \icol{\delta(\lambda_{\ell,m},1 \bmod{2^{m-1}})  \\ \delta(\lambda_{\ell,m},2 \bmod{2^{m-1}})  \\\vdots\\\\ \delta(\lambda_{\ell,m},(2^{r}-1) \bmod{2^{m-1}}) } - \icol{\delta(\lambda_{m,\ell},1 \bmod{2^m}  \\ \delta(\lambda_{\ell,m},2 \bmod{2^m}  \\\vdots\\\\ \delta(\lambda_{\ell,m},(2^{r}-1) \bmod{2^m} }
\end{equation}

With this we can calculate the components of the basis vector $b_{0,1}^\ell$ of colour $\ell$:

\begin{equation}
	b_{0,1}^\ell = \icol{\ell^0\\ \ell^{\sum_{i=1}^{1}  \Delta_i} \\  \ell^{\sum_{i=1}^{2} \Delta_i}  \\ \vdots \\ \ell^{\sum_{i=1}^{2^r-1} \Delta_i} } 
\end{equation}

To get the complete set of $\frac{2^{n^n}-2^n}{2}$ basis vectors having colour $2^n$, we can apply the following function, where $x,y$ are the possible colour pairs, where $[b_{0,1}^\ell]_{2^r}$ means the $2^r$ component of vector $b_{0,1}^\ell$:
\begin{equation}
	b_{x,y}^\ell = x ([b_{0,1}^\ell]_{2^r} -1)+(y-x) b_{0,1}^\ell - (y-x-1), \: \text{with} \: x,y \in \{0,n-1\} \: \text{and} \: x < y
\end{equation}

In the case of block size $n=2$ we have $\ell=2^2=4$, and we get:

\begin{equation}
	\lambda_{4,0}=1; \lambda_{4,1}=3; \lambda_{4,2}=11
\end{equation}

\begin{equation}
\vec{\Delta} = \icol{1 \\ 1 \\1 \\1 \\1 \\1 \\1}-\icol{0 \\ 1 \\0 \\1 \\0 \\1 \\0}+\icol{0 \\ 3 \\0 \\3 \\0 \\3 \\0}-\icol{0 \\ 0 \\0 \\11 \\0 \\0 \\0} = \icol{1 \\ 3 \\1 \\11 \\1 \\3 \\1}
\end{equation}

\begin{equation}
	b_{0,1}^\ell = \icol{ \ell^0\\ \ell^{\sum_{i=1}^{1} \Delta_i} \\ \ell^{\sum_{i=1}^{2} \Delta_i} \\ \ell^{\sum_{i=1}^{3} \Delta_i} \\ \ell^{\sum_{i=1}^{4} \Delta_i} \\ \ell^{\sum_{i=1}^{5} \Delta_i} \\ \ell^{\sum_{i=1}^{6} \Delta_i} \\ \ell^{\sum_{i=1}^{7} \Delta_i}}=
	\icol{4^0\\4^1\\4^4\\4^5\\4^{16}\\4^{17}\\4^{20}\\4^{21}} 
\end{equation}

In the case of block size $n=3$ we have $\ell=2^3=8$ and we get:

\begin{equation}
	\lambda_{8,0}=1; \lambda_{8,1}=7; \lambda_{8,2}=55
\end{equation}

\begin{equation}
\vec{\Delta} = \icol{1 \\ 1 \\1 \\1 \\1 \\1 \\1}-\icol{0 \\ 1 \\0 \\1 \\0 \\1 \\0}+\icol{0 \\ 7 \\0 \\7 \\0 \\7 \\0}-\icol{0 \\ 0 \\0 \\55 \\0 \\0 \\0} = \icol{1 \\ 7 \\1 \\55 \\1 \\7 \\1}
\end{equation}

\begin{figure}[h!]
\centering 
  \includegraphics[width=125mm]{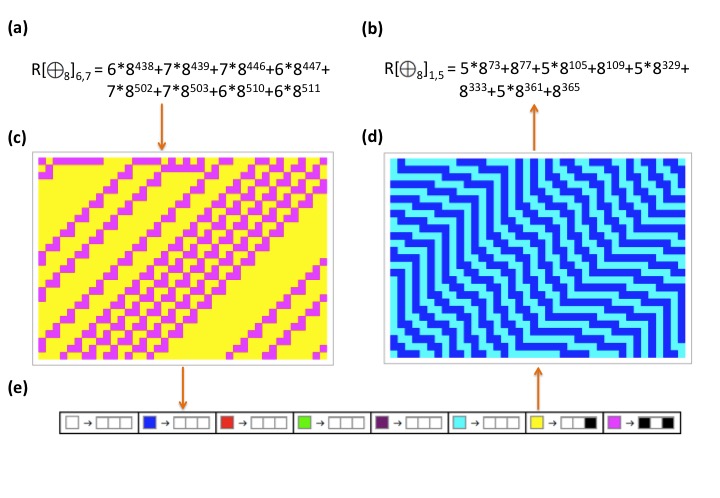}
\caption{\label{fig_5} 
\label{fig_basis_transform}
Here is shown the 8-colour CA rule number which is equivalent to ECA rule 38 (a) and the 8-colour CA rule number which is equivalent to emulated ECA 85 of ECA rule 38 via the block transformation (b). Space-time diagram (c) shows CA (a) under initial condition just using the base digits (6,7) and space-time diagram (d) shows the output of every 3rd step of the emulated CA (with the same initial condition) (b). In contrast to (c) the space-time diagram only contains digits (1,5). Therefore the emulation maps the 2-colour basis $b_{6,7}^8$ to the 2-colour basis $b_{1,5}^8$. 
}
\end{figure}

\begin{equation}
	b_{0,1}^8 = \icol{8^0\\8^1\\8^8\\8^9\\8^{64}\\8^{65}\\8^{72}\\8^{73}} 
\end{equation}

As an example for the general case we now show a block emulation of ECA 38 with block size 3. This is an example of a 2-colour CA which projects via the block transformation of block size 3 to an $8$-colour CA. Since $n=3$ and colour $k=2$ we have $\ell=2^3=8$. We choose the following block transformation:
$\Box  \rightarrow \Box \Box\Box \Box, \mathcolor{blue}{\blacksquare}  \rightarrow \Box \Box\Box \Box, \mathcolor{red}{\blacksquare}  \rightarrow \Box \Box\Box \Box,  \mathcolor{green}{\blacksquare}  \rightarrow \Box \Box \Box \Box, \mathcolor{purple}{\blacksquare}   \rightarrow \Box \Box\Box \Box, \mathcolor{cyan}{\blacksquare}  \rightarrow \Box \Box\Box \Box, \mathcolor{yellow}{\blacksquare}  \rightarrow \Box \Box \Box \blacksquare,  \mathcolor{magenta}{\blacksquare}  \rightarrow \blacksquare \Box \Box \blacksquare$.

This block transformation maps ECA rule 38 to an $8$-colour CA which is equivalent to ECA rule 85. More precisely, the block transformation maps from the $8$ colour basis $b_{6,7}^8$ to the basis $b_{1,5}^8$. This can be shown as follows:\\

First, we take the basis vector $b_{0,1}^8$ and transform it into the basis vector $b_{1,5}^8$

\begin{equation}
	b_{1,5}^8 = 73+4 b_{0,1}^8-4=\icol{8^{73}\\8^{77}\\8^{105}\\8^{109}\\8^{329}\\8^{333}\\8^{361}\\8^{365}}
\end{equation}

Then we transform the rule number vector in binary representation $\Re^{-1}[\oplus_{2}]_{0,1}$ into its quarternary representation $\Re[\oplus_{2}]_{x,y}$ with colours $x,y$, where $\Re^{-1}[\oplus_{2}]_{0,1}$ is the binary inverse of $\Re^{-1}[\oplus_{2}]_{0,1}$ 
 \begin{equation}
	\Re[\oplus_{2}]_{x,y}  = x \Re^{-1}[\oplus_{2}]_{0,1}  + y \Re[\oplus_{2}]_{0,1} 
\end{equation},
or in this particular case choosing $x=1$ and $y=5$
\begin{equation}
	\Re[\oplus_{2}]_{1,5}  = 1 \Re^{-1}[\oplus_{2}]_{0,1} + 5 \Re[\oplus_{2}]_{0,1} 
\end{equation}

Using the general formula for calculating the $\ell$-ary rule number $R[\oplus_{\ell}]_{x,y}$
\begin{equation}
	R[\oplus_{\ell}]_{x,y}  = \Re[\oplus_{2}]_{x,y} \bullet b_{x,y}^\ell
\end{equation}
we calculate the quarternary rule number $R[\oplus_{8}]_{1,5} $ in basis $1,5$
\begin{equation}
	R[\oplus_{8}]_{1,5}  = \icol{5\\ 1\\ 5\\ 1\\ 5\\ 1\\ 5\\ 1} \bullet \icol{8^{73}\\8^{77}\\8^{105}\\8^{109}\\8^{329}\\8^{333}\\8^{361}\\8^{365}}=5*8^{73}+8^{77}+5*8^{105}+8^{109}+5*8^{329}+8^{333}+5*8^{361}+8^{365}
\end{equation}

Fig.~\ref{fig_basis_transform} shows the above described emulation of ECA rule 38 emulating ECA rule 85 with a block transformation of size $n$. The 8-colour CA rule number which is equivalent to ECA rule 38 (a) and the 8-colour CA rule number which is equivalent to emulated ECA 85 of ECA rule 38 via the block transformation (b). Space-time diagram (c) shows CA (a) under initial conditions, simply using the base digits (6,7), and space-time diagram (d) shows the output of every 3rd step of the emulated CA (with the same initial condition) (b). In contrast to (c) the space-time diagram only contains digits (1,5). Therefore the emulation maps the 2-colour basis $b_{6,7}^8$ to the 2-colour basis $b_{1,5}^8$. 
We now conclude with the following observation:

\begin{obs}
There exists only one valid block transformation of size $n$ for 2-colour CA A to emulate 2-colour CA B, both with range $r$ if there exists a mapping from one 8-colour basis $b_{x,y}^8$ to another 8-colour basis $b_{x',y'}^8$ for every $n$th step of the emulated space-time.
\end{obs}

We have thus shown how to interpret a 2-colour CA block transformation as $2^n$-colour CA with rule functions only containing 2-colour tuples forming a 2-colour basis of the $2^n$-colour rule space. One could, in a similar fashion, expand the results to investigate the even more general case of a $k$-colour CA block transformation mapping to a $k^n$-colour rule space. One may also devise an algebraic approach to finding all possible block transformations of a $k$-colour CA with range $r$.

\end{document}